%% file: scifile.tex
\documentclass[12pt]{article}

\usepackage{scicite}

\usepackage{times}
\usepackage{comment}
\usepackage{graphics}
\usepackage{graphicx}

\newcommand{\SM}{Supplementary Materials}
\newcommand{\apj}{\textit{Astrophys. J.}}

\newcommand{\mnras}{\textit{Mon. Not. R. Astron. Soc.}}

\newcommand{\kepler}{\emph{Kepler}}
\newcommand{\gaia}{\emph{Gaia}}
\newcommand{\luna}{{\tt LUNA}}
\newcommand{\multi}{{\sc MultiNest}}
\newcommand{\ttvfaster}{{\tt TTVfaster}}
\newcommand{\exotransmit}{{\tt Exo-Transmit}}
\newcommand{\forecaster}{{\tt forecaster}}
\newcommand{\emcee}{{\tt emcee}}
\newcommand{\cofiam}{{\tt CoFiAM}}

\newcommand{\local}{{\tt local}}

\newcommand{\pdc}{{\tt PDC}}
\newcommand{\sap}{{\tt SAP}}
\newcommand{\isochrones}{{\tt isochrones}}

\newenvironment{sciabstract}{%
\begin{quote} \bf}
{\end{quote}}

\newcounter{lastnote}
\newenvironment{scilastnote}{%
\setcounter{lastnote}{\value{enumiv}}%
\addtocounter{lastnote}{+1}%
\begin{list}%
{\arabic{lastnote}.}
{\setlength{\leftmargin}{.22in}}
{\setlength{\labelsep}{.5em}}}
{\end{list}}

\title{Evidence for a Large Exomoon Orbiting Kepler-1625b} 

\author
{Authors: Alex Teachey,$^{1\ast}$ David M. Kipping$^{1}$\\
\\
\textbf{Affiliations:}
\normalsize{$^{1}$Department of Astronomy, Columbia University in the City of New York}\\
\\
\normalsize{$^\ast$E-mail: ateachey@astro.columbia.edu.}
}

\date{}

\begin{document} 


\baselineskip24pt


\maketitle 

\textbf{Short title:} Evidence for a Large Exomoon Orbiting Kepler-1625b

\textbf{One Sentence Summary:} Hubble Space Telescope observations show a timing offset and an exomoon-like transit associated with a Jupiter-sized planet.

\begin{sciabstract}
\textbf{Abstract:} Exomoons are the natural satellites of planets orbiting stars outside our solar system, of which there are currently no confirmed examples. We present new observations of a candidate exomoon associated with Kepler-1625b using the Hubble Space Telescope, to validate or refute the moon's presence. We find evidence in favor of the moon hypothesis, based on timing deviations and a flux decrement from the star consistent with a large transiting exomoon. Self-consistent photodynamical modeling suggests that the planet is likely several Jupiter masses, while the exomoon has a mass and radius similar to Neptune. Since our inference is dominated by a single but highly precise Hubble epoch, we advocate for future monitoring of the system to check model predictions and confirm repetition of the moon-like signal.
\end{sciabstract}


\clearpage




\section*{Introduction}
The search for exomoons remains in its infancy. To date, there are no confirmed
exomoons in the literature, although an array of techniques have been proposed
to detect their existence, such as microlensing \cite{han:2002,han:2008,
liebig:2010}, direct imaging \cite{cabrera:2007,agol:2015}, cyclotron radio
emission \cite{noyola:2014}, pulsar timing \cite{lewis:2008} and transits
\cite{sartoretti:1999,kipping:2009a,kipping:2009b}. The transit method is
particularly attractive however since many small planets down to lunar radius
have already been detected \cite{barclay:2013}, and transits afford repeated
observing opportunities to further study candidate signals.

Previous searches for transiting moons have established that Galilean-sized
moons are uncommon at semimajor axes of 0.1 to 1 astronomical unit (AU) \cite{teachey:2018}. This
result is consistent with theoretical work that has shown that the shrinking
Hill sphere \cite{namouni:2010} and potential capture into evection resonances
\cite{spalding:2016} during a planet's inward migration could efficiently
remove primordial moons. Nevertheless, amongst a sample of 284 transiting
planets recently surveyed for moons, one planet did show some evidence for a
large satellite, Kepler-1625b \cite{teachey:2018}. The planet is a
Jupiter-sized validated world \cite{morton:2016} orbiting a solar-mass star
\cite{mathur:2017} close to 1\,AU in a likely circular path
\cite{teachey:2018}, making it a prime \textit{a priori} candidate for moons.
On this basis, and the hints seen in the three transits observed by \kepler, we
requested and were awarded time on the Hubble Space Telescope (HST) to
observe a fourth transit expected on 28 to 29 October 2017. In this work, we
report on these new observations and their impact on the exomoon
hypothesis for Kepler-1625b.


\section*{Materials and Methods}
Our original analysis was the product of a multiyear survey and thus utilized
an earlier version of the processed photometry released by the \kepler\ Science
Operations Center (SOC). In that study \cite{teachey:2018}, we used the simple
aperture photometry (\sap) from SOC pipeline version 9.0 \cite{jenkins:2010},
but the most recent and final data release uses version 9.3. In this work we
reanalyzed the \kepler\ data using the revised photometry, which includes
updated aperture contamination factors that also affect our analysis. During
this process, we also investigated the effect of varying the model used to
remove a long-term trend present in the \kepler\ data.

\begin{figure}
\begin{center}
\includegraphics[width=16.0cm]{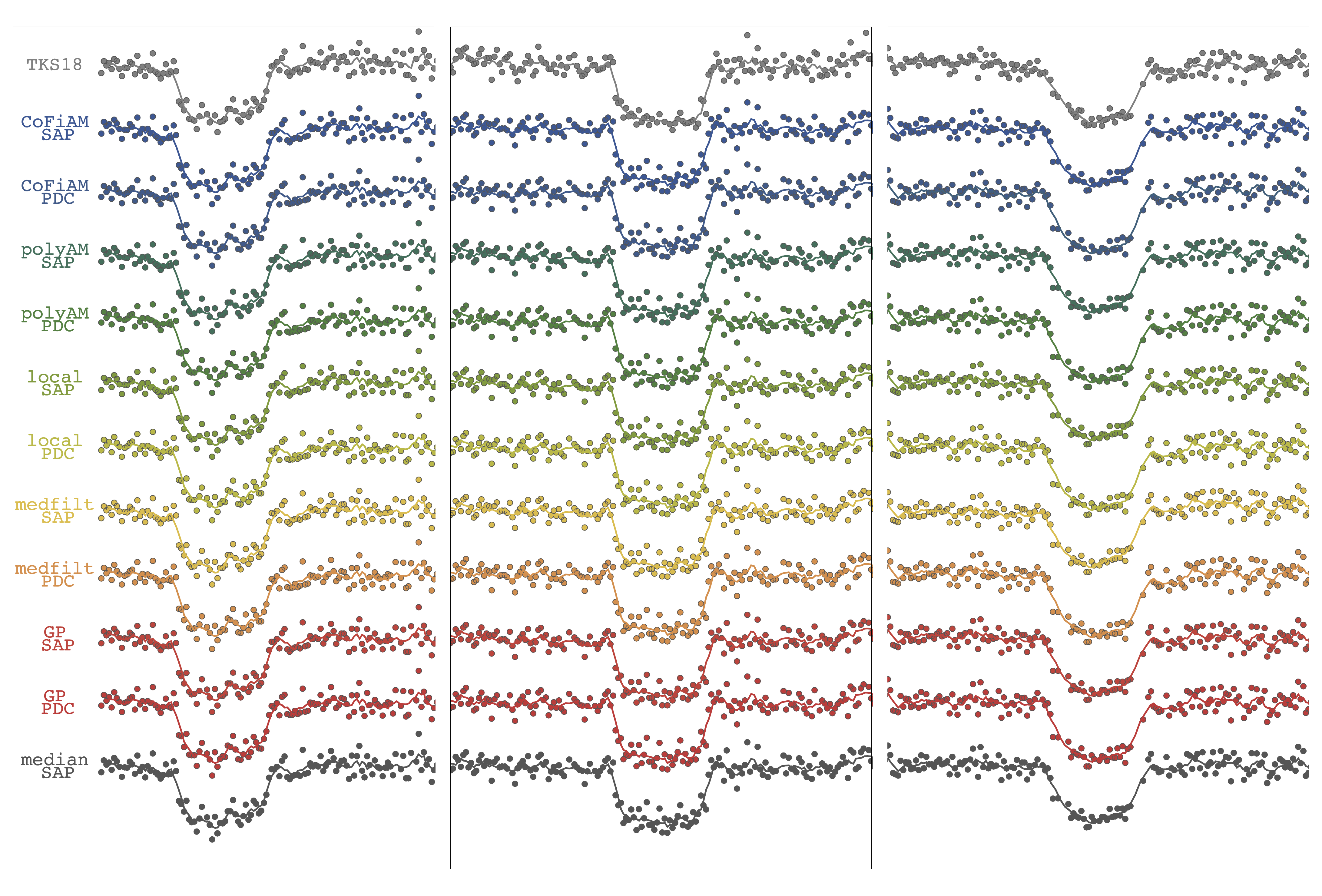}
\caption{\textbf{Method marginalized detrending.}
Comparison of five different detrending methods on two different data \kepler\ products. Top curve shows the \kepler\ reduction used in \cite{teachey:2018} and the bottom curve shows the method marginalized product used in this work.
}
\label{fig:kepdetrend}
\end{center}
\end{figure}

We detrended the revised \kepler\ photometry using five independent methods.
The first method is the \cofiam\ (Cosine Filtering with Autocorrelation Minimization)
algorithm \cite{HEK2} which was the approach
used in the original study, since it was specifically designed with
exomoon detection in mind. In addition, we considered four other popular
approaches: a polynomial fit, a local line fit, a median filter, and a Gaussian
process (see the \SM\ for a detailed description of each). The detrended
photometry is stable across the different methods (see
Fig.~\ref{fig:kepdetrend}), with a maximum standard deviation (SD) between any two
SAP time series of 250 parts per million (ppm), far below the median formal uncertainty of
$\sim 590$\,ppm. Although we verified that the Presearch Data Conditioning (\pdc)
version of the photometry \cite{stumpe:2012,smith:2012} produces similar
results (as evident in Fig.~\ref{fig:kepdetrend}), we ultimately only used
the five \sap\ reductions in what follows. We produced a ``method marginalized''
final time series by taking the median of the $i$th datum across
the five methods and propagating the variance between them into a revised
uncertainty estimate (see the \SM\ for details). In this way, we produced a
robust correction of the \kepler\ data accounting for differences in model
assumptions.

We fit photodynamical models \cite{luna:2011} to the revised \kepler\ data,
using the updated contamination factors from SOC version 9.3,
before introducing the new HST data. Bayesian model selection revealed
only a modest preference for the moon model, with the Bayes factor ($K$),
going from $2 \log K = 20.4$ in our original study down to just $1.0$ now. 
Detailed investigation revealed that this is not due to our new detrending approach,
as we applied our method marginalized detrending to the original version 9.0 data and 
recovered a similar result to our original analysis (see the \SM\ for details). 
Instead, it appeared that the reduced evidence was largely caused by the changes 
in the \sap\ photometry going from version 9.0 to 9.3, and to a lesser degree by the 
new contamination factors. This can be seen in Fig.~\ref{fig:kepdetrend}, where 
the third transit in particular experienced a pronounced change between the 
two versions, and it was this epoch that displayed the greatest evidence 
for a moon-like signature in the original analysis.

With a much larger aperture than \kepler, HST is expected to provide several times
more precise photometry. Accordingly, the question as to whether Kepler-1625b
hosts a large moon should incorporate this new information and in what
follows we describe how we processed the HST data and then combined them with the
revised \kepler\ photometry.

HST monitored the transit of Kepler-1625b occurring on 28 to 29 October, 2017
with Wide Field Camera 3 (WFC3). A total of 26 orbits, amounting to some 40 hours, 
were devoted to observing the event. The observations consisted
of one direct image and 232 exposures using the G141 grism, a slitless
spectroscopy instrument that projects the star's spectrum across the charge-coupled device (CCD). This
provides spectral information on the target in the near-infrared from
about 1.1 to 1.7 $\mu$m. Of these 232 exposures, only three were
unusable, as they coincided with the spacecraft's passage through the South
Atlantic Anomaly, at which time HST was forced to use its less-accurate gyroscopic guidance
system. Each exposure lasted roughly 5 min, resulting in about 45 min
on target per orbit. Images were extracted using standard tools made available
by the Science Telescope Space Institute (STScI) and are described in the \SM.

Native HST time stamps, recorded in the Modified Julian Date
system, were converted to Barycentric Julian Date (BJDUTC)
for consistency with the \kepler\ time stamps. The BJDUTC
system accounts for light travel time based on the position of the 
target and the observer with respect to the solar system barycenter
at the time of observation. As the position of HST is constantly changing 
we set the position of the observer to be the center of the Earth at 
the time of observation, for which a small discrepancy of $\pm 23$ ms 
is introduced. This discrepancy can be safely ignored for our purposes.

\begin{figure*}
\centering
\includegraphics[width=16.0cm]{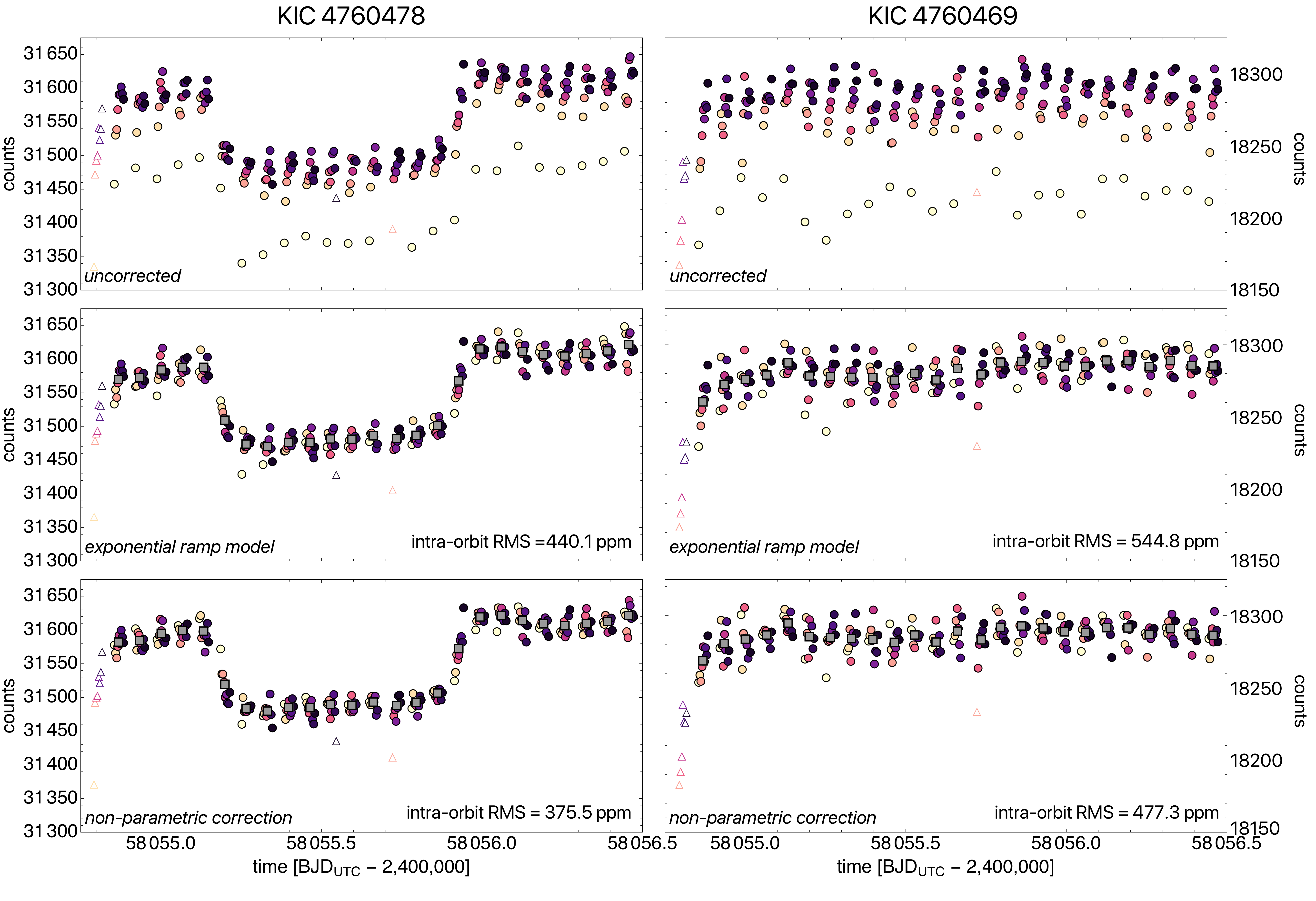}
\caption{\textbf{Hook corrections}. (Top) The optimal aperture photometry of
our target (left) and the best comparison star (right),
where the hooks and visit-long trends are clearly
present. Points are colored by their exposure number
within each HST orbit (triangles represent outliers).
(Middle) A hook-correction using the
common exponential ramp model on both stars. 
(Bottom) The result from an alternative and novel
hook-correction approach introduced in this work.
}
\label{fig:hooks}
\end{figure*}

While the telescope performed nominally throughout the observation, three
well-documented sources of systematic error were present in our data that
required removal. First, thermal fluctuations due to the spacecraft's orbit
led to clear brightness changes across the entire CCD (sometimes referred to as
``breathing''), which were corrected for by subtracting image median fluxes (see
the \SM\ for details). After computing an optimal aperture for the target, we
observed a strong intra-orbit ramping effect (also known as the ``hook'') in the
white light curve (see Fig.~\ref{fig:hooks}), which has been previously
attributed to charge trapping in the CCD \cite{agol:2010,berta:2012}. We
initially tried a standard parametric approach for correcting these ramps using
an exponential function, but found the result to be suboptimal. Instead, we
devised a new nonparametric approach described in the \SM\ that substantially
outperformed the previous approach.

We achieved a final mean intra-orbit precision of 375.5\,ppm (versus 440.1\,ppm
using exponential functions), which was about 3.8 times more precise
than \kepler\ when correcting for exposure time. The transit of Kepler-1625b was
clearly observed even before the hook correction. After removal of the hooks,
an apparent second decrease in brightness appeared towards the end of the
observations, which was evident even in the noisier exponential ramp corrected
data (see Fig.~\ref{fig:hooks}). Repeating our analysis for the only other
bright star fully on the CCD, KIC 4760469, revealed no peculiar behavior at this
time indicating that the dip was not due to an instrumental common mode.
Similarly, the centroids of both the target and the comparison star showed
no anomalous change around this time (see Fig. S6 in the \SM). 
A detailed analysis of the centroid variations of both the target
and the comparison star revealed that the 10 millipixel motion
observed was highly unlikely to be able to produce the
${\sim}500$\,ppm dip associated with the moon-like signature.
Further, we found that the signal was achromatic
appearing in two distinct spectral channels, which was consistent with
expectations for a real moon. Finally, a detailed analysis of the
photometric residuals revealed that the fits including a moon-like transit were
consistent with uncorrelated noise equal to the value derived from
our hook correction algorithm. These three tests, detailed in the
\SM, provide no reason to doubt that the moon-like dip is astrophysical
in nature and thus we treat it as such in what follows.

Upon inspection of the HST images we identified a previously uncataloged point
source within 2 arcseconds of our target. The star resides at position angle
8.5$^\circ$ east of north, with a derived \kepler\ magnitude of 22.7.
We attribute its new identification to the fact that it is both
exceptionally faint and so close to the target that it was always lost in the
glare in other images. Using a \gaia-derived distance to the target we found
that, were this point source to be at the same distance, it would be within
4500\,AU of Kepler-1625. However, it is not known whether the two sources are physically
associated, however. Its faintness means that it produces negligible contamination 
to our target spectrum. We estimated that the source has a variability of 0.33\% and 
contributes less than 1 part in 3000 to our final WFC3 white light curve, which means that 
the net contribution to our target is 1\,ppm and can be safely ignored.

In addition to the breathing and the hooks, a third well-known source of WFC3
systematic error we see is a visit-long trend (apparent in
Fig.~\ref{fig:hooks}). These trends have not yet been correlated to any
physical parameter related to the WFC3 observations \cite{wakeford:2016},
and thus the conventional approach is a linear slope (for example, 25-27)
although a quadratic model
has been used in some instances (for example, 28,29) 
The time scale of the variations is comparable to the transit itself and thus
cannot be removed in isolation; rather, any detrending model is expected to be
covariant with the transit model. For this reason, it was necessary to perform
the detrending regression simultaneous to the transit model fits. We
considered three possible trend models; linear, quadratic and exponential. All
models include an extra parameter describing a flux offset between the
14\textsuperscript{th} and 15\textsuperscript{th} orbits. This is motivated by
the fact that the spacecraft performed a full guide star acquisition at the
beginning of the 15\textsuperscript{th} orbit (a new ``visit''), and ended up
placing the spectrum ${\sim}$0.1\,pixels away from where it appeared during the
first 14 orbits. Although the white light curve shows no obvious flux change
at this time, the reddest channels display substantial shifts motivating this
offset term.

Finally, we extracted light curves in nine wavelength bins across the spectrum in
an attempt to perform transmission spectroscopy. As a planet transits its host
star, the atmosphere may absorb different amounts of light depending on the
constituent molecules and their abundances \cite{seager:2000}. This makes the
planet's transit depth wavelength-dependent. An accurate measurement of these
transit depths not only provides the potential to characterize the atmosphere's
composition; it is also potentially useful in providing an independent
measurement of the planet's mass \cite{dewit:2013}. While a low surface gravity
planet will show very pronounced molecular features and a steep slope at short
wavelengths due to Rayleigh scattering, a high surface gravity world will yield
a substantially flatter transmission spectrum.

With the HST WFC3 data prepared, we are ready to combine them with the revised
\kepler\ data to regress candidate models and compare them.
We considered four different transit models, which, when combined with
three different visit-long trend models, leads to a total of 12 models to
evaluate. The four transit models here were designated as P, for the
planet-only model; T, for a model that fits the observed transit timing
variations (TTVs) in the system agnostically; Z, for the zero-radius moon model,
which may produce all the gravitational effects of an exomoon without the
flux reductions of a moon transit; and M, which is the full planet plus moon
model. Models were generated using the \luna\ photodynamical software package
\cite{luna:2011} and regression was performed via the multimodal nested sampling
algorithm \multi\ \cite{feroz:2008,feroz:2009}. For each model, we derived 
not only the joint \textit{a posteriori} parameter samples, but also a Bayesian evidence
(also known as the marginal likelihood) enabling direct calculation of the
Bayes factor between models.

\section*{Results}
One clear result from our analysis is that the HST transit of Kepler-1625b
occurred 77.8 min earlier than expected, indicating TTVs in the system. 
Bayes factors between models P and T support
the presence of significant TTVs for any choice of detrending model (see Table~\ref{tab:evidences}), 
with the T fits returning a $\chi^2$ decreased by 17 to 19 (for 1048 data points).
Further, if we fit the \kepler\ data in isolation and make predictions for the
HST transit time, the observed time is $>3$\,$\sigma$ discrepant (see Fig. S12 in the \SM).
For reference, each \kepler\ transit midtime has an uncertainty on the order of 10 min
and the SD on linear ephemeris predictions is 25.2 min derived
from posterior samples.
Identifying TTVs was among the first methods proposed
to discover exomoons \cite{sartoretti:1999}, but certainly perturbations from
an unseen planet could also be responsible. We find that the
${\simeq}25$\,min amplitude TTV can be explained by an external perturbing
planet (see the \SM), although with only four transits on hand it is not possible
to constrain the mass or location of such a planet, and no other planet has
been observed so far in the system.

We also found that model Z consistently outperforms model T, though
the improvement to the fits is smaller at $\Delta \chi^2 \simeq 2$-$5$
(see Table~\ref{tab:evidences}). This suggests that the evidence for 
the moon based on timing effects alone goes beyond the TTVs, providing 
modest evidence in favor of additional dynamical effects such as 
duration changes \cite{kipping:2009a} and/or impact parameter variation 
\cite{kipping:2009b}, both expected consequences of a moon present in 
the system. This by itself would not constitute a strong enough case 
for a moon detection claim, but we consider it to be an important 
additional check that a real exomoon would be expected to pass.

The most compelling piece of evidence for an exomoon would be an exomoon
transit, in addition to the observed TTV. If Kepler-1625b's early transit were
indeed due to an exomoon, then we should expect the moon to transit late on the
opposite side of the barycenter. The previously mentioned existence of an
apparent flux decrease towards the end of our observations is therefore
where we would expect it to be under this hypothesis. Although we have
established that this dip is most likely astrophysical, we have not yet
discussed its significance or its compatibility with a self-consistent
moon model.

\begin{figure*}
\centering
\includegraphics[width=15.5cm]{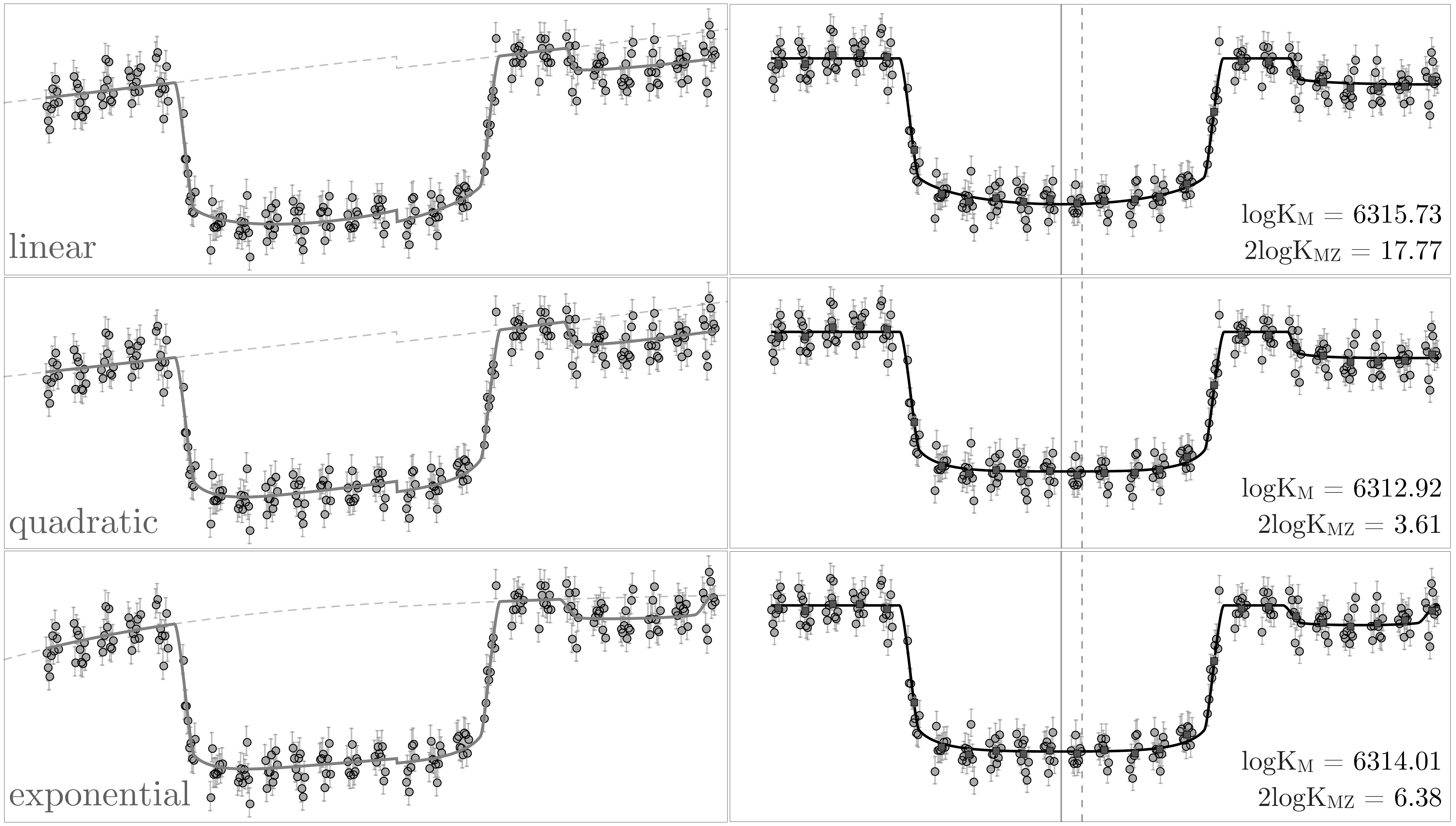}
\caption{\textbf{HST detrending.} The HST observations with three proposed trends fit to the data
(left) and with the trends removed (right). Bottom-right numbers
in each row give the Bayes factor between a planet plus moon model
(model $\mathrm{M}$) and a planet plus moon model where the moon radius
equals zero (model $\mathrm{Z}$), which tracks the significance of
the moon-like dip in isolation.
}
\label{fig:trends}
\end{figure*}

We find that our self-consistent planet plus moon models (M) always
outperform all other transit models in terms of maximum likelihood and
Bayesian evidences (see Table~\ref{tab:evidences}). The moon signal is 
found to have a signal-to-noise ratio of at least 19. The presence of a TTV
and an apparent decrease in flux at the correct phase position together
suggest that the exomoon is the best explanation. However, as is
apparent from Fig.~\ref{fig:trends}, the amplitude and shape of the
putative exomoon transit vary somewhat between the trend models, leading
to both distinct model evidences and associated system parameters.

\section*{Discussion}
Although the overall preference of the moon model is arguably best framed by
comparison to model P, the significance of the moon-like transit alone
is best framed by comparing M and Z alone. Such a comparison reveals a
strong dependency of the implied significance on the trend model used.
In the worst case, we have the quadratic model with $2\log K \simeq 4$,
corresponding to ``positive evidence'' \cite{kass:1995} - although we note
that the absolute evidence $\mathcal{Z}_M$ is the worst amongst the three.
The linear model is far more optimistic yielding $2\log K \simeq 18$,
corresponding to ``very strong evidence'' \cite{kass:1995}, whereas
the exponential sits between these extremes. The question then
arises, which of our trend models is the correct one?

Because the linear model is a nested version of the quadratic model,
and both models are linear with respect to time, it is more
straightforward to compare these two. The quadratic model essentially
recovers the linear model, apparent from Fig.~\ref{fig:trends},
with a curvature within 1.5\,$\sigma$ of zero, and yields almost
the same best $\chi^2$ score to within 1.2. This lack of meaningful
improvement causes the log evidence to drop by 2.8, since evidences
penalize wasted prior volume. The exponential model appears more
competitive with a log evidence of 1.72 lower, but a direct comparison
of two different classes of models, such as these, is muddied by the
fact that these analyses are sensitive to the choice of priors. The most
useful comparison here is simply to state that the maximum likelihoods
are within $\Delta \chi^2=0.68$ of one another and thus are likely
equally justified from data-driven perspective.

Another approach we considered is to weigh the trend models using the
posterior samples. Given a planet or moon's mass, there is a probabilistic
range of expected radii based on empirical mass-radius relations
\cite{forecaster}. Although we exclude extreme densities in our fits,
parameters from model M can certainly lead to improbable solutions with
regard to the photodynamically inferred \cite{weigh:2010} masses and radii.

To investigate this, we inferred the planetary mass using two methods for each
model and evaluated their self-consistency. The first method combines the
photodynamically-inferred planet-to-star mass ratio \cite{weigh:2010} with a
prediction for the mass based on the well-constrained radius using \forecaster;
an empirical probabilistic mass-radius relation \cite{forecaster}. The second
method approaches the problem from the other side, taking the moon's radius and
predicting its mass with \forecaster\ and then calculating the planetary mass via
the photodynamically-inferred moon-to-planet mass ratio. Our analysis
(discussed in more detail in the \SM) reveals that all three models have
physically plausible solutions and generally converge at
${\sim}10^3$\,$M_{\oplus}$ for the planetary mass, with the exception of the
quadratic model that had broader support extending down to Saturn-mass. We
ultimately combined the two mass estimates to provide a final best-estimate
for each model in Table~\ref{tab:params}.

As a consistency check, we used our derived transmission spectrum to constrain
the allowed range of planetary masses for a cloudless atmosphere
\cite{dewit:2013}. Using an MCMC (Markov chain Monte Carlo) with \exotransmit\ \cite{kempton:2017},
we find that masses in the range of $>0.4$ Jupiter masses (to 95\% confidence) are
consistent with the nearly flat spectrum observed, assuming a cloudless
atmosphere (see the \SM\ for details).

\begin{figure*}
\centering
\includegraphics[width=15.5cm]{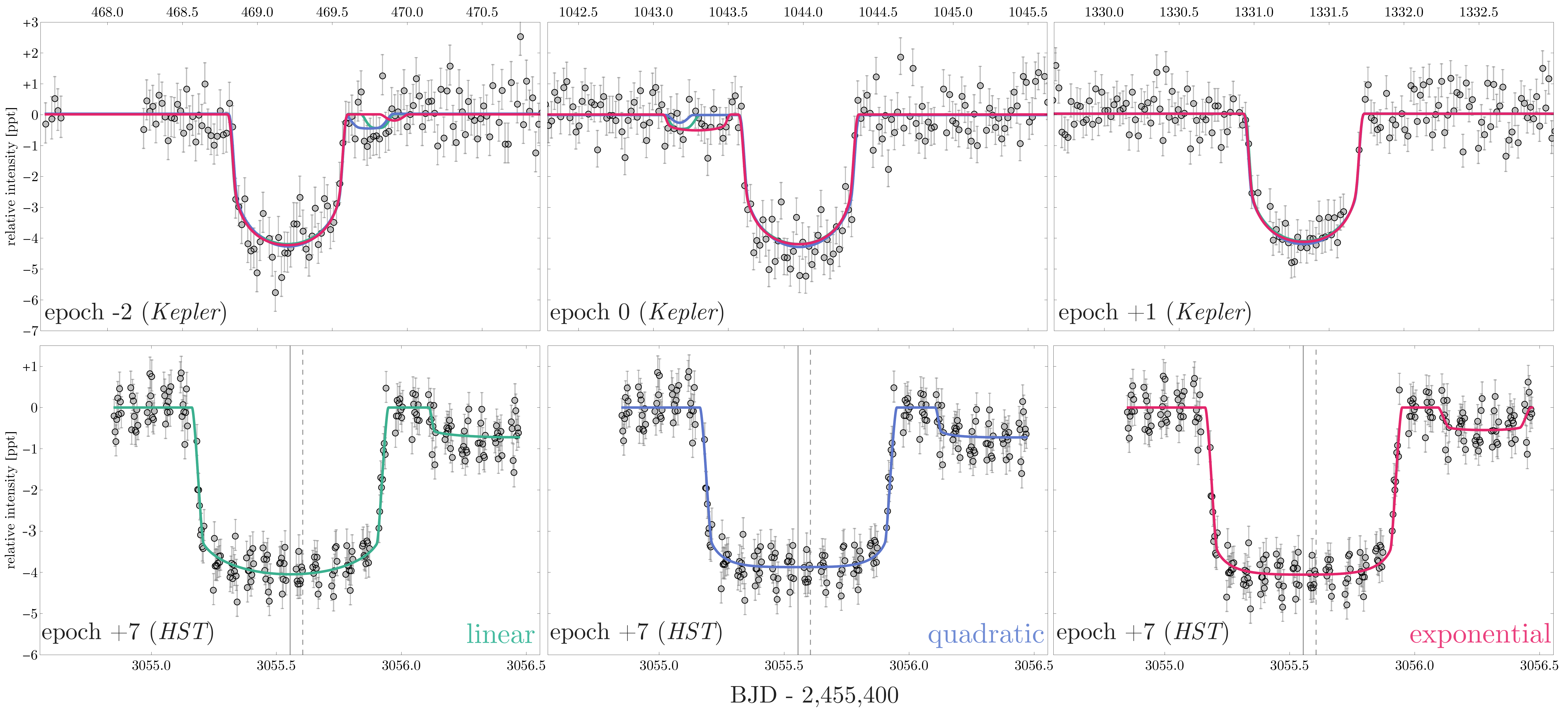}
\caption{\textbf{Moon solutions.} The three transits in \kepler\ (top) and the October 2017 transit
observed with HST (bottom) for the three trend model solutions. The three
colored lines show the corresponding trend model solutions for model
M, our favored transit model. The shape of the HST transit
differs from that of the \kepler\ transits owing to limb darkening differences between
the bandpasses.
}
\label{fig:fits}
\end{figure*}

In conclusion, the linear and exponential models appear to be the most
justified by the data and also lead to slightly improved physical
self-consistency, although we certainly cannot exclude the quadratic model
at this time. For this reason, we elected to present the associated system
parameters resulting from all three models in Table~\ref{tab:params}.
The maximum \textit{a posteriori} solutions from each, using model M, are
presented in Fig.~\ref{fig:fits} for reference.

We briefly comment on some of the inferred physical parameters for
this system. First, we have assumed a circular moon orbit throughout due to
the likely rapid effects of tidal circularization. However, we did
allow the moon to explore three-dimensional orbits and find some
evidence for noncoplanarity. Our solution somewhat favors a moon orbit tilted
by about 45$^\circ$ to the planet's orbital plane, with both pro- and
retrograde solutions being compatible. The only comparable known large moon
with such an inclined orbit is Triton around Neptune, which is generally
thought to be a captured Kuiper Belt object \cite{agnor:2006}. 
However, we caution that the constraints here are weak, reflected by the posterior's
broad shape, and thus it would be unsurprising if the true answer is 
coplanar.

One jarring aspect of the system is the sheer scale of it. The exomoon
has a radius of ${\simeq}4$\,$R_{\oplus}$, making it very similar to Neptune
or Uranus in size. The measured mass, including the \forecaster\ constraints,
comes in at $\log(M_S/M_{\oplus})=(1.2\pm0.3)$, which is again
compatible with Neptune or Uranus (although note that this solution is in part
informed by an empirical mass-radius relation). This Neptune-like moon orbits a
planet with a size fully compatible with that of Jupiter at
$(11.4\pm1.5)$\,$R_{\oplus}$, but most likely a few times more massive.
Finally, although the moon's period is highly degenerate and multimodal, we
find the semimajor axis is relatively wide at ${\simeq}40$ planetary
radii. With a Hill radius of $(200 \pm 50)$ planetary radii, this is well within
the Hill sphere and expected region of stability (see the \SM\ for further discussion).

The blackbody equilibrium temperature of the planet and moon, assuming
zero albedo, is ${\sim} 350$\,K. Adopting a more realistic albedo can drop this
down to $\sim 300$\,K. Of course, as a likely gaseous pair of objects there
is not much prospect of habitability here, although it appears that the
moon can indeed be in the temperature zone for optimistic definitions of
the habitable zone.

What is particularly interesting about the star is that it appears
to be a solar-mass star evolving off the main sequence. This inference
is supported by a recent analysis of the \gaia\ DR2 parallax by
\cite{berger:2018}, as well as our own isochrone fits (see the \SM).
We find that the star is certainly older than the Sun, at
${\simeq}$9\, gigayears in age, and that insolation at the location of the
system was thus lower in the past. The luminosity was likely close to
solar for most of the star's life, making the equilibrium temperature
drop down to ${\sim}250$\,K for Jovian albedos for most of its existence.
The old age of the system also implies plenty of time for tidal evolution,
which could explain why we find the moon at a fairly wide orbital
separation.

The origins of such a system can only be speculated upon at this time. A
mass ratio of 1.5\% is certainly not unphysical from in-situ formation
using gas-starved disk models, but it does represent the very upper end of what
numerical simulations form \cite{cilibrasi:2018}. In such a scenario, a
separate explanation for the tilt would be required. Impacts between gaseous
planets leading to captured moons are not well-studied but could be worth
further investigation. A binary exchange mechanism would be challenged by the
requirement for a Neptune to be in an initial binary with an object of comparable mass, 
such as a super-Earth \cite{agnor:2006}. Formation of an initial binary
planet, perhaps through tidal capture, seems improbable due to the tight orbits
simulation work tends to produce from such events \cite{ochiai:2014}.
If confirmed, Kepler-1625b-i will certainly provide an interesting puzzle for
theorists to solve.

\section*{Conclusion}
Together, a detailed investigation of a suite of models tested in this
work suggests that the exomoon hypothesis is the best explanation for the available 
observations. The two main pieces of information driving this
result are (i) a strong case for TTVs, in particular a 77.8\,min early transit
observed during our HST observations and (ii) a moon-like transit signature
occurring after the planetary transit. We also note that we find a modestly
improved evidence when including additional dynamical effects induced by moons
aside from TTVs.

The exomoon hypothesis is further strengthened by our analysis that
demonstrates that (i) the moon-like transit is not due to an instrumental common
mode, residual pixel sensitivity variations, or chromatic systematics; (ii) the moon-like
transit occurs at the correct phase position to also explain the observed TTV;
and (iii) simultaneous detrending and photodynamical modeling retrieves a
solution that is not only favored by the data, but is also physically
self-consistent.

Together, these lines of evidence all support the hypothesis of an exomoon 
orbiting Kepler-1625b. The exomoon is also the simplest hypothesis 
to explain both the TTV and the post-transit flux decrease, since other
solutions would require two separate and unconnected explanations for these 
two observations.

There remain some aspects of our present interpretation of the data that give us pause.
First, the moon's Neptunian size and inclined orbit are peculiar, though it is difficult to assess
how likely this is \textit{a priori} since no previously known exomoons exist. Second, the
moon's transit occurs towards the end of the observations and
more out-of-transit data could have more cleanly resolved this signal.
Third, the moon's inferred properties are sensitive
to the model used for correcting HST's visit-long trend and thus some
uncertainty remains regarding the true system properties.
However, the solution we deem most likely, a linear visit-long trend,
also represents the most widely agreed upon solution for the visit-long trend
in the literature.

Finally, it is somewhat ironic that the case for observing Kepler-1625b with
HST was contingent on a previous data release of the \kepler\ photometry that
indicated a moon \cite{teachey:2018}, while the most recent data release only
modestly favors that hypothesis when treated in isolation. Despite this, we
would argue that planets like Kepler-1625b -- Jupiter-sized planets on wide,
circular orbits around solar-mass stars -- were always ideal targets exomoon follow-up. 
There are certainly hints of the moon present even in the revised
\kepler\ data, but it is the HST data -- with a precision four times superior
to \kepler -- that are critical to driving the moon as the favored model. These
points suggest that it would be worthwhile to pursue similar \kepler\ planets
for exomoons with HST or other facilities, even if the \kepler\ data alone do not
show large moon-like signatures. Furthermore, our work demonstrates how impactful
the changes to \kepler\ photometry were, at least in this case, as it suggests
other results over the course of the \kepler\ mission may be similarly
affected, particularly for small signals.

All in all, it is difficult to assign a precise probability to the reality
of Kepler-1625b-i. Formally, the preference for the moon model over the planet-only 
model is very high, with a Bayes factor exceeding 400,000. On the other hand, this is
a complicated and involved analysis where a minor effect unaccounted for, or
an anomalous artifact, could potentially change our interpretation. In short, it
is the unknown unknowns that we cannot quantify. These reservations exist
because this would be a first-of-its-kind detection -- the first exomoon.
Historically, the first exoplanet claims faced great skepticism because
there was simply no precedence for them. If many more exomoons are detected in
the coming years with similar properties to Kepler-1625b-i, it would hardly
be a controversial claim to add one more. Ultimately, Kepler-1625b-i cannot
be considered confirmed until it has survived the long scrutiny of many
years, observations and community skepticism, and perhaps the detection
of similar such objects. Despite this, it is an exciting reminder of how
little we really know about distant planetary systems and the great spirit
of discovery that exoplanetary science embodies.


\clearpage


\clearpage


\begin{scilastnote}
\section*{Acknowledgements}
\item We wish to thank STScI staff scientists Bill Januszewski and
Kevin Stevenson for their critical contributions during the planning and
execution of the HST observation. 
We also thank Jon Jenkins at NASA and Paul Dalba at Boston University
for useful discussions regarding source contamination in the \kepler\ data. 
Members of the Cool Worlds Lab at Columbia University (Ruth Angus,
Jingjing Chen, Jorge Cortes, Tiffany Jansen, Moiya McTier, Emily Sandford, and
Adam Wheeler) provided valuable feedback at every stage of this analysis.
We are also grateful to members of the Hunt for Exomoons with Kepler project for their continued support
throughout the early years of our program.
Finally, we thank Travis Berger and collaborators for sharing their
\textit{Gaia}-derived posteriors for the target's radius.

\item \textbf{Funding:} Analysis was carried out in part on the NASA Supercomputer PLEIADES
(grant no. HEC-SMD-17-1386).
A.T. is supported through the NSF Graduate Research Fellowship (DGE 16-44869). 
D.M.K. is supported by the Alfed P. Sloan Foundation Fellowship.
This work is based in part on observations made with the NASA/ESA HST,
obtained at the Space Telescope Science Institute, which is operated by the
Association of Universities for Research in Astronomy, Inc., under NASA contract
NAS 5-26555. These observations are associated with program no. GO-15149.
Support for program no. GO-15149 was provided by NASA through a grant from the
Space Telescope Science Institute, which is operated by the Association of
Universities for Research in Astronomy, Inc., under NASA contract NAS 5-26555.
This paper includes data collected by the \kepler\ Mission. Funding for the
\kepler\ Mission was provided by the NASA Science Mission directorate.
This research has made use of the Exoplanet Follow-up Observation Program
website, which is operated by the California Institute of Technology, under
contract with NASA under the Exoplanet Exploration Program.

\item \textbf{Author Contributions:} A.T. was responsible for the proposal,
planning, and data reduction of the October 2017 HST observation. In addition,
A.T. modeled source blending in the \kepler\ data, investigated the possibility
of an external perturbing planet as the source of TTVs, and analyzed the transmission spectrum.
D.M.K. led the detrending of the \kepler\ and HST light curves,
and performed the joint fits to the data. 
D.M.K. also carried out the color, centroid, and residual 
analyses, as well as the planetary mass inference and isochrone fitting. 
All aspects of these tasks were executed through joint consultation, and 
the paper was written collaboratively by the two authors.

\item \textbf{Competing interests:} The authors declare that they have no competing interests.

\item \textbf{Data and materials availability:} The raw data from both the \kepler\
and HST observations are freely available for download at the Mikulski Archive
for Space Telescopes (https://archive.stsci.edu). All relevant information
required for replication of these results and to evaluate the conclusions in the paper 
are present in the paper and/or the Supplementary Materials. Additional data related 
to this paper may be requested from the authors. This work made use of Numpy, Scipy, Pandas, Matplotlib, 
Astropy, \ttvfaster, \exotransmit, \forecaster, \luna\ and \multi.
\end{scilastnote}

\section*{List of Supplementary Materials}
\begin{enumerate}
    \item Materials and Methods
    \item Tables S1-S3
    \item Figures S1-S18
    \item References (42-72)
\end{enumerate}

\clearpage

\begin{table*}
\label{tab:combined}
\caption{\textbf{Model performance.}
Bayesian evidences ($\mathcal{Z}$) and maximum likelihoods ($\hat{\mathcal{L}}$)
from our combined fits using \kepler\ and new HST data.
\kepler plus HST fits. The subscripts are P for the planet model, T for the planetary TTV
model, Z for the zero-radius moon model and M for the moon model.
The three columns are for each trend model attempted. The primed values correspond to those derived
the \kepler\ data in isolation.
}
\begin{tabular}{llll} 
\hline
 & linear & quadratic & exponential \\ [0.5ex] 
\hline
$\log\mathcal{Z}_{\mathrm{P}}$ & $6302.79 \pm 0.11$ & $6306.68 \pm 0.11$ & $6308.41 \pm 0.11$ \\
$\log\mathcal{Z}_{\mathrm{T}}$ & $6304.86 \pm 0.11$ & $6308.81 \pm 0.12$ & $6310.71 \pm 0.11$ \\
$\log\mathcal{Z}_{\mathrm{Z}}$ & $6306.84 \pm 0.11$ & $6311.12 \pm 0.12$ & $6310.82 \pm 0.12$ \\
$\log\mathcal{Z}_{\mathrm{M}}$ & $6315.73 \pm 0.12$ & $6312.92 \pm 0.12$ & $6314.01 \pm 0.12$ \\
\hline \hline
$2\log K(\mathcal{Z}_{\mathrm{M}}'/\mathcal{Z}_{\mathrm{P}}')$ & $1.00 \pm 0.22$ \\
\hline
$2\log(\mathcal{Z}_{\mathrm{M}}/\mathcal{Z}_{\mathrm{P}})$ & $25.88 \pm 0.32$ & $12.47 \pm 0.33$ & $11.19 \pm 0.32$ \\
$2\log(\mathcal{Z}_{\mathrm{M}}/\mathcal{Z}_{\mathrm{T}})$ & $21.72 \pm 0.33$ &  $8.21 \pm 0.34$ & $17.81 \pm 0.33$ \\
$2\log(\mathcal{Z}_{\mathrm{M}}/\mathcal{Z}_{\mathrm{Z}})$ & $17.77 \pm 0.33$ &  $3.61 \pm 0.33$ &  $6.38 \pm 0.34$ \\
\hline \hline
$\Delta\chi_{\mathrm{PM}}'^2=2\log(\hat{\mathcal{L}}_{\mathrm{M}}'/\hat{\mathcal{L}}_{\mathrm{P}}')$ & $18.66$ \\
\hline
$\Delta\chi_{\mathrm{PM}}^2=2\log(\hat{\mathcal{L}}_{\mathrm{M}}/\hat{\mathcal{L}}_{\mathrm{P}})$ & $54.93$ & $41.04$ & $41.57 $\\
$\Delta\chi_{\mathrm{TM}}^2=2\log(\hat{\mathcal{L}}_{\mathrm{M}}/\hat{\mathcal{L}}_{\mathrm{T}})$ & $35.69$ & $23.97$ & $23.97$ \\
$\Delta\chi_{\mathrm{ZM}}^2=2\log(\hat{\mathcal{L}}_{\mathrm{M}}/\hat{\mathcal{L}}_{\mathrm{Z}})$ & $33.68$ & $19.59$ & $19.22$ \\ [1ex]
\hline 
\label{tab:evidences}
\end{tabular}
\end{table*}
\clearpage

\begin{table*}
\label{tab:kepler}
\caption{\textbf{System parameters.}
Median and $\pm 34.1$\% quantile range of the \textit{a posteriori} model parameters
from model M, where each column defined a different visit-long trend model.
The top panel gives the credible intervals for the actual parameters used
in the fit, and the lower panel gives a selection of relevant derived
parameters conditioned upon our revised stellar parameters.
The quoted inclination of the satellite is the inclination modulo 90$^\circ$.
}
\begin{tabular}{llll} 
\hline
Parameter & Linear & Quadratic & Exponential \\ [0.5ex] 
\hline
Photodynamics only \\
$R_{P,\mathrm{Kep}}/R_{\star}$ & $0.06075_{-0.00065}^{+0.00062}$ & $0.06061_{-0.00073}^{+0.00068}$ &
$0.06072_{-0.00063}^{+0.0062}$ \\
$R_{P,\mathrm{HST}}/R_{P,\mathrm{Kep}}$ & $0.998_{-0.013}^{+0.013}$ & $1.009_{-0.017}^{+0.019}$ &
$1.006_{-0.014}^{+0.014}$ \\
$\rho_{\star,\mathrm{LC}}$\,[g\,cm$^{-3}$] & $424_{-16}^{+9}$ & $424_{-15}^{+9}$ &
$425_{-14}^{+9}$ \\
$b$ & $0.104_{-0.066}^{+0.084}$ & $0.099_{-0.063}^{+0.088}$ &
$0.096_{-0.058}^{+0.078}$ \\
$P_P$\,[days] & $287.37278_{-0.00065}^{+0.00075}$ & $287.3727_{-0.0015}^{+0.0022}$ &
$287.37269_{-0.00076}^{+0.00074}$ \\
$\tau_0$\,[BJD$_{\mathrm{UTC}}$] & $2456043.9587_{-0.0027}^{+0.0027}$ & $2456043.9572_{-0.0093}^{+0.0033}$ &
$56043.9585_{-0.0029}^{+0.0025}$ \\
$q_{1,\mathrm{Kep}}$ & $0.45_{-0.14}^{+0.19}$ & $0.44_{-0.15}^{+0.19}$ &
$0.45_{-0.14}^{+0.18}$ \\
$q_{2,\mathrm{Kep}}$ & $0.31_{-0.15}^{+0.19}$ & $0.32_{-0.16}^{+0.20}$ &
$0.31_{-0.15}^{+0.19}$ \\
$q_{1,\mathrm{HST}}$ & $0.087_{-0.041}^{+0.057}$ & $0.096_{-0.045}^{+0.064}$ &
$0.087_{-0.040}^{+0.056}$ \\
$q_{2,\mathrm{HST}}$ & $0.25_{-0.15}^{+0.25}$ & $0.21_{-0.14}^{+0.23}$ &
$0.22_{-0.14}^{+0.22}$ \\
$P_S$\,[days] & $22_{-9}^{+17}$ & $24_{-11}^{+18}$ &
$22_{-9}^{+15}$ \\
$a_{SP}/R_P$ & $45_{-5}^{+10}$ & $36_{-13}^{+10}$ &
$42_{-4}^{+7}$ \\
$\phi_S$\,[$^{\circ}$] & $179_{-70}^{+136}$ & $141_{-65}^{+161}$ &
$160_{-60}^{+150}$ \\
$i_S$\,[$^{\circ}$] & $42_{-18}^{+15}$ & $49_{-22}^{+21}$ &
$43_{-19}^{+15}$ \\
$\Omega_S$\,[$^{\circ}$] & $0_{-83}^{+142}$ & $12_{-113}^{+132}$ &
$8_{-81}^{+136}$ \\
$(M_S/M_P)$ & $0.0141_{-0.0039}^{+0.0048}$ & $0.0196_{-0.0071}^{+0.0294}$ &
$0.0149_{-0.0038}^{+0.0052}$ \\
$(R_S/R_P)$ & $0.431_{-0.036}^{+0.033}$ & $0.271_{-0.099}^{+0.150}$ &
$0.363_{-0.079}^{+0.048}$ \\
$\Delta a_0$\,[ppm] & $330_{-120}^{+120}$ & $180_{-210}^{+170}$ & $220_{-140}^{+130}$ \\
\hline
+ Stellar properties \\
$R_{\star}$\,[$R_{\odot}$] & $1.73_{-0.22}^{+0.24}$ & $1.73_{-0.22}^{+0.24}$ & 
$1.73_{-0.22}^{+0.24}$ \\
$M_{\star}$\,[$M_{\odot}$] & $1.04_{-0.06}^{+0.08}$ & $1.04_{-0.06}^{+0.08}$ & 
$1.04_{-0.06}^{+0.08}$ \\
$\rho_{\star,\mathrm{iso}}$\,[kg\,m$^{-3}$] & $0.29_{-0.09}^{+0.13}$ & 
$0.29_{-0.09}^{+0.13}$ & $0.29_{-0.09}^{+0.13}$ \\
$e_{\mathrm{min}}^{\dagger}$ & $0.13_{-0.09}^{+0.11}$ & $0.13_{-0.09}^{+0.11}$ &
$0.13_{-0.09}^{+0.11}$ \\
$R_P$\,[$R_{\oplus}$] & $11.4_{-1.5}^{+1.6}$ & $11.4_{-1.4}^{+1.6}$ &
$11.4_{-1.4}^{+1.6}$ \\
$\log_{10}(M_P/M_{\oplus})$ & $2.86_{-0.50}^{+0.48}$ & $2.40_{-0.72}^{+0.70}$ &
$2.75_{-0.54}^{+0.53}$ \\
$a_P$\,[AU] & $0.98_{-0.13}^{+0.14}$& $0.98_{-0.12}^{+0.14}$ &
$0.98_{-0.12}^{+0.14}$ \\
$R_S$\,[$R_{\oplus}$] & $4.90_{-0.72}^{+0.79}$ & $3.09_{-1.19}^{+1.71}$ &
$4.05_{-1.01}^{+0.86}$\\
$\log_{10}(M_S/M_{\oplus})$ & $1.00_{-0.48}^{+0.46}$ & $0.74_{-0.52}^{+0.56}$ &
$0.93_{-0.50}^{+0.49}$ \\
$S_{\mathrm{eff}}$\,[$S_{\oplus}$] & $2.65_{-0.16}^{+0.19}$ & $2.64_{-0.16}^{+0.18}$ &
$2.64_{-0.16}^{+0.18}$ \\
\hline
+\,\,forecaster \\
$\log_{10}(M_P/M_{\oplus})$ & $3.12_{-0.27}^{+0.26}$ & $2.65_{-0.52}^{+0.50}$ &
$3.01_{-0.30}^{+0.26}$ \\
$\log_{10}(M_S/M_{\oplus})$ & $1.27_{-0.30}^{+0.29}$ & $1.11_{-0.58}^{+0.55}$ &
$1.20_{-0.34}^{+0.32}$ \\
$M_P$\,[$M_{J}$] & $[1.2,12.5]$ & $[0.2,9.0]$ &
$[0.6,10.5]$\\
$M_S$\,[$M_{\oplus}$] & $[4.4,68]$ & $[1.0,140]$ &
$[2.6,76]$ \\
$K$\,[m/s] & $[35,380]$ & $[6,280]$ & 
$[18,320]$ \\ [1ex]
\hline 
\label{tab:params}
\end{tabular}
\end{table*}
\clearpage


\clearpage

\clearpage

\clearpage

\clearpage

\input{SM_input.tex}

\bibliographystyle{Science}

\end{document}

%% file: SM_input.tex
\section*{Supplementary Materials}

\clearpage

\section{Materials and Methods}

\subsection{\kepler\ Re-analysis}
\label{sec:kepler}

\input{sections/kepler_analysis.tex}

\subsection{Hubble Observations}
\label{sec:hubble}

\input{sections/hubble_obs.tex}


\subsection{Joint Fits}
\label{sec:fits}

\input{sections/fits.tex}




\clearpage

\section*{Supplementary Figures and Tables}
\clearpage

\renewcommand{\thefigure}{S\arabic{figure}}

\begin{figure}
\centering
\includegraphics[width=16cm]{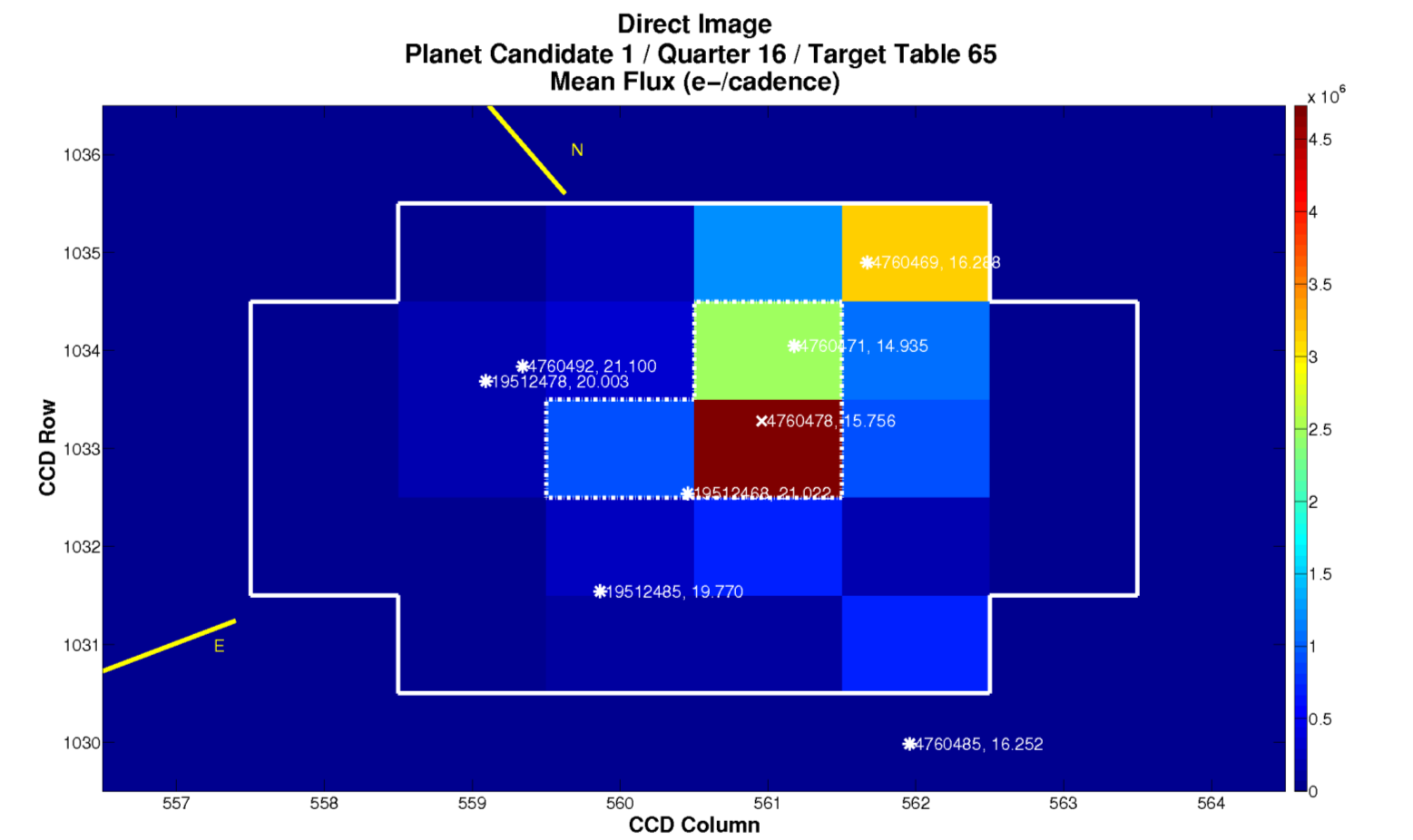}
\caption{\textbf{The ``Phantom'' Star.} Model of the \kepler\ optimal aperture
taken from the Data Validation Report (Q16
aperture), with the model star field overlaid. KIC 4760471 is clearly marked
within the green pixel, but the star apparently does not exist.}
\label{fig:DV_optimal_aperture_diagram}
\end{figure}
\clearpage 

\begin{figure*}
\begin{center}
\includegraphics[width=16.0cm,angle=0,clip=true]{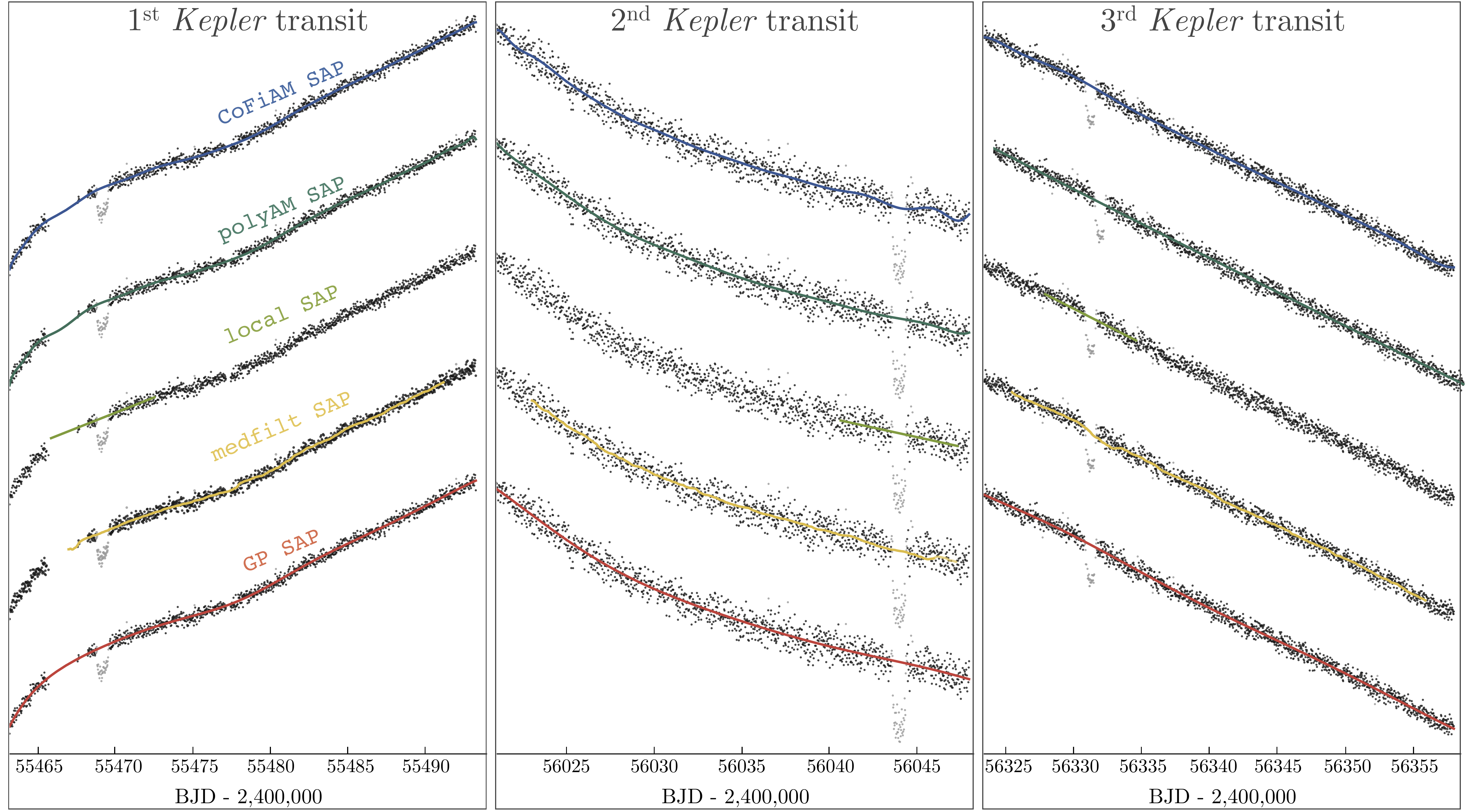}
\caption{
\textbf{\kepler\ detrending.} Comparison of five different methods used for detrending the \sap\ \kepler\
data. Baselines shown represent the full training set used, except for
the \local\ method which is trained on only data immediately surrounding the
transits of interest.
}
\label{fig:detrending_SAP}
\end{center}
\end{figure*}
\clearpage

\begin{figure}
\begin{center}
\includegraphics[width=16cm,angle=0,clip=true]{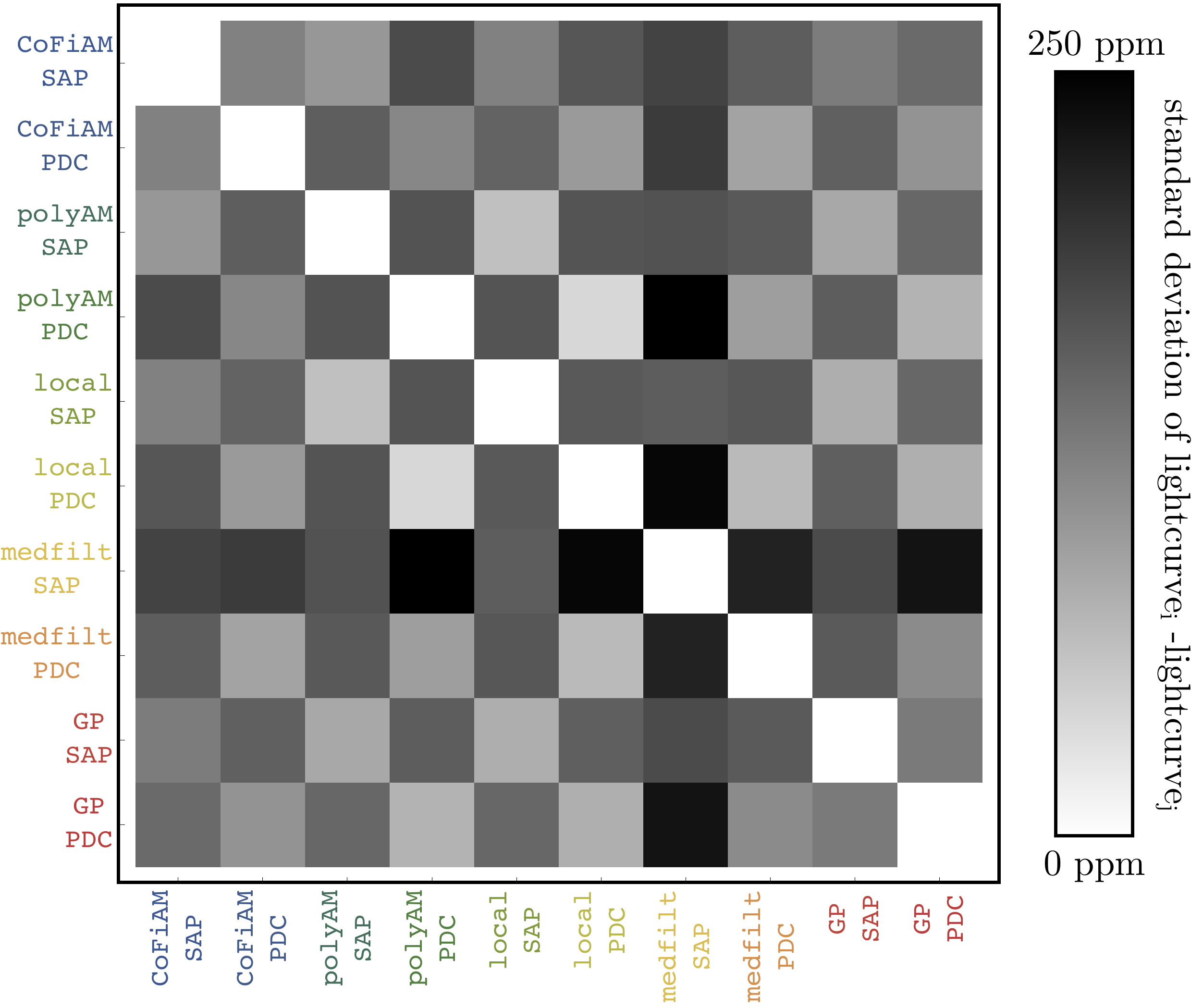}
\caption{
\textbf{\kepler\ detrending comparison.} Matrix plot of the standard deviations obtained when taking the differences
between the resulting fluxes from each detrending approach. The plot is scaled
such that white equals 0\,ppm and black equals 250\,ppm. For comparison the
typical photon noise uncertainty in 600\,ppm.
}
\label{fig:residuals_matrix}
\end{center}
\end{figure}
\clearpage

\begin{figure}
    \centering
    \includegraphics[width=16cm]{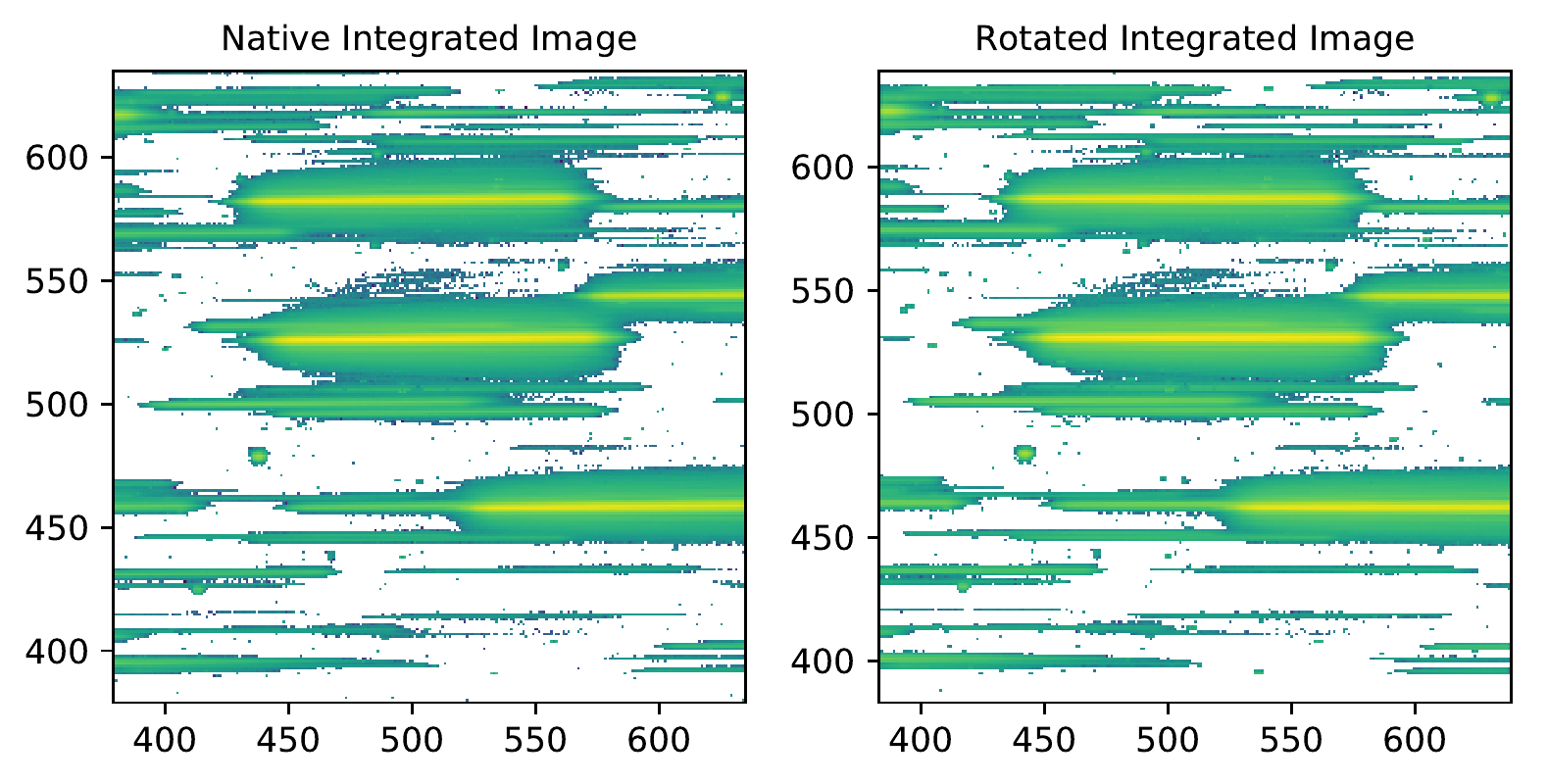}
    \caption{\textbf{HST image rotation.} \textit{Left:}Integrated HST image with the spectra natively inclined 
    0.5 degrees with respect to the $x$-axis. \textit{Right:} the rotated image for 
    simplifying the spectral extraction. Pixel values are logarithmic to show the full 
    extent of the spectra; white space indicates backgrounds integrating to 
    values $<0$, for which the logarithm is undefined.}
    \label{fig:native_and_rotated}
\end{figure}
\clearpage

\begin{figure*}
\centering
\includegraphics[width=16.0cm]{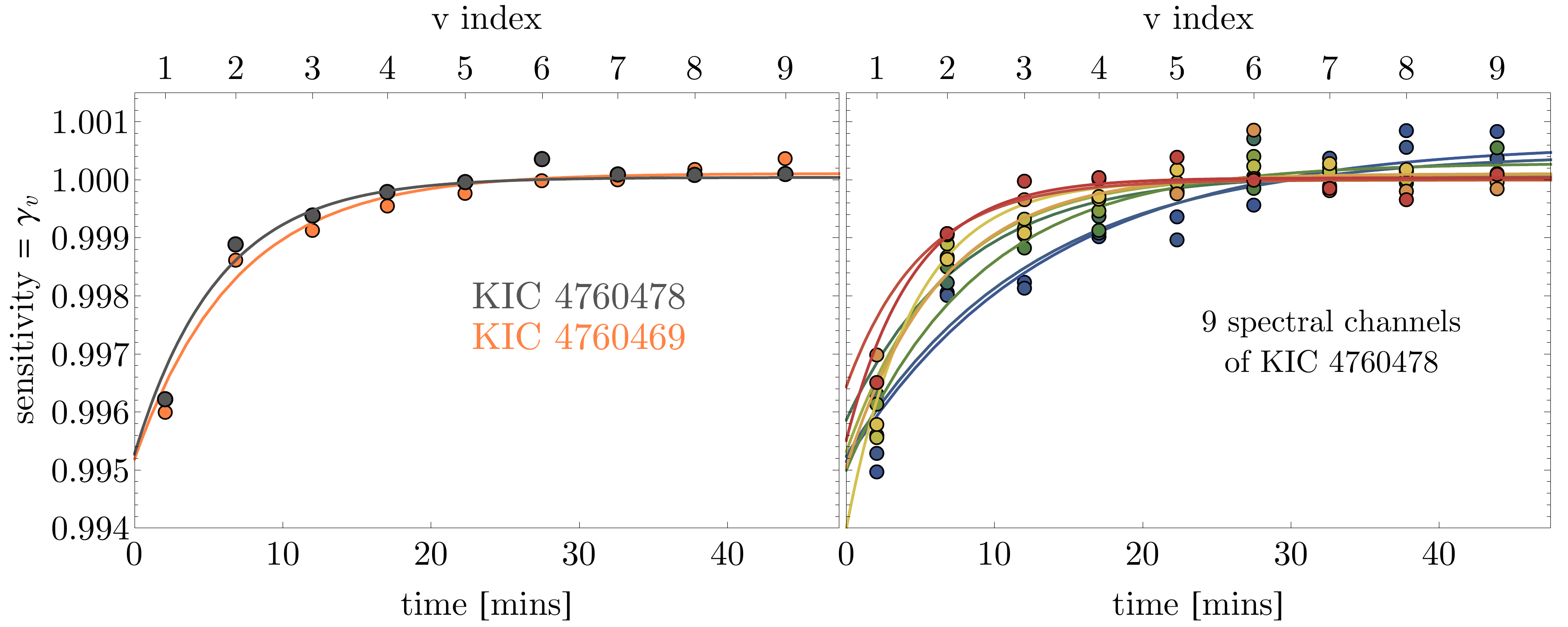}
\caption{\textbf{HST hook model comparison.} \textit{Left}: Comparison of the two models for the WFC3 hook; an exponential 
ramp fit (solid) and a novel discrete model introduced in this work (points). 
The left panel shows the results from the white light curve of the target and 
a comparison star. \textit{Right:} Same as left except we show the 9 different spectral 
elements used in this work for the target. The v index is the exposure number within
a given orbit.}
\label{fig:ramps}
\end{figure*}
\clearpage

\begin{figure*}
\centering
\includegraphics[width=16.0cm]{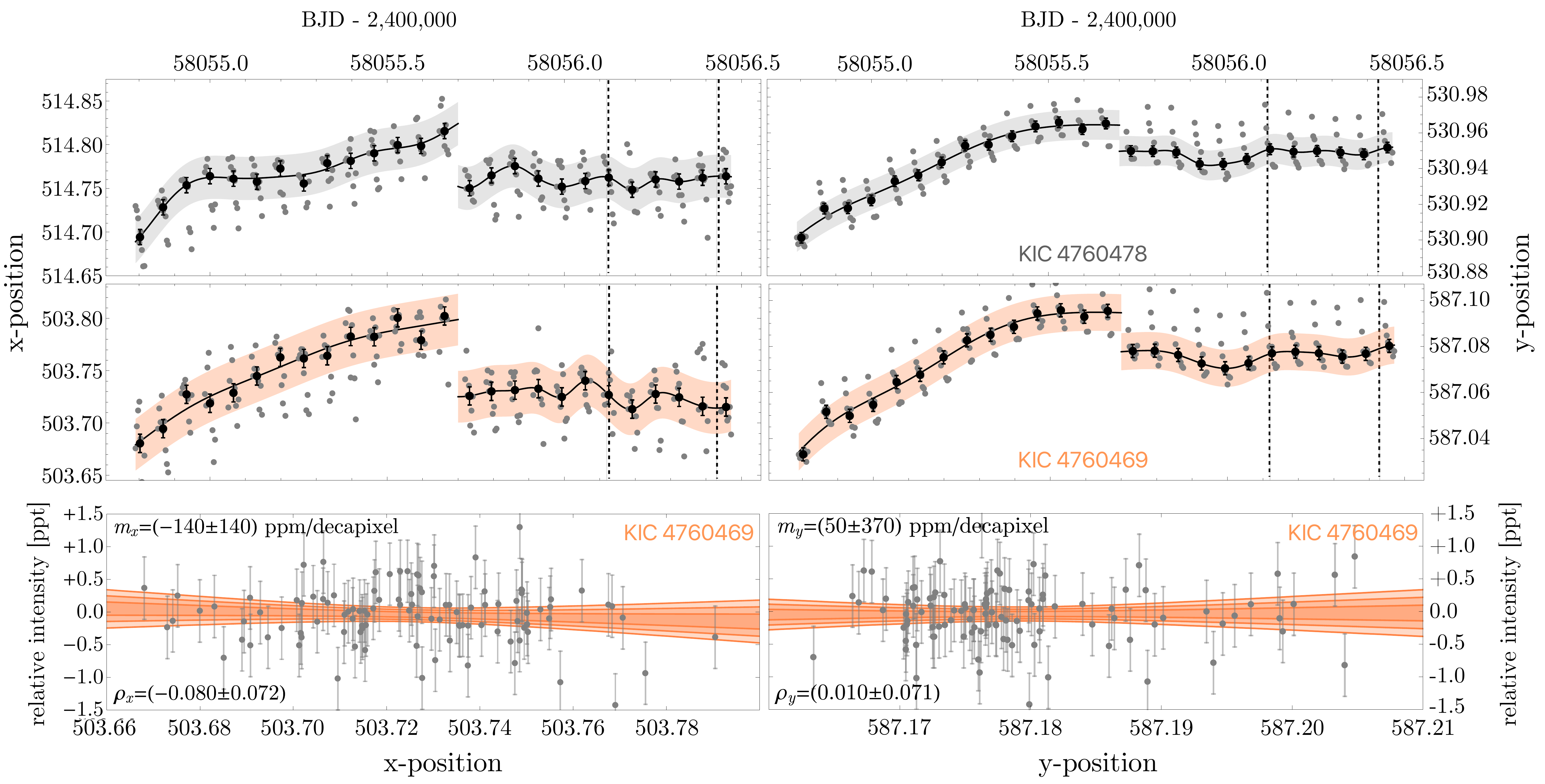}
\caption{\textbf{HST centroids.} Top row shows the centroid position of
our target in both row (left) and column (right) pixel
index, with a GP model overlaid (shaded region). Middle
row shows that the same but for our best comparison star
(right). The column positions have been offset by 0.1
pixels after the visit change to more easily fit them on
a single scale. The vertical grid lines mark the location
of the moon-like dip seen in the photometry of the target,
where we note that no peculiar behavior is evident. The
lowest row shows a correlation plot of intensity versus
centroid position for the comparison star (second visit),
where no clear dependency is evident either.
}
\label{fig:centroids}
\end{figure*}
\clearpage

\begin{figure*}
\centering
\includegraphics[height=18cm]{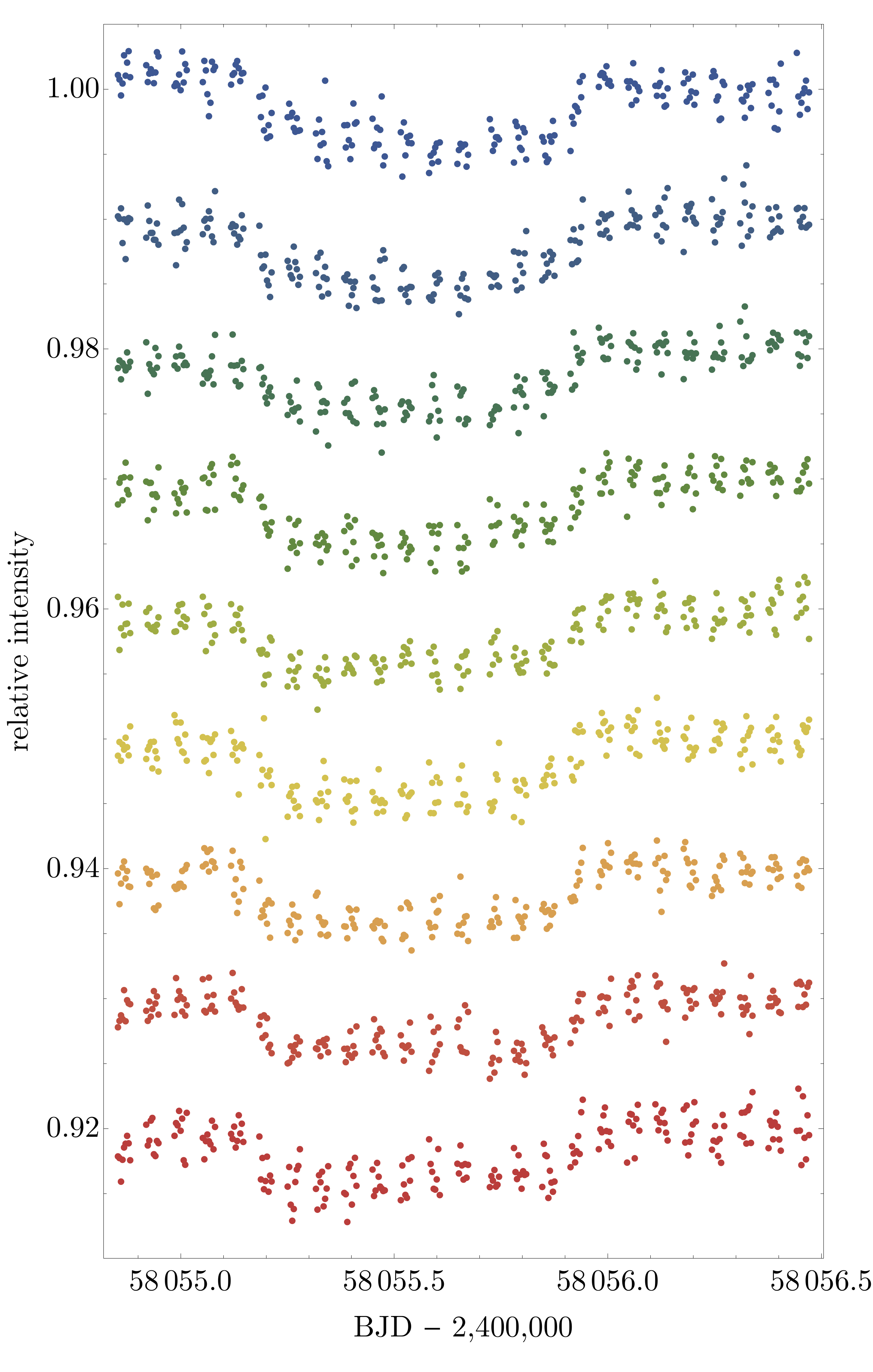}
\caption{\textbf{Spectral analysis}. Nine spectral channels, color-coded in wavelength from
the bluest channel (top) to the reddest (bottom), extracted from
our WFC3 photometry of Kepler-1625b. Naturally the noise in
each channel is considerably higher than the white light curve.
}
\label{fig:spectral_channels}
\end{figure*}
\clearpage

\begin{figure}
    \centering
    \includegraphics{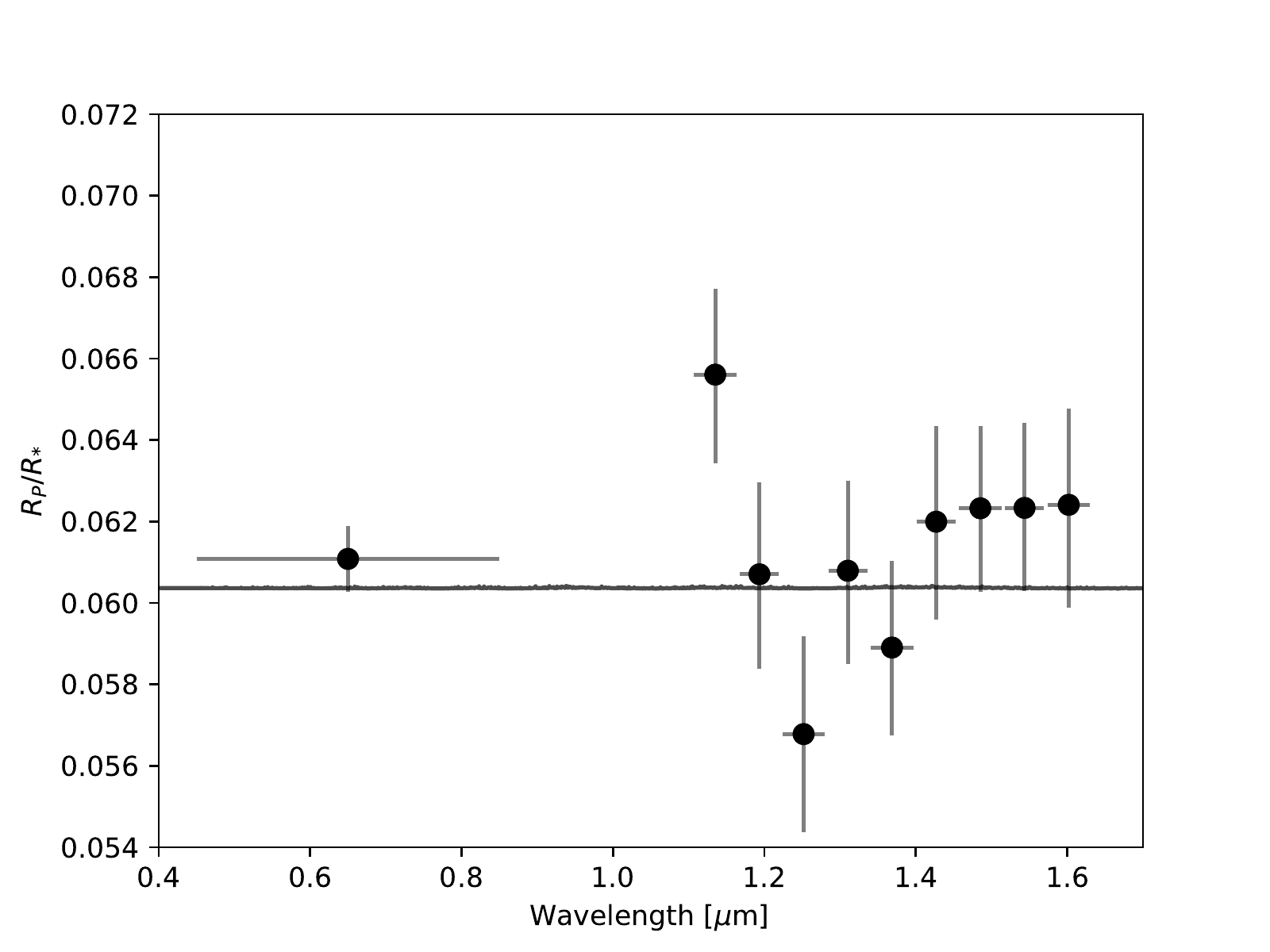}
    \caption{\textbf{Transmission spectrum.} The transmission spectrum measured as the ratio of 
    the planet radius to the stellar radius, utilizing the \kepler\
    bandpass at far left and the spectral channels extracted from 
    the WFC3 grism photometry. For reference a model spectrum 
    assuming $M_P = M$\textsubscript{Jup} is also plotted.}
    \label{fig:transmission_spectrum}
\end{figure}

\begin{figure}
    \centering
    \includegraphics[width=16cm]{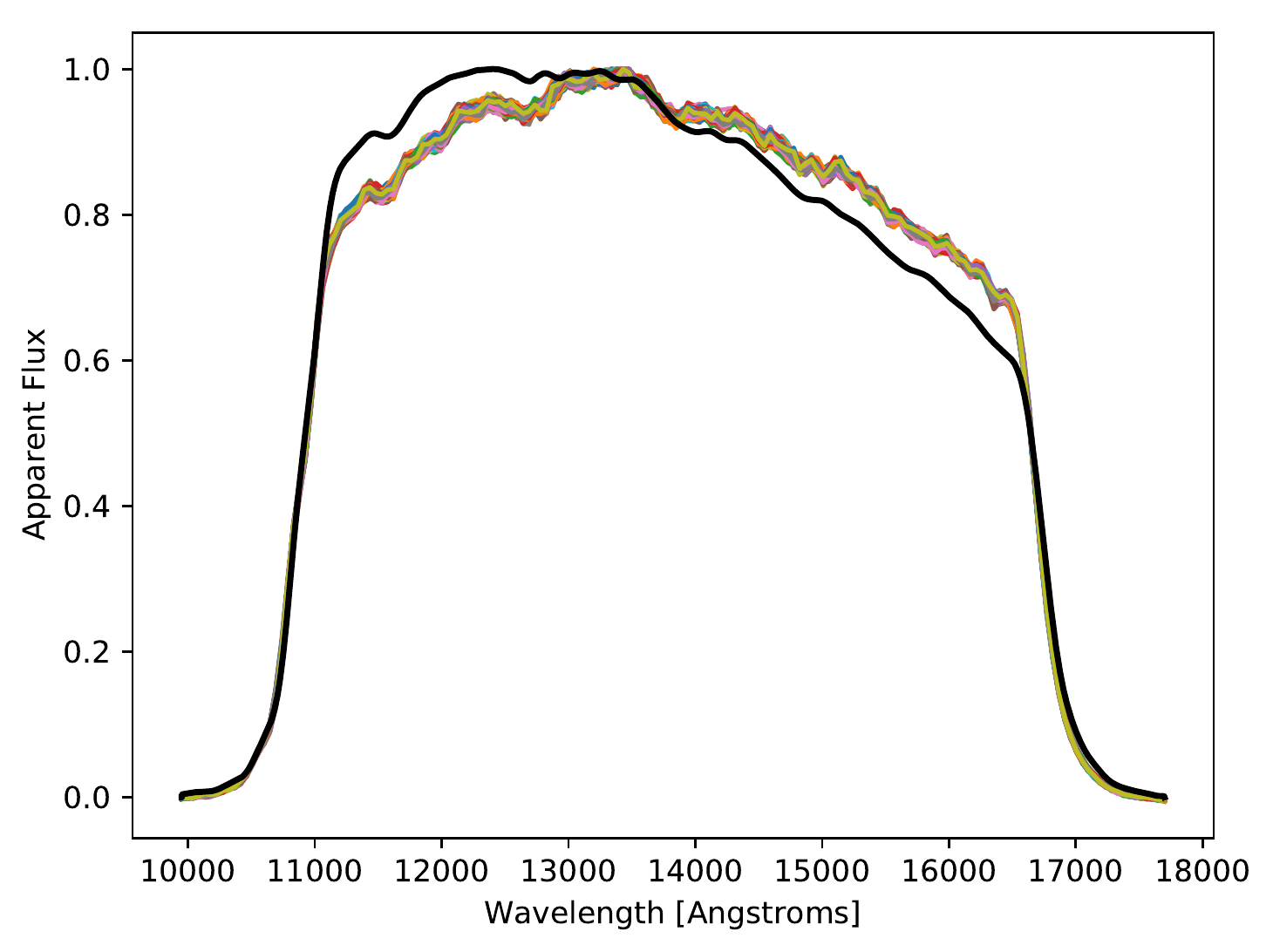}
    \caption{\textbf{Wavelength solution.} Calculated blackbody for Kepler-1625 multiplied by the G141 response 
    function (black). The spectrum of Kepler1625 extracted from each HST image is 
    overlaid in multiple colors (for each exposure). Both curves are normalized by 
    dividing out their maximum values. Note the overprediction of the model at 
    shorter wavelengths and underprediction at longer wavelengths.}
    \label{fig:kepler1625_response_function}
\end{figure}
\clearpage

\begin{figure}
    \centering
    \includegraphics[width=16cm]{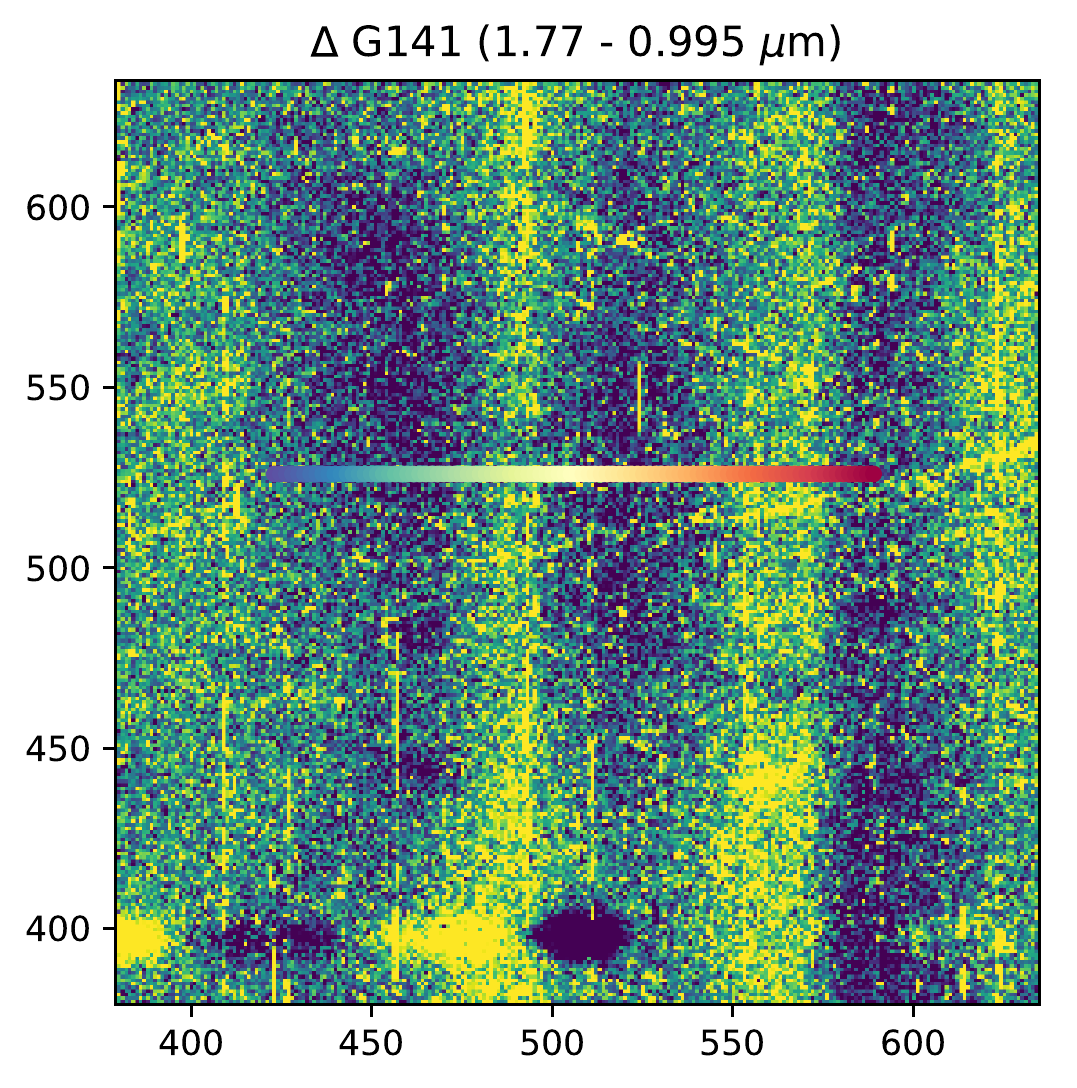}
    \caption{\textbf{Wavelength-dependent pixel sensitivity}. Change in pixel sensitivity across the G141 wavelength range. 
    The color range is $\pm$ 1\%. Towards the purple end short wavelengths are 
    more sensitive than longer wavelengths, while at the yellow end longer 
    wavelengths are more sensitive. The full width of the target spectrum's 
    response function is overplotted.}
    \label{fig:flat_field_delta_sensitivity}
\end{figure}
\clearpage

\begin{figure}
    \centering
    \includegraphics[width=16cm]{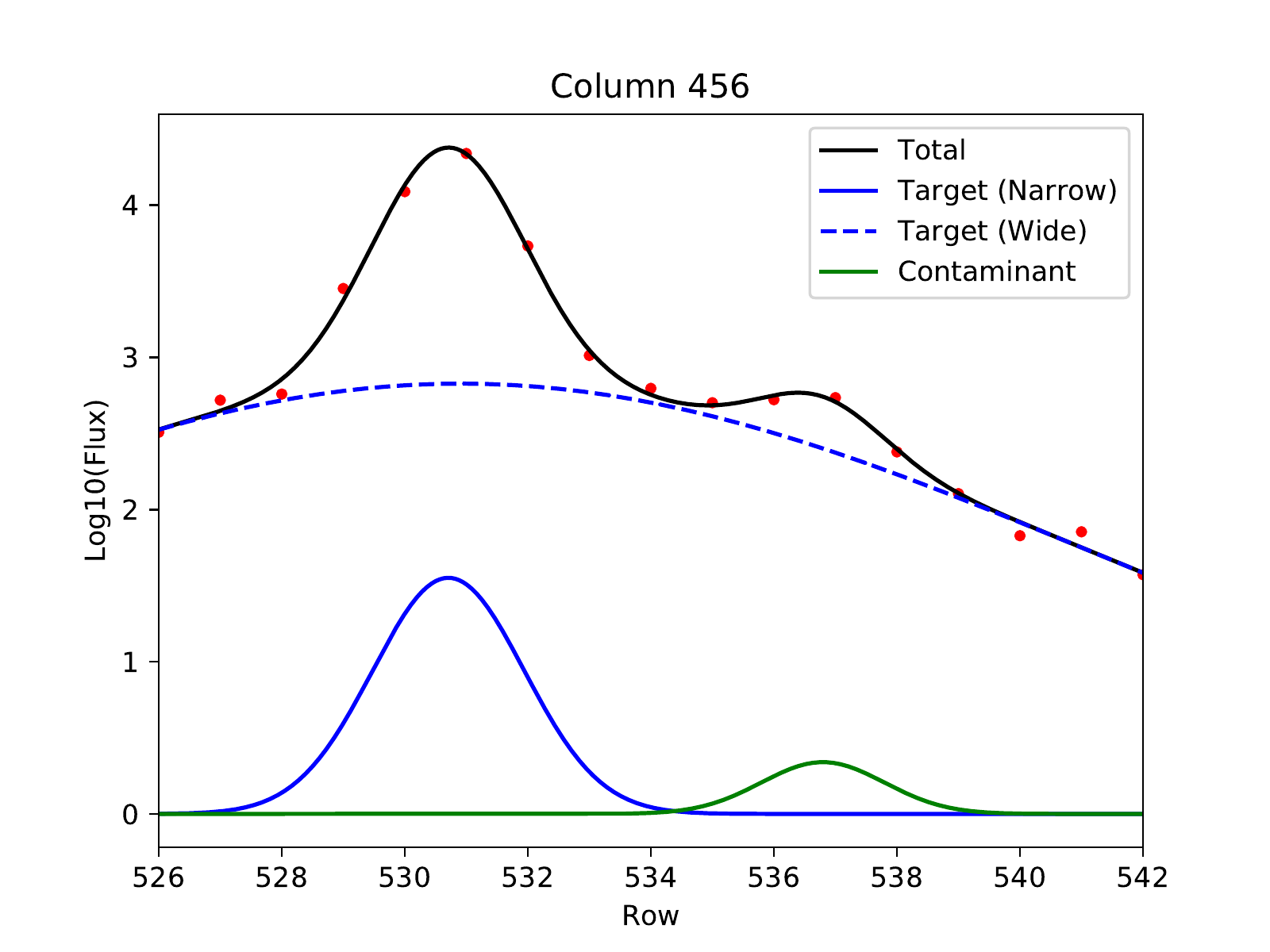}
    \caption{\textbf{Modeling the uncatalogued source contamination.} A single column within the HST optimal aperture, fitting three Gaussians 
    to the source and the contaminating uncatalogued source. Pixel fluxes are shown 
    by the red data points. The blending is calculated by taking the inverse of the 
    starlight fraction within the optimal aperture originating from the target star.}
    \label{fig:HST_OA_blend}
\end{figure}

\begin{figure*}
\centering
\includegraphics[width=16cm]{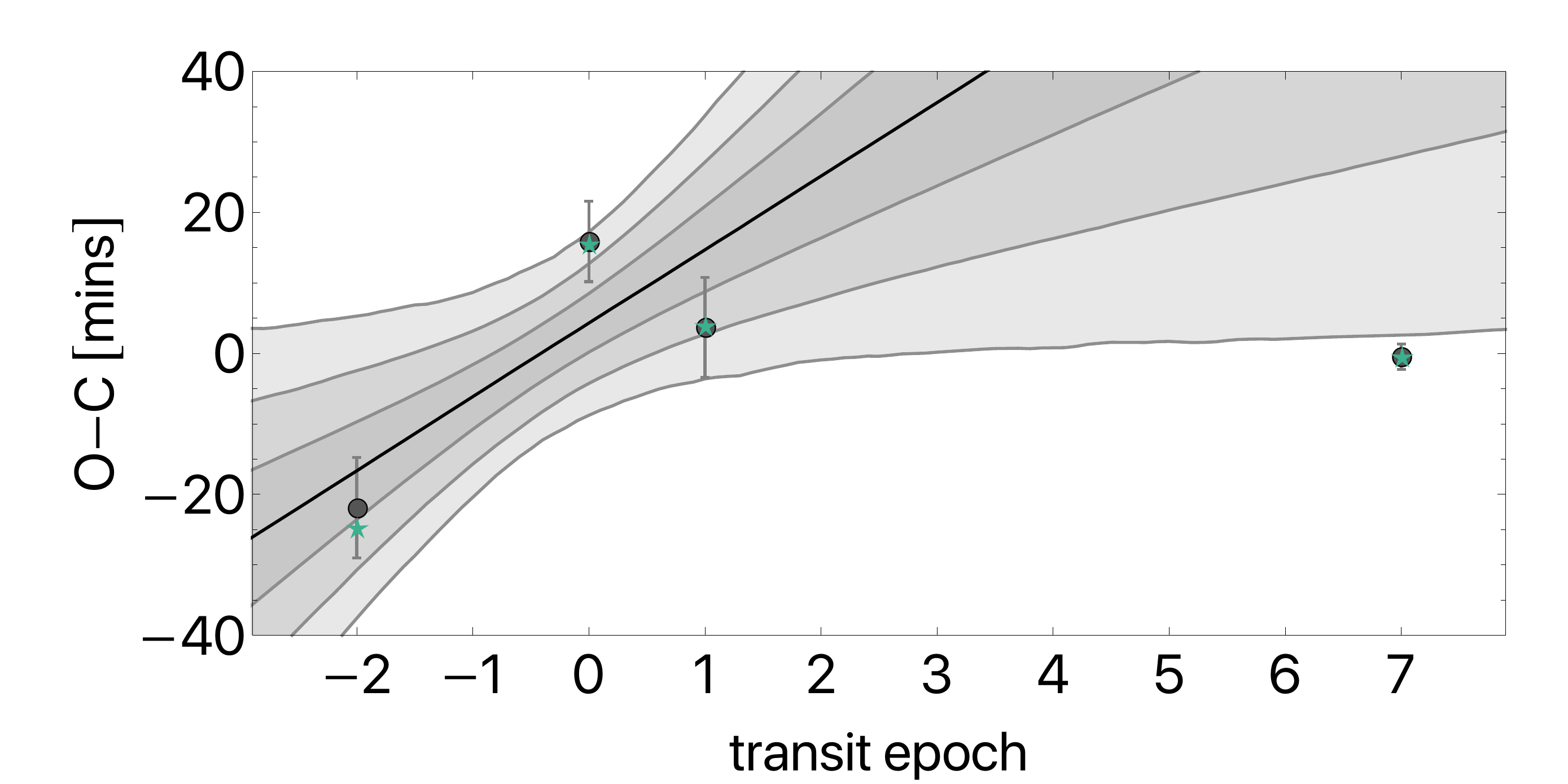}
\caption{\textbf{Transit timing variations.} TTVs for Kepler-1625b, defined as observed times minus calculated
times, where calculated times come from a linear ephemeris fit (model P) to
all of the data marginalized over the various visit-long trend models
attempted. The sloped line and
three shaded regions represent the median, one-, two- and three-sigma credible
intervals for the \textit{a posteriori} linear ephemeris when conditioned upon the
\kepler\ data alone, which reveals how deviant the HST epoch is. The
green stars indicate the O-C values produced by the moon model, M.}
\label{fig:TTVs}
\end{figure*}
\clearpage

\begin{figure*}
    \centering
    \includegraphics[width=15cm]{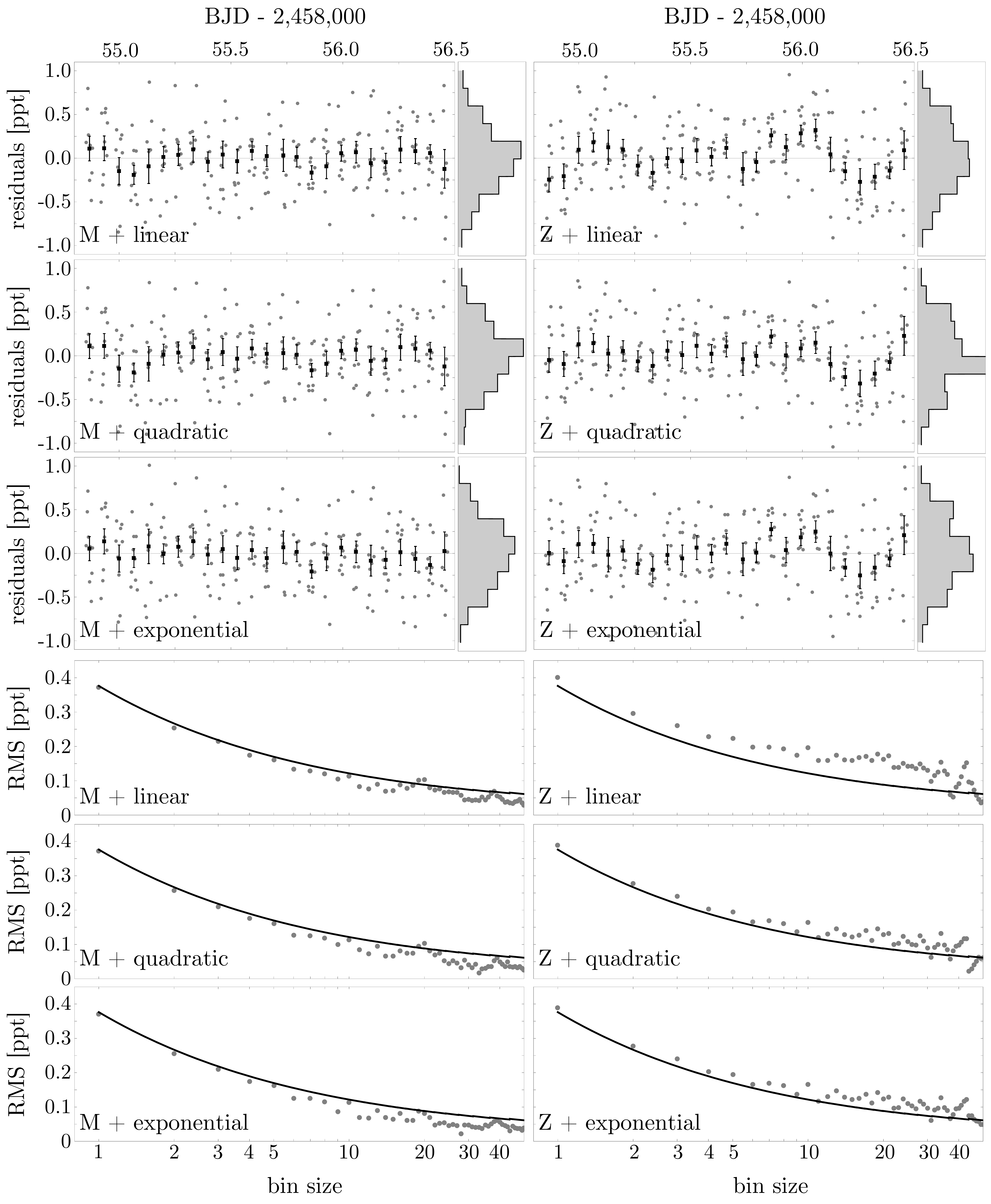}
    \caption{\textbf{Residual analysis.} Analysis of the white light curve from WFC3 residuals for six different models.
    Upper six panels show the photometric residuals, with orbit binned points, alongside 
    a histogram of the unbinned scatter. Lower panels show the root mean square
    (RMS) as a function of bin size, where the solid line is that expected for
    pure Gaussian noise equal to the assumed photometric noise in our fits. In both
    sets, excess noise in the Z-models is visible, caused by the moon-like dip being
    ignored in those fits.}
    \label{fig:residuals}
\end{figure*}

\begin{figure*}
\centering
\includegraphics[width=16cm]{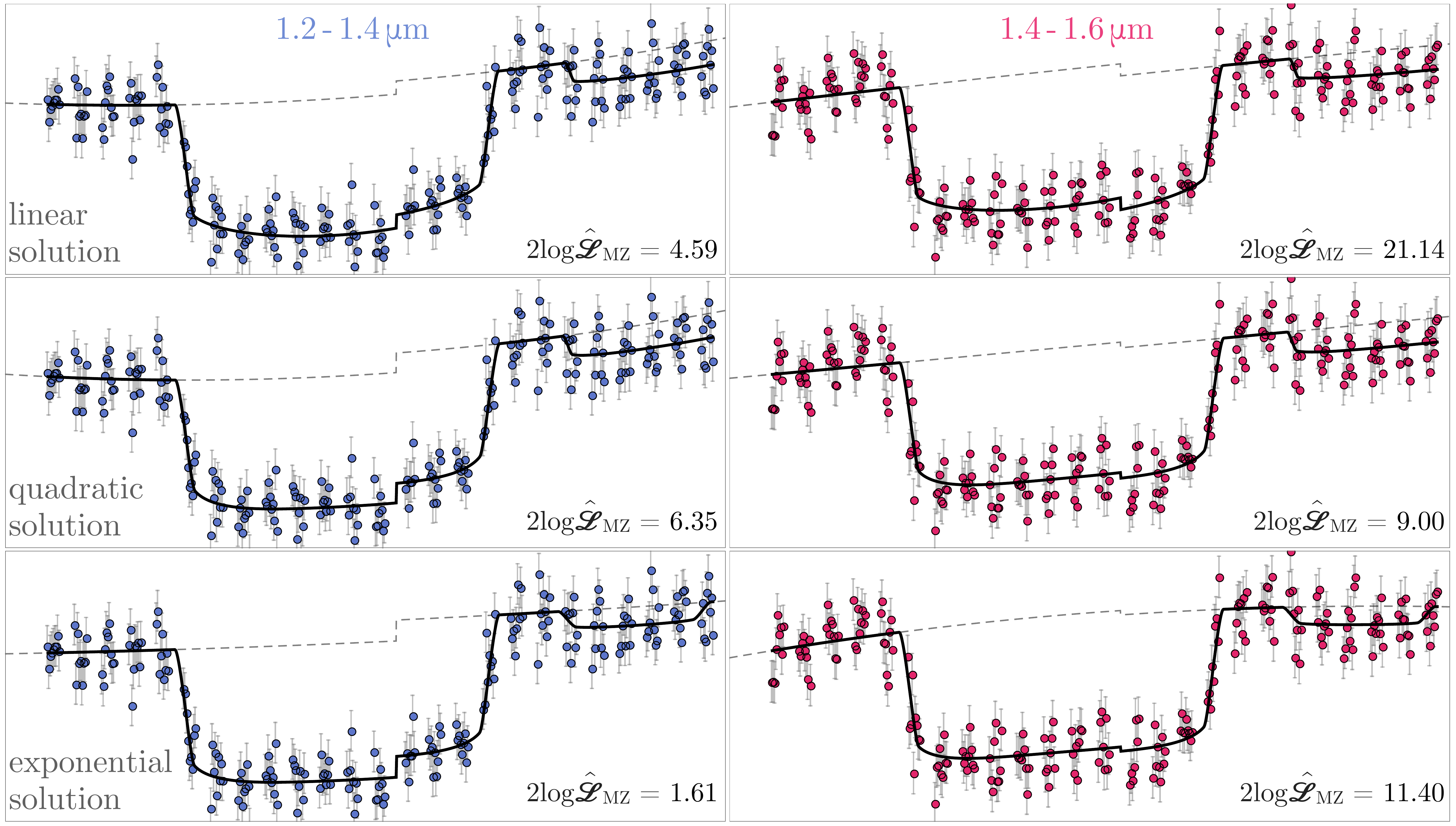}
\caption{\textbf{Chromatic test.} 
Tests to see if the moon-like dip is present in two independent
spectral regions of the WFC3 bandpass (each column). Each row
shows our maximum \textit{a posteriori} moon model (model M) plotted in
black, multiplied by a simple trend model regressed to each channel.
Since we have three different moon models depending on which trend model
is used on the white light curve, we show all three (one per row).
In every case, these templates give a closer match to the data than
those resulting from model Z (no moon transit).
}
\label{fig:chromatic}
\end{figure*}
\clearpage

\begin{figure*}
\centering
\includegraphics[width=16cm]{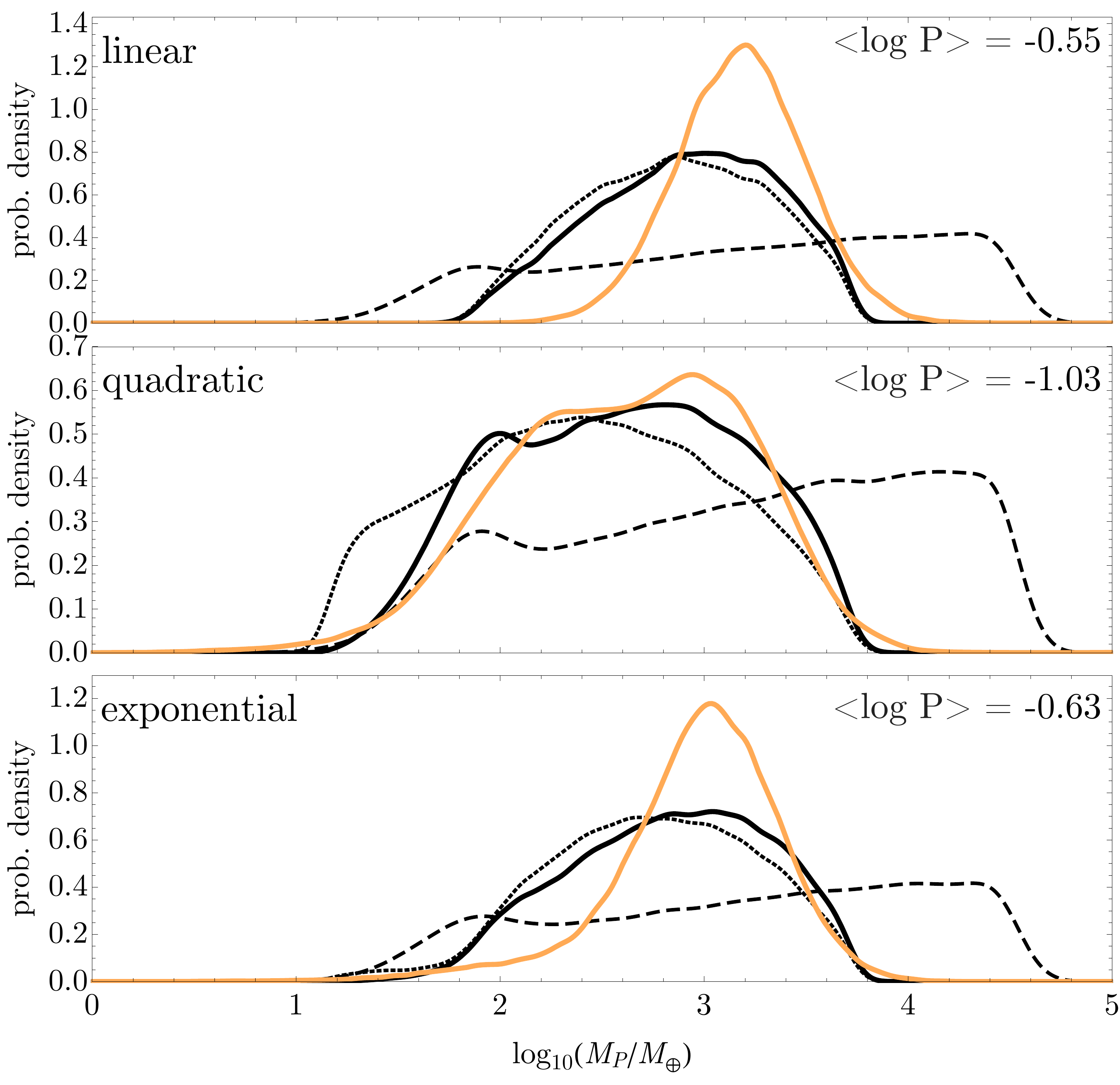}
\caption{\textbf{Mass constraints.} Mass solutions for the planet from the various HST detrendings. The
dotted lines represent the photodynamical posterior probability distribution
while the dashed lines are posteriors generated by \forecaster\
\cite{forecaster}. The solid black is the product of these probabilities. The orange
lines represent the mass solution for the planet derived from that of the
moon, which is well constrained based on the inferred radius. The mean likelihood,
tracking the compatibility of the two solid curves, is shown in the upper-right corner.}
\label{fig:masses}
\end{figure*}
\clearpage

\begin{figure*}
    \centering
    \includegraphics[width=16cm]{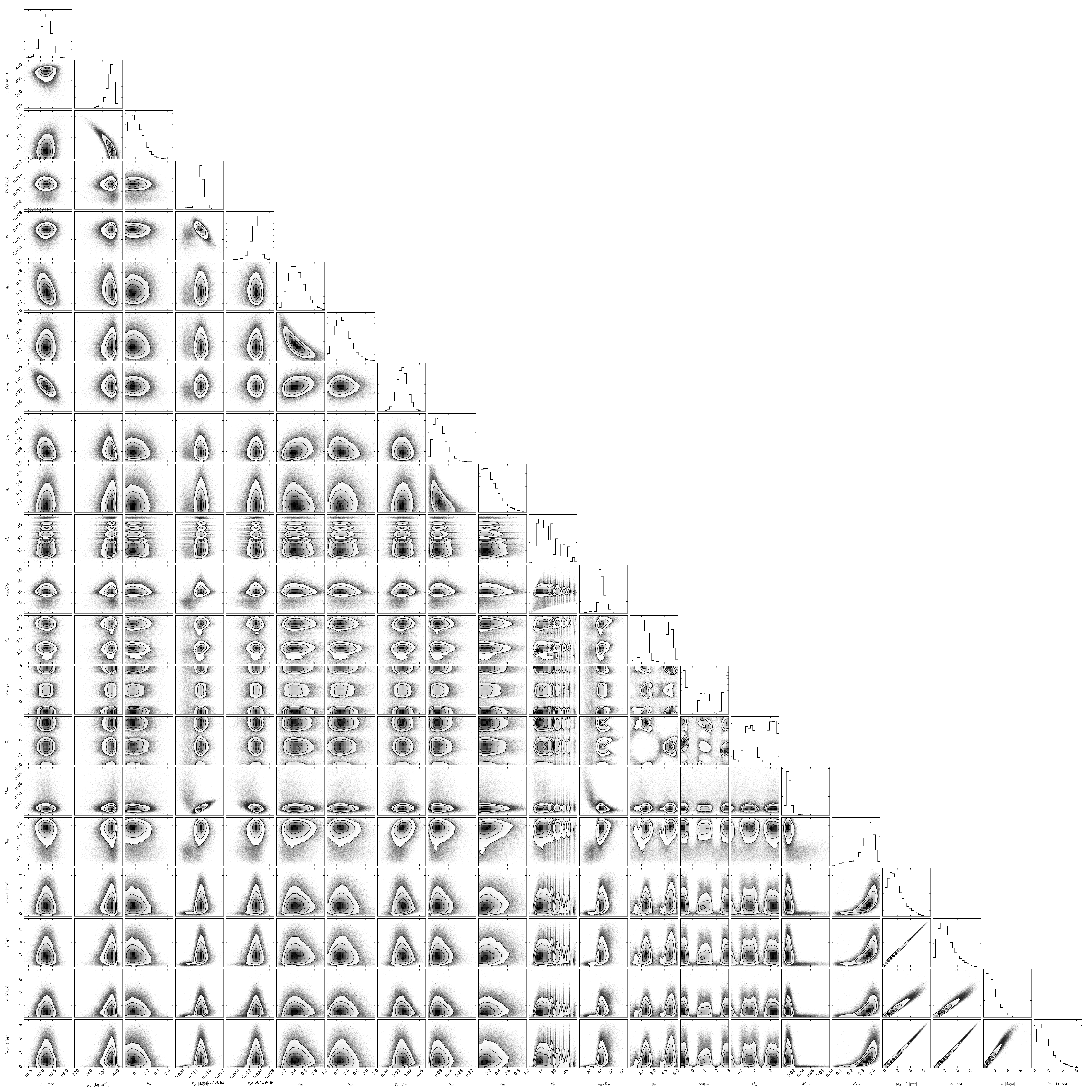}
    \caption{\textbf{Model posteriors.} Model posteriors of the parameters explored in the moon model.
    Shown here the results from the exponential trend model.}
    \label{fig:raw_posteriors}
\end{figure*}
\clearpage 

\begin{figure}
    \centering
    \includegraphics[width=16cm]{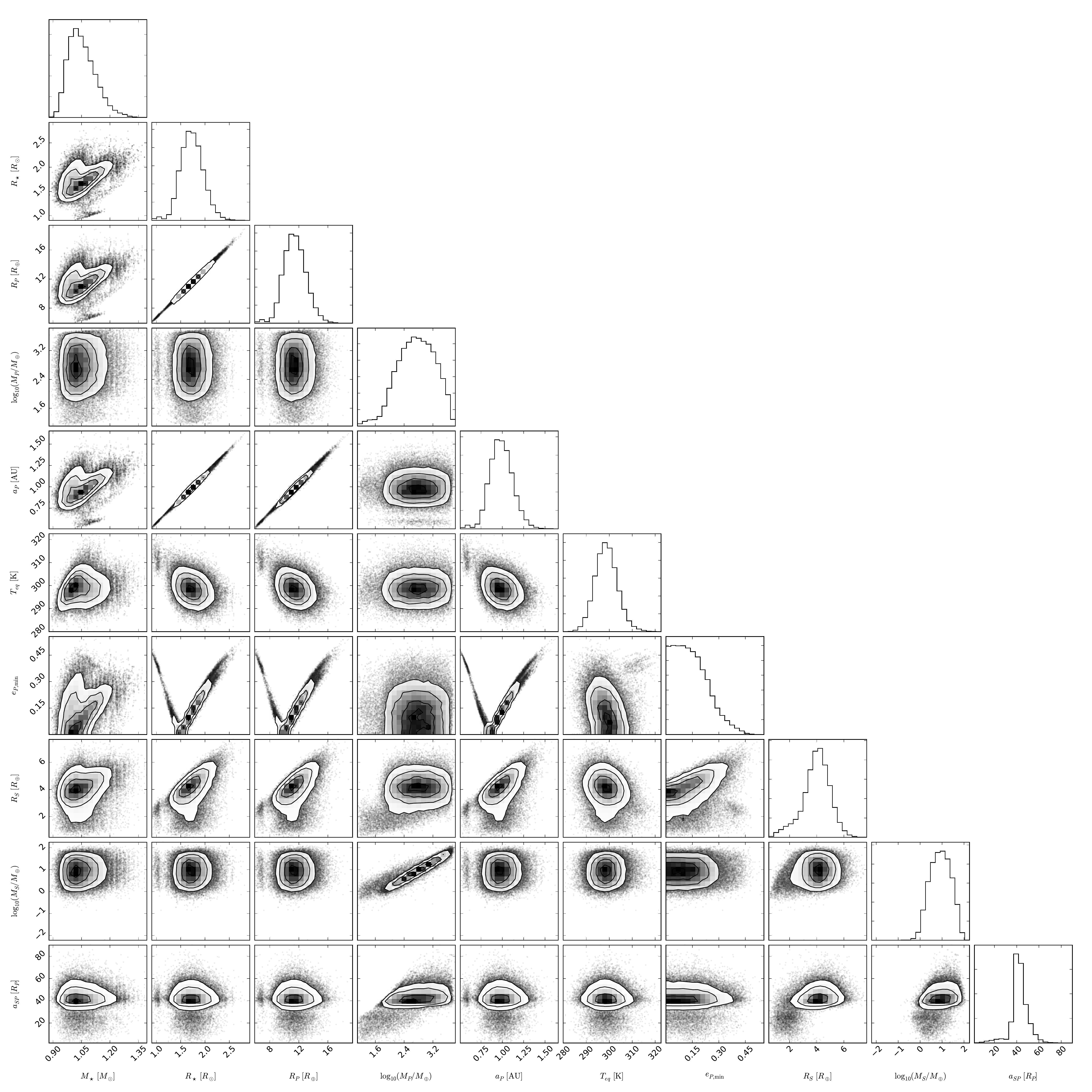}
    \caption{\textbf{Physical posteriors.} Physical system parameter posteriors derived from the exponential model results.}
    \label{fig:phys_posteriors}
\end{figure}

\begin{figure*}
\centering
\includegraphics[width=16cm]{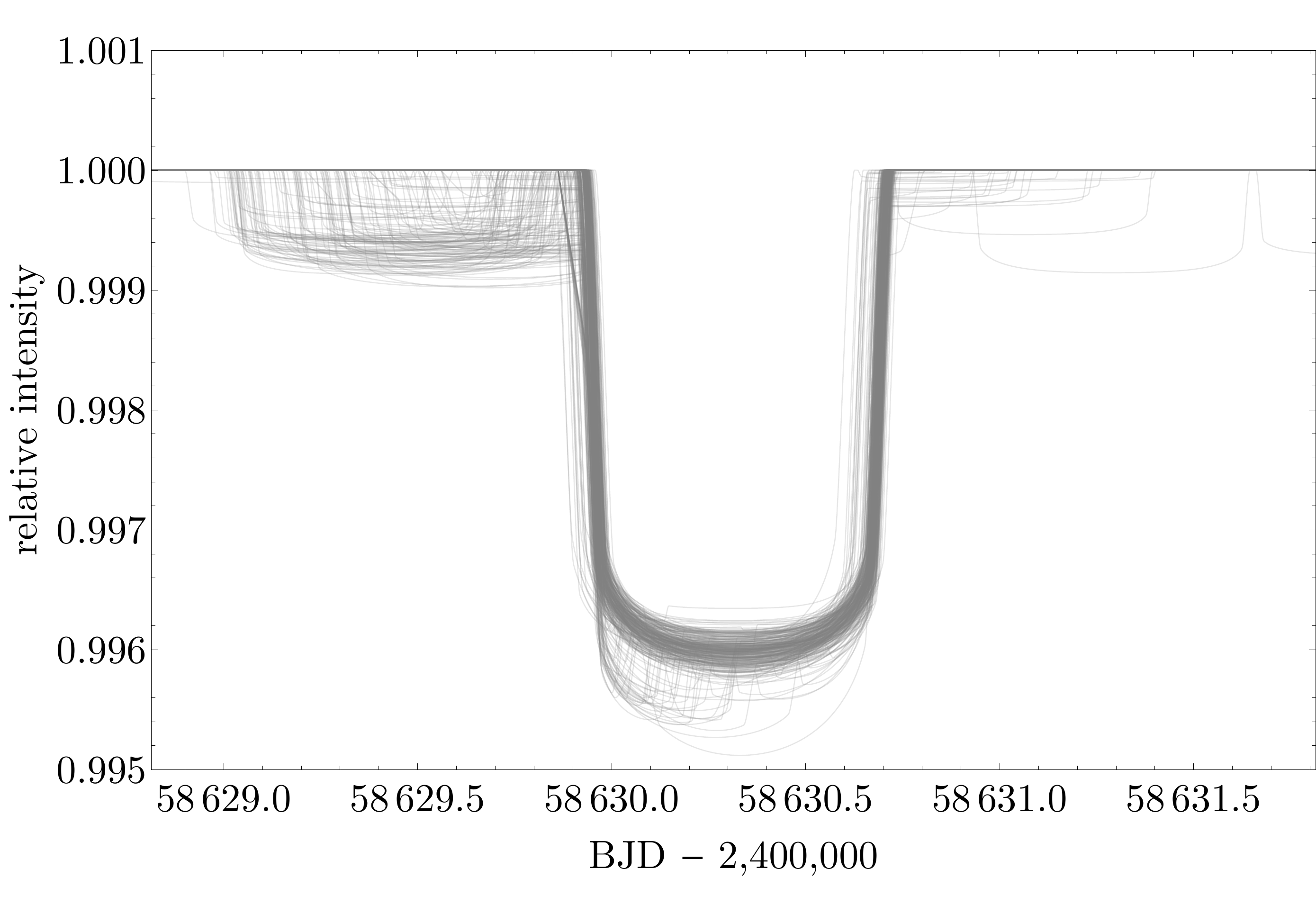}
\caption{\textbf{The May 2019 transit.}
Predictions for the May 2019 transit of Kepler-1625
assuming the planet-moon model is correct. 100 random
draws from each of the three instrumental trend model
posteriors are overlaid showing broad consistency between 
the three model predictions.
}
\label{fig:future}
\end{figure*}
\clearpage

\setcounter{figure}{1}
\renewcommand{\thetable}{S\arabic{figure}}
\setcounter{table}{0}

\input{tables/keplertable.tex}
\clearpage
\setcounter{figure}{2}

\input{tables/spectrum.tex}
\clearpage

\setcounter{figure}{3}
\input{tables/ttvtable.tex}
\clearpage

%% file: sections/kepler_analysis.tex
\subsubsection{Background}

The original analysis of the candidate exomoon signal is described extensively
in Section~8 of \cite{teachey:2018}. We refer readers to that work for the
details of our original interpretation of the data. We briefly recap here the
case for Kepler-1625b as a candidate exomoon host.

The analysis in \cite{teachey:2018} was the largest and most ambitious search
for exomoons to date, requiring many years to complete. Consequently, the data
used in that study were first downloaded on November 10th 2014. Since the
\kepler\ Science Processing Pipeline (see \cite{jenkins:2010}), built by the
\kepler\ Science Operations Center (SOC), evolved over the years of the primary
mission, the data analyzed in \cite{teachey:2018} does not represent the most
up-to-date data product at the time of writing. The three quarters in which
Kepler-1625b is observed to transit are quarters 7, 13 and 16. In
\cite{teachey:2018}, the simple aperture photometry (SAP) time series used were
produced by SOC pipeline v9.0.3 (corresponding to data release DR21) for
quarters 7 and 13 and v9.0.7 (DR22) for quarter 16. At the time of writing, the
most up-to-date (and the final) data release is DR25, for which quarters 7, 13
and 16 were processed by SOC v9.3.24, v9.3.29 and v9.3.31.

Based on our prior experience with \kepler\ data products, we did not
anticipate the \kepler\ photometry or case for the exomoon candidate would
significantly change as a result of going from SOC v9.0 to v9.3. Nevertheless,
in this more detailed and focused study, we decided to revisit the \kepler\
photometry and verify that this was true, as well as ensure that our results
are robust against choice of detrending method.

Joint modeling of the original data had suggested the presence of a large moon
in the system, with statistically significant drops in flux appearing on the
wings of these transit events. Based on DR25 estimates of the star's radius
($1.793 \pm 0.263 \, R_{\odot}$), the planet is approximately the size of
Jupiter. Meanwhile the photodynamical fits to the data, which require a
self-consistent moon model for all transit events, suggested the moon's radius
was comparable to that of Neptune (although with sizable uncertainty). No bad
data flags (such as reaction wheel zero crossing events) or anomalous pixel
behavior (such as that seen a previous candidate Kepler-90g; \cite{kepler90})
could explain the candidate signal at the time, nor could we identify any
astrophysical explanation besides the moon hypothesis that accorded well with
the data in hand.

Despite this, we argued that three transits were insufficient to claim
strong evidence for an exomoon detection, and so an additional validation observation was
sought and awarded on the Hubble Space Telescope (HST). Twenty-six orbits amounting to ${\sim}$40 hours were
awarded and the observations were executed on 28-29 October 2018. 

In our analysis of the HST data, it was necessary to perform a joint fit with
the \kepler\ photometry. To that end, we conducted a revised analysis of the
\kepler\ data, now with the SOC v9.3 data product. Upon initial inspection of
these results it was found that significant differences do exist between the
revised \kepler\ data and the earlier release used for this target. In
addition to the photometry, which has undergone noticeable modification, one
changing value in particular -- the crowding or blend factor
(\texttt{CROWDSAP}), a measure of aperture contamination by nearby sources --
appeared to play an important role in determining whether the moon model is
favored. This is because blending can introduce transit depth variations which
could be explained by a moon in proximity to the planet. Because this single
value was such an important part of the moon fit we sought to determine the
cause of this change from SOC v9.0 to SOC v9.3.


\subsubsection{KIC 4760471 - The Phantom Star}

In investigating the source of the modified crowding values we examined the
Data Validation (DV) report for Kepler-1625b closely. This examination revealed
that there is star included in the optimal aperture sky maps that does not exist.
The source, KIC 4760471, is clearly present in the DV maps, situated almost
directly between our target and the neighboring KIC 4760469, and in fact, the
star is purportedly brighter than our target by about half a magnitude (see
Figure~\ref{fig:DV_optimal_aperture_diagram}). However, we find no evidence of
this source in images from 2MASS, UKIRT, Pan-STARRS, \gaia, or our HST images. Nor
is the star included in the catalogue of nearby sources available on the
\kepler\ Exoplanet Follow-Up Observing Program (ExoFOP) website. We must
conclude that this star is a spurious inclusion in the KIC, and this raises
the question of how many other such stars there may be in the catalogue that
may be producing erroneous contamination estimates. Indeed, this is not the
first ``phantom star'' to be identified in the KIC; \cite{dalba:2017} also
reported the discovery of a spurious KIC inclusion that resulted a
$\sim 50\%$ transit depth change for Kepler-445c.

After modeling what \kepler\ should be seeing based on the published Pixel
Response Function \cite{bryson:2010}, we introduced KIC 4760471 into the
model to test whether the star was actually included in the models for
SOC v9.3 and earlier releases. We find that it cannot have been a part of the
SOC model, despite its inclusion in the DV report, as the brightness and
proximity of this phantom star would contaminate the optimal aperture by
$\sim50\%$ or more, and this is not reflected in the CROWDSAP numbers
published by the SOC. This ultimately led us to conclude that the SOC
v9.3 were accurate even if the DV report appears in error.

\subsubsection{Method-Marginalized Detrending the \kepler\ Data}

A critical aspect of light curve analysis is the removal of systematic trends
in the data. For the exomoon search this is especially important, as we must be
careful to neither produce nor remove the subtle signatures of the moon which,
unlike planetary transits, will not show morphological repetition from one
epoch to the next. Various approaches to detrending have been utilized in the 
literature, but no single method is considered the gold standard.

The Hunt for Exomoons with Kepler (HEK) project
developed the Cosine Filtering with Autocorrelation Minimization
(\cofiam) method \cite{HEK2}, which was specifically designed for the moon search, as it
focuses on preserving small / short duration features (possible exomoon transit
signals) in the vicinity of the planetary transit. The algorithm represents
the long-term trends as a sum of harmonic cosine functions with the longest period
component being equal to the twice the baseline. The algorithm forbids
components with periods (i.e. timescales) less than twice the transit duration,
meaning that the Fourier decomposition of a transit is not disturbed by
\cofiam. This removes any long-term trends in the data, be they
instrumental or astrophysical, while preserving short-duration events like
those expected from an exomoon transit. We refer the reader to \cite{HEK2}
for a more in-depth discussion of the method.

In addition, it is worthwhile to explore other detrending approaches to see
what effect they may have on the final results, precisely because the signal we
seek is so subtle. To that end we examined the results from a number of fairly
standard detrending approaches: a polynomial fit, a local line fit, a median
filter, and a Gaussian process. The polynomial method is identical to \cofiam\
except we replace the basis set from cosines to polynomials, exploring up to
twentieth order and selecting the locally-minimized autocorrelation result.
The local line fit is a polynomial fit up to twentieth order but only training
on data immediately surrounding (we used $\pm 80$\,hours) the transits and
selecting the order which minimizes the Bayesian Information Criterion (BIC).
The median filter uses a bandpass equal to five times the transit duration
to remove long-term trends. The Gaussian Process regression adopts a square
exponential kernel and trains on the entire quarter masking the transits.

The regressed trend models for the SAP photometry are shown on
Figure~\ref{fig:detrending_SAP}. We detrended both the SAP and PDC data with
all five methods, giving a total of ten light curves as shown in
Figure~1 of the main text. A visual examination of these results suggests no
clear favorite or obviously problematic detrending, but we may compare
residuals from the difference of two methods to gain a sense of where and to
what extent they differ. A matrix representation of these residuals can be seen
in Figure~\ref{fig:residuals_matrix}, where the maximum standard deviation
peaks at $\sim250$\,ppm, much less than the formal photometric errors on the
\kepler\ data of $\sim590$\,ppm. This suggests that the result is robust
against choice of detrending method.

We decided to marginalize over the different detrendings to create a robust
light curve to work with in what follows. To do this, we took the five
SAP light curves and computed the median flux at each time stamp. The PDC
data was not used here because the data have been corrected for contamination
effects already and thus are expected to be slightly offset from the SAP
results. The expectation then is that any anomalous features produced by one
detrending method will be mitigated. The uncertainties on each data point are
appropriately scaled up by quadrature addition of 1.4286 multiplied by the
median deviation across the different methods at each time step (which
had a median value of 34.6\,ppm), thereby yielding uncertainties accounting
for detrending differences. This effectively imposes a more stringent
requirement for the moon to make its presence obvious in the data, since larger
uncertainties provide more flexibility for the planet-only model. The final
light curve is presented in Figure~1 in the main text.

\subsubsection{Revised Photodynamical Fits of the \kepler-only Data}

Although we have new HST data in hand, we decided to re-analyze the revised
\kepler\ data in isolation before combining the data sets. The main
purpose of this exercise was that later results can be placed in better
context to assess which data set is primarily responsible for any interesting
signals found. In \cite{teachey:2018}, Bayesian photodynamical fits were
conducted on Kepler-1625b to test for the presence of a possible moon. This
led to a Bayes factor between the planet+moon (M) and planet-only (P) model
which would be classified as ``very strong'' evidence on the \cite{kass:1995}
scale ($2 \log K = 20.4$). The third transit in particular seemed to dominate
the signal and the strong dependency of our inference on just a single
epoch led us to conclude that the case for an exomoon remained ambiguous
despite the high Bayes factor (and ultimately led us to pursue follow-up
observations).

Using the revised SOC v9.3 data, we performed the same fits for models M
and P, as well as a transit timing variation model, T, and a zero-radius
moon model, Z. The evidences, derived using \multi\ \cite{feroz:2009},
are presented in Table~\ref{tab:kepler}.


From this table, it is immediately apparent that the case for an exomoon
is dramatically weakened using the revised \kepler\ data, going from
$2\log K \simeq 20$ to $2 \log K \simeq 1$. This naturally raises the
question as to what exactly caused such a large decrease. In total,
four aspects of the analysis have changed:

\begin{enumerate}
\item Revised analysis used a slightly contracted baseline
\item Revised analysis used method marginalized detrending, rather than
just relying on \cofiam
\item Revised analysis used updated and thus different crowding factors
\item Revised analysis used the SOC v9.3 data products, rather than the
SOC v9.0
\end{enumerate}

In principle, one or more of these must be responsible for the decrease
in the evidence for the exomoon hypothesis and here we attempt to distill
by elimination which one(s) it is.

The baseline (which is the temporal window around each transit used in the
\multi\ fits) in our revised analysis was contracted from the original
analysis. Because \cite{teachey:2018} were blindly searching for moons in an
ensemble out to 100 planetary radii from the host, transits were detrended with
a baseline equal to an estimated 150 planetary radii. That calculation,
detailed in \cite{teachey:2018}, was not only based on now out-dated system
parameters, but also is excessive for the purpose of validating the exomoon
candidate Kepler-1625b-i which has a hypothesized semi-major axis in the range
of 20 planetary radii. We estimated that 80 hours on either side of the transit
would easily accommodate the features of interest in our new analysis
(whereas 98.8\,hours was used in \cite{teachey:2018}). Nevertheless, this
difference could somehow explain the change and so we conducted a controlled
experiment where we fit the exact original data from \cite{teachey:2018},
but contracted the baseline to $\pm 80$\,hours. Rather than reducing the
evidence, this actually boosted it slightly, with $2 \log K$ increasing from 20.6 to 24.4 
(here and in what follows we quote the Bayes factor between the M and P models only). 
This increase can be understood when considering the fact that a larger fraction
of the data are affected by the moon-like signals. We also verified we
recovered the same evidence as \cite{teachey:2018} with all inputs left the
same in two independent fits, which both gave the same result to within 0.6
in $2 \log K$. These tests demonstrate that the contracted baseline is
certainly not responsible for the decreased evidence, as we might reasonably
expect.

Our second hypothesis was that it was the different detrending algorithms used
that caused the difference. We therefore compared the fit that used the
original data but truncated baseline (from the previous paragraph), with a
fit that was identical except that the photometry was detrended using the method
marginalized approach, rather than \cofiam\ alone (but still using the SOC
v9.0 data products and blend factors in both cases). This again only led to a
higher evidence for the moon, with $2\log K$ now reaching 29.2. This therefore
establishes that the decreased evidence for the moon model described at the
beginning of this subsection, is not caused by either the baseline or
detrending method used by \cite{teachey:2018}.

This leaves two remaining possibilities: the blending factors and/or the actual
photometric data products. We took the method marginalized light curve produced
from SOC v9.0 and fit it in two ways; one using the crowding factors produced
by v9.0 and one using those from v9.3 - but everything else kept the same. The
former of these two fits corresponds to the $2\log K = 29.2$ case of the
previous paragraph. The latter case gives a greatly reduced $2\log K = 14.8$.
Thus, the contamination factors must be, in part, responsible - likely as a result
of transit depth variations.

The other possibility was investigated by repeating the previous experiment
but instead comparing to a fit where the blend factors are those of v9.0 but
the data input to the method marginalized detrending algorithm comes from the
SAP light curve of v9.3. This causes the Bayes factor to decrease from
$2\log K = 29.2$ to $2\log K = 6.8$, a decrease even greater than that due to
blending.

We are therefore able to deduce that our revised analysis of the \kepler\
data alone leads to decreased evidence for an exomoon as a result of two
changes introduced between SOC v9.3 and v9.0: i) the new blend factors
and ii) the new SAP photometry. The choice of baseline and/or detrending
method are certainly not responsible. Comparing the light curve from
\cite{teachey:2018} and/or the revised method marginalized version on
the same v9.0 data products reveals that the largest difference, versus
the v9.3 product, is the third transit in quarter 16. This transit appears
distorted and asymmetrical in the original analysis, explained by a large
moon carving out flux around the planetary transit. The new data shows
a much cleaner signal more closely resembling the other two transits.

%% file: sections/hubble_obs.tex
\subsubsection{Observation Design}
\label{obs:design}

In a planet-moon transit event, an in-situ observer would see the moon either 
leading or trailing its host planet, and in rarer cases the event could begin 
with both bodies passing in front of the star simultaneously. To adequately observe 
an exomoon transit it is therefore imperative that observations begin well before the 
anticipated planetary ingress and conclude well after planetary egress, as we expect 
to see flux reductions due to the presence of a moon in these parts of the light curve. 
The separation in time between the planet and moon ingresses / egresses is directly 
connected to their sky-projected separation, which will be a function of the moon's 
semi-major axis.

With a photodynamical moon model in hand from fits to the \kepler\ data it was possible 
to run the model forward in time to generate expected light curves for the October 2017 
transit, though the morphology of the transit was poorly constrained when projected five 
epochs into the future. To determine the best start and end times for the HST observation 
we drew from the model posteriors and generated forward models based on these inputs, 
generating a range of possible outcomes. We set our start and end times such that exomoon 
ingress and egress features would be captured in the observing window to 95\% confidence. 
We requested a start time as close as possible to Barycentric Julian Date 
(BJD\textsubscript{UTC}) 2458054.8 and ending no earlier than BJD\textsubscript{UTC}= 2458056.5. 

For the observation we selected the G141 grism on Wide Field Camera 3 (WFC3), which is 
sensitive from $\sim$ 1.1 to 1.7 microns.  This choice was motivated in part by the 
expectation that stellar variability, to the extent it is present, should be suppressed 
towards the infrared.  In addition, because the observation would amount to some 40 hours on 
target, we knew that HST would pass through the South Atlantic Anomaly (SAA) for a significant 
fraction of this time, and WFC3 is one of the few instruments aboard HST that may be used 
during passage through the SAA. The grism creates a dispersion spectrum such that the light 
from each source is spread across the CCD, providing spectral information on the target and 
its neighbors. Use of the grism also allows for longer exposures, as it takes much longer 
for a bright target to saturate. It is worth pointing out that time on target is of paramount 
interest in carrying out such an observation; we wish to minimize telescope overheads, which 
can include data readouts that interrupt the observation, and the RMS of the observation is 
directly tied to the time on target.

The drawback to using the grism is primarily the fact that a suitable roll angle for the 
telescope must be selected, one that minimizes overlap between the target spectrum and 
neighboring spectra. With each spectrum illuminating roughly 135 pixels in the direction 
of dispersion and several pixels on either side of the spectrum's central line, crowded 
fields must be modeled with care. Depending on the time of observation and the orientation 
of HST with respect to the Sun at that time, suitable roll angles may not be available. 
For the field containing Kepler-1625 we found only $\sim$ 20 of 360 degrees suitable for 
a grism observation.

We used the HST Exposure Time Calculator to plan the observations 
(ETC ID WFC3IR.sp.912655). 
Inputs were 
1) the star's J-band magnitude (14.364$\pm$0.029) from 2MASS; 
2) $E(B-V)$ = 0.19, taking line of sight extinction $A_V$ to target = 0.594 (NED value taken 
from \cite{schlafly:2011}, and assuming $R_V = 3.1$; 
3) a built-in Pickles model spectrum for a G0III star with $T_{eff}$ = 5610 K (the closest 
match available to Kepler-1625); and 
4) zodiacal light, Earth shine and air glow models for the date of observation, also built-in 
options in the ETC. 
We found that a 300 second exposure would fall well short of the time to saturation, which the
ETC calculated to be 508.74 seconds. We opted not to utilize HST's spatial scanning mode, which 
moves the spacecraft perpendicular to the direction of spectral dispersion during each exposure, 
thereby spreading the light onto more pixels. Spatial scanning was inappropriate for our observation, 
both because we were not close to reaching saturation (typically the primary motivation for spatial 
scans), and because the crowded field meant the spatial scan would have to be extremely short,
and potentially unachievable by the spacecraft given the length of each exposure. We note also that 
the software provided by STScI for reduction of the HST data does not currently support spatial 
scanning mode. 

The standard data reduction pipeline distributed through STScI, aXe \cite{axe} requires at least 
one direct image of the target (i.e. without the grism) so that spectral calibrations can be made.
Using source locations derived from the direct image using SourceExtractor \cite{SExtractor}, the 
aXe software is designed to calculate the position of each spectrum, the wavelength solution for 
each pixel, and contamination from nearby sources. A single direct image using the F130N filter 
was made at the start of the observing run with an exposure time of 103.129 seconds, followed by a 
total of 232 grism exposures. All data were taken in the 256$\times$256 subarray to reduce overheads.
Each grism exposure lasted 290.776 seconds, well short of the saturation time. \texttt{SAMP-SEQ} was 
set at SPARS25 and \texttt{NSAMP} was 14, meaning there were 14 non-destructive readouts of each 
exposure at roughly 25 second intervals. These multiple readouts are useful in part for rejecting 
cosmic rays within the front-end data processing pipeline \texttt{calwf3}.

Roughly 3100 seconds, or 51.67 minutes, were available for observing the target during 
each orbit. The remaining time in each orbit (about 44 minutes) were unusable as the target 
was occulted by the Earth. For all but two orbits, there were only 72 seconds of unused visibility 
time. For the first orbit 130 seconds went unused, and for the 15\textsuperscript{th} orbit, 
270 seconds went unused. This was due to increased overheads, namely a full guide star acquisition, 
which takes longer than re-acquisition and reduces the number of exposures that can be made in an orbit. 
Each exposure was long enough that data dumps could be made in parallel; therefore, there were no gaps 
between the first and last exposure in a given orbit.

It was found that only three of our exposures would occur during passage through the SAA, 
which was fortuitous. For most of the observation SAA passage was restricted to times when 
the target was occulted by the Earth, when no data could be taken. For these SAA-affected 
exposures the Fine Guidance System (FGS) cannot be used; instead, HST must use gyro control 
to stay on target. The gyros are known to have a pointing drift on the order of 1.5 mas 
per second, meaning that for our ${\sim}$290 second exposure we could expect the target 
to drift by approximately half an arcsecond.

\subsubsection{Execution}
\label{obs:execution}

Observations began at 6:52:15 (UTC) on 28th October 2017 and ended at 23:20:08 on 29th October 2017. 
No malfunctions of the spacecraft were detected. Three of the 232 exposures were indeed affected 
by passage through the SAA, in line with expectations. While the SAA had no discernible effect 
on the photometry itself, pointing drifted considerably during these exposures, as expected. 
The spectra from these images, landing on neighboring pixels and significantly dilluted by smearing, 
showed significantly different flux levels, even with appropriate pixel sensitivity corrections, 
and were therefore left out of our final analysis.

The final orientation of the spacecraft (provided in the header as the position angle of 
HST's V3 axis) was $\sim249$ East of North and produced a clean spectrum of the target 
with minimal contamination from nearby stars.   

\subsubsection{Data Preparation}
\label{obs:preparation}

For initial processing of the raw data files from STScI we followed the WFC3 IR Data Reduction 
Cookbook, using updated configuration and reference files where 
available, and making minor updates to source code when deprecated packages were still 
in use and caused fatal errors. We refer interested readers to the Cookbook for details
on the data reduction. We utilize the \texttt{.flt} files produced by the front-end 
\texttt{calwf3} pipeline, which account for a number of known systematics and perform 
cosmic-ray rejection. We point out that while the \texttt{calwf3} performs a flat 
field correction, for grism observations the flat field division is unity everywhere, 
as the flat fielding for grism images is intended to be performed later as a part 
of the spectral extraction process.

The standard aXe software is unable to handle images taken in subarray mode, 
so each $256 \times 256$ image must be embedded in a larger $1014 \times 1014$ array 
before processing. It is important that the images be embedded in the correct position, 
as subsequent flat fielding and background removal are performed on a pixel-by-pixel basis.
We modeled our own embedding code after a python script provided by STScI, which we 
were unable to run on our machines. We verified the embedding was done correctly by eye, 
matching obvious artifacts in the master sky image and the flat field cube to the 
same artifacts that are readily apparent in the images pre-correction. With this step 
completed we could follow the instructions of the cookbook.

Unfortunately we were unable to follow the prescribed reduction beyond the 
\texttt{axeprep} stage of the cookbook, as we experienced a persistent breakdown with 
the \texttt{axecore} task which we were never able to resolve. The \texttt{axecore} 
tool is responsible for automated spectral extraction. From here on we wrote our own 
code to analyze the data. 

Following the STScI prescription, the background flux levels are calculated by 
finding the median flux of the image, masking out pixels with greater than 5 
electron counts. The Master Sky Image \cite{master_sky}, calculated from in-orbit 
observations, is then scaled up by the median background flux level at each 
time-step and subtracted. This step removes the well-documented ``breathing'' 
effect that appears in HST time-series observations, which has been attributed 
to thermal fluctuations in the telescope as it orbits the Earth. An additional 
flat-fielding must then be performed, using the flat field data cube produced 
from pre-flight laboratory testing. The flat field cube encodes four polynomial 
coefficients at each pixel location to model the pixel sensitivity as a function 
of wavelength.

The \texttt{axecore} task is designed to calculate a wavelength solution for 
each pixel in the spectrum, and with this information a wavelength-specific 
flat-fielding can be performed. However, due to the code breakdown we did not 
have wavelength solutions in hand until later in the analysis. While the pixels 
show significant sensitivity variation across the CCD, we found there was minimal 
variation across the wavelength range for any given pixel (typically less than 1\%), 
so we opted to compute the median sensitivity for each pixel and we used these 
values to perform the flat-fielding. We are thus marginalizing over intra-pixel 
sensitivity variations, but we point out that STScI documentation also suggests 
G141 grism observations may be flat-fielded using the F140W flat field which also does 
not encode intra-pixel sensitivity variations. We compared our median G141 flat 
field image to the F140W flat and determined that flattening the G141 cube was a more 
faithful rendering of the wavelength-sensitivity information for the grism. 

\subsubsection{Imputation of Bad Pixels}
\label{obs:imputation}

Following the flat field corrections there remained $\mathcal{O}(10^3)$ pixels
that were clearly outliers, as evident from a median stack of the images. Wherever 
possible, we elected to perform imputation of all time stamps associated with these
outlier pixels. The tasks of outlier identification and imputation are treated
independently.

To identify the outliers we start by calculating median pixel values across our
observations, which should reveal pixels that behave anomalously in more than
50\% of the exposures. This superstack is done on all images grism images, except for
numerous images which were identified during preliminary analyses as behaving
in a non-representative way: namely cadences 107, 116, 125 and 126; as well as
orbits 1 (telescope settling), 7 (transit ingress) and 18 (transit egress).
Each superstack pixel has an uncertainty equal to 1.4286 multiplied by the
median absolute deviation (MAD) across all times.

Since the grism spreads the target spectrum along the pixel rows, we expect (and indeed
observe) that the observed flux follows a smoother pattern along the rows than
the columns. Exploiting this fact, we extract each row's flux vs column index
from the superstacked image and look for outliers along each row independently.

We do this by constructing a 3-point moving median and then training a Gaussian
process (GP) with a squared exponential kernel on the result. This GP is then
evaluated on all column indices, computing residuals as we move along. A
median-version of the reduced $\chi^2$ is computed to re-scale. We define
residuals as the difference from the GP normalized by the pixel’s uncertainty. 
We also scale the uncertainties such that the median version of the residual's
reduced $\chi^2$ equals unity. Outlier pixels are then flagged as those which
depart more than 10\,$\sigma$ (chosen after some experimentation) away
from the GP model. Note that our GP lacks predictive power on the edge columns
and this process does not consider column pixels during the search.

In addition to this procedure, we flag pixels as an outlier if the pixel's
derived uncertainty across all images exceeds 20\,$\sigma$ of the median
error of the row’s pixels, where $\sigma$ is again coming from another MAD. 

If a pixel is flagged as an outlier but has an immediate neighbor of similar
flux, we remove the outlier flag. This is done by first computing the maximum
deviation of each superstacked pixel with its row-wise precursor and successor
and then seeing if the candidate outlier is less than 10\,$\sigma$ away from
the median deviations seen in that row (where again $\sigma$ comes from the
MAD). This was necessary to avoid killing zeroth-order spectrum features which
look like islands of outliers in a single row (though we do not attempt to use
zeroth-order features for analysis, as the target's zeroth-order is off the CCD).

Imputation is not performed using medians as this essentially represents a
zeroth order polynomial which is not sufficient to capture the gradients
observed across the pixels. Instead, for each pixel in the image we produce a
predicted flux based on a 1-dimensional spline interpolation with two pixels
preceding and two pixels following it in the row. We found that a 1-D row
interpolation is superior to a 2-D interpolation, as the latter does not
adequately handle pixel replacement across the spectrum peaks, that is,
perpendicular to the direction of dispersion. If the pixel has been
flagged as an outlier it will be replaced with the predicted flux. An
exception to this rule is if one of these four training pixels is itself
an outlier. In such cases, we flag the pixel as an irreplaceable outlier.

After the first round of outlier identification and imputation, we ran
the algorithm a second time. This process led to the identification of
1756 outlier pixels, of which 634 were irreplaceable whose fluxes were
simply set to NaN after this point to mask them.

Finally, we note that the spectra produced by the grism are inclined 0.5
degrees with respect to the pixel grid, per WFC3 Grism documentation. 
We therefore use a standard SciPy package \cite{scipy} to rotate the image
clockwise by 0.5 degrees, performing a 3rd-order spline interpolation, thereby
aligning the spectra with the $x$-axis. This simplifies the extraction of the
spectra considerably, as the optimal aperture may be neatly aligned with the
image grid, and produces no discernible artifacts in the spectra. The pixel
errors must also be rotated, which is potentially problematic if the errors
across the image were random. However, since the errors scale predictably with
flux levels this rotation is also well behaved, and the resulting distribution
of errors across the image is unchanged from the native images.

\subsubsection{Optimal Aperture}
\label{obs:optimalaperture}

The target's point spread function is centered in the rotated images at
approximately $x=515$ and $y=531$ (where $x$ represents the column index
and $y$ the row index). To extract photometric time series, we
elected to employ simple aperture photometry rather than modeling the complex 
point spread function (PSF) observed. This is well justified since the high
angular resolution of HST, combined with our observational design, means
that we do not see significant overlap of neighboring sources with the target.

In choosing an aperture, we could simply draw a broad box around the target by
hand, but instead we elected to choose an optimal aperture which minimizes the
scatter in the final target light curve. The optimal aperture was found in a
two-step process. First, we setup a grid of 105,840 candidate apertures where
each permutation has a unique aperture defined by four parameters,
$\{x_{\mathrm{min}},x_{\mathrm{max}},y_{\mathrm{min}},
y_{\mathrm{max}}\}$, such that the aperture is bounded by $x_{\mathrm{min}}
\leq x \leq x_{\mathrm{max}}$ and $y_{\mathrm{min}} \leq y 
\leq y_{\mathrm{max}}$. The grid of candidates spans the range 
$426 \leq x_{\mathrm{min}} \leq 480$, $481 \leq x_{\mathrm{min}} \leq 599$,
$515 \leq x_{\mathrm{min}} \leq 530$ and $532 \leq x_{\mathrm{min}} \leq 544$,
where we step between the extrema in 2-pixel intervals. We remind the
reader that these pixel values do not correspond to the native images from
HST, but to our rotated image, for which there is an offset. In each of these
candidate apertures, we extract a white light curve for the target and
correct for the hook effect using a simple exponential ramp (explained in
detail in Section~\ref{obs:hooks}). While we eventually developed our own
approach to removing the hook trends, this simple model does reasonably well
correcting for charge trapping, and thus observations are expected to
be stable within each orbit, although visit-long trends have not been
corrected for at this point.

As visit-long trends persist, and there are of course flux decreases caused
by Kepler-1625b's transit, as well, comparing the raw root-mean-square (RMS) of each
candidate aperture's is not an appropriate cost function to score the different
apertures. Instead, we reasoned that if we mask the times during the ingress and
egress of the planetary transit (which take up one HST orbit each; orbits
7 and 18 respectively), then we should expect the photometry to be stable
within each orbit (but not necessarily between each orbit). We further mask the
first orbit, which appears to represent a settling-orbit for the photometric
behavior and is typically discarded in similar studies. Finally, we also mask
time stamps 107, 116, 125 and 126 where we later came to suspect outlier
behavior. The remaining 202 points (of the original 232) are then grouped into
their respective orbits and the RMS of each is computed. We then define a cost
function as the mean of these RMS values (23 in total).

This process identified an optimal aperture defined by $456 \leq x \leq 581$
and $526 \leq y \leq 542$. However, the grid search used a resolution of
2-pixels in its search and further more used a fast but sub-optimal hook
correction method. As described in Section~\ref{obs:hooks}, we found a novel
non-parametric hook correction method is able to out-perform the exponential
model and better capture the sharp hook morphology. We therefore performed
a second-stage in our search where we essentially walk the aperture in
1-pixel intervals away from the previously found solution. Each bound
(i.e. $x_{\mathrm{min}}$, $x_{\mathrm{max}}$, $y_{\mathrm{min}}$,
$y_{\mathrm{max}}$) is perturbed by $\pm 1$ pixel to create 8 candidate
grids, as well as the original solution to give a ninth. Across these
9 possibilities, we extract photometry as before but this time perform the more
computationally intensive non-parametric hook correction described in
Section~\ref{obs:hooks}. If a better aperture is found amongst the 9 options,
we walk to that solution and repeat, else we stop.

In practice, we perturbed the optimal aperture from stage one adding a random
integer between -2 and +2 to each bound and walked from that position, in
order to test if the walker would return to the same solution. Indeed this
is what happened and the final optimal aperture returned to $456 \leq x
\leq 581$ and $526 \leq y \leq 542$, which has mean intra-orbit RMS of
375.5\,ppm. In what follows, we set value that value, 375.5\,ppm, as the standard
photometric error for this optimal time series.

\subsubsection{Modeling the Hooks}
\label{obs:hooks}

A well-known feature of time-series observations on HST are the exponential
ramps or hooks, e.g. \cite{berta:2012, deming:2013}; a phenomenon that's also
been observed in \textit{Spitzer} data, e.g. \cite{deming:2006, knutson:2007,
charbonneau:2008, agol:2010}. As the observation begins, the flux readings ramp
up with each subsequent exposure towards a saturation asymptote. This is
thought to be due to charge trapping in the CCD \cite{agol:2010}. Once the
observations are interrupted, either due to occultation of the target by the
Earth or through a non-parallel data dump, the ramps resume.

Common previous approaches for removing this systematic include templating the
ramps from the out-of-transit orbits to detrend them in all other orbits
\cite{berta:2012}, and assuming a parametric model fit (typically using
exponential functions) to each ramp for removal \cite{agol:2010}. The templating
approach is not ideal for our observations since we do not know which orbits
are in- or out-of-transit \textit{a priori}, due to the candidate exomoon. For this
reason we initially pursued an exponential ramp model. The results from this
approach were certainly reasonable and provided a clear transit recovery.
Despite this, in our quest to extract as much information as possible from
these observations, we devised an alternative strategy that ultimately provided
a superior correction.

The inspiration behind our new approach can be seen in
Figure 
2 of the main text. Each orbit is typically comprised of 9
exposures, which are shown with distinct colors in the figure, 
and can be labelled with the index $v=1,2,...,9$. If one considers 
just the $v=1$ points, the light curve appears remarkably clean, and the 
same argument holds true for any specific choice of $v$. This is 
understandable if we consider the fact that each observation $v$ shares 
a common observational history; that is, the $v=1$ exposures all occur 
immediately following target acquisition after occultation,  the $v=2$ 
exposures all have a common history following $v=1$ observations, with 
all the charge trapping associated with those exposures, and so on. These 
common histories will then act as a baseline flux level for each $v$ that may 
be independently corrected. We therefore hypothesized that these nine light 
curves could be combined by simply scaling them independently to create one 
coherent light curve. We assign 9 scaling factors, $\gamma_v$, to each and 
treat these as unknown parameters to be solved for.

Following our earlier argument in Section~\ref{obs:optimalaperture}, we
expect the hook-corrected light curve to exhibit stable intra-orbit
photometry (but not inter-orbit). We therefore use the same cost function
as used earlier, namely the mean intra-orbit RMS. We iteratively optimize
for the $\gamma_v$ terms until the cost function improves negligibly. The
final $\gamma_v$ terms are shown in Figure~\ref{fig:ramps}, where we also
overplot the optimized exponential ramp model for comparison. This plot
reveals that the exponential ramp model is not able to fully capture the
very sharp turn-on of the hook. The exponential hook correction light curve is
also shown in Figure 
2 of the main text, where the mean intra-orbit
RMS is considerably higher at 440.1\,ppm (versus 375.5\,ppm). Nevertheless, 
an exomoon-like decrease in brightness is observed in both versions following
the planetary transit.

We highlight that our non-parametric approach is somewhat guaranteed to
out-perform the ramp model due to more degrees of freedom. However, the
ability to capture sharper hooks, combined with the more agnostic nature
of the method's assumptions ultimately led to us to use this method for our
final hook-correction. We highlight that 375.5\,ppm per 300\,seconds
corresponds to 154.8\,ppm per \kepler\ long-cadence, which is
3.8 times lower than the median \kepler\ uncertainty resulting from our
method marginalized detrending (589.9\,ppm). Thus, HST greatly out-performs
\kepler\ on this target. For this reason, one might reasonably expect that
the HST data will be the dominant transit epoch for constraining putative
moons.

From Figure 
2 in the main text, an apparent decrease in brightness
is evident with both versions of the hook correction for Kepler-1625, occurring
a few hours after the primary transit has finished. The precise shape of
this moon-like dip appears dependent upon the trend assumed in the data,
which is discussed in detail in the main text.

\subsubsection{Wavelength Solution}
\label{obs:wavelengthsolution}

To derive a wavelength solution for each pixel we extracted the spectra for
the target and comparison star and found the best fits (minimizing $\chi^2)$ to a
model spectral profile (Figure \ref{fig:kepler1625_response_function}). 
The model is produced by multiplying the G141 response function by a blackbody
curve, the latter produced using published values for the stars' effective
temperatures. By examining the wings of the model curve (where sensitivity
falls off rapidly) and comparing to the extracted spectra the fits are
excellent. We note however that this simple model overpredicts the flux at
shorter wavelengths and underpredicts at longer wavelengths, as can be seen
in Figure \ref{fig:kepler1625_response_function}. We examined whether this
discrepancy could be due to our marginalizing over the wavelength
information in the flat field, as there is a distinctive wave-like structure
in the pixel sensitivities that propagates across the CCD in the direction
of dispersion. In some places on the CCD shorter wavelengths are more sensitive
than longer wavelengths, where in other places the reverse is true (see
Figure~\ref{fig:flat_field_delta_sensitivity}). As we marginalize over the
wavelength information in the flat field, this information is lost.

However, we find that this intra-pixel flat-field structure cannot explain the 
spectrum-model discrepancy in Figure \ref{fig:kepler1625_response_function}, 
as flux errors across the spectrum average out to be less than 0.2\% for any 
flux bin when accounting for the wavelength dependence, whereas the discrepancies 
are clearly well in excess of that. We speculate that the discrepancy merely
arises from the fact that stars are not perfect blackbodies. In any case, 
these discrepancies do not invalidate the wavelength solution, as the rapid 
fall-off in sensitivity at either side of the spectrum clearly matches the 
sensitivity curve very well.

\subsubsection{Nearby Uncatalogued Source}
Upon inspection of the HST images it became clear that a previously 
uncatalogued point source is present in close proximity to the target. 
The object shows up in every image obtained by HST and cannot be an artifact 
on the CCD, as it moves like all the other point sources during the 
three SAA-affected exposures. By every indication it is a another spectrum 
for a nearby point source. 

To estimate its position on the sky we simply take the first pixel along the 
spectrum for which Flux $>$ 0. Note that this is \textit{not} the sky position 
of the star. However, we may do the same for the target star and nearby KIC 4760469, 
for which the sky separation and position angle is known, and thereby orient 
ourselves to calculate the sky separation and position angle of the uncatalogued source. 

To calculate the magnitude of the new source, we step through each column 
along the spectra and fit a three-Gaussian model to the data. One Gaussian is fit 
to the new (contaminating) source while two Gaussians are fit to the target -- 
one narrow, and another wide. The combination of these two Gaussians do an 
excellent job fitting the peaky-but-broad profile of the target star, with 
the wider component accounting for what appears to be flux bleeding into neighboring 
pixels (see Figure \ref{fig:HST_OA_blend}). Having stepped through every column, the areas under these three Gaussians 
can be computed, and the flux of the uncatalogued source may be compared to that 
of the target. With this information the relative magnitude with respect to the 
target may be computed. 

We find that the star is located $\sim$ 1.78$''$ away from the target at Position 
Angle 8.5 degrees East of North. We compute a \kepler\ bandpass magnitude of 22.7. 
To compute a blend factor we take the same approach as before, i.e. modeling the target 
and the contaminant with three Gaussians, only now we restrict the window to the 
optimal aperture see Figure \ref{fig:HST_OA_blend}. We compute a blend factor of 1.000328, 
indicating an extremely small contribution from the uncatalogued source on our light curve.
Furthermore, an extracted light curve from the uncatalogued source shows a variability 
of 0.33\%, so its contribution to the target light curve is 1\,ppm. We therefore 
ignore it in subsequent analysis.

Using a \gaia-derived distance of $(2460 \pm 220)$\,pc to the target, we may 
calculate the physical separation of this uncatalogued source from Kepler-1625, 
under the assumption that the two objects are at the same distance. We find that separation 
would be ${\sim}$4400\,AU, placing it well within the gravitational 
influence of Kepler-1625. It is however impossible with the data in hand to determine 
whether these objects are in fact physically associated.

\subsubsection{Visit-long trends}
\label{obs:trends}

In addition to the breathing and hook effects, one other well-known source of
systematic error requires correction - the visit long trend
\cite{wakeford:2016}. Visit-long trends with WFC3 are typically modeled as a
linear slope (e.g. see \cite{huitson:2013,ranjan:2014,knutson:2014}).
However, our observations are unusual in that they span 40 hours, far more
than the few hours used when observing transiting planet on short orbital
periods. These trends have not yet been correlated to any physical parameter
related to the WFC3 observations \cite{wakeford:2016}, and indeed not all
observations appear affected. Simple inspection of our white light curve,
using either the exponential or non-parametric hook correction, show clear
evidence for a visit-long trend (see Figure 
2 in the main text).

Although a linear trend is the most common approach (e.g. see
\cite{huitson:2013,ranjan:2014,knutson:2014}), we note that 
\cite{stevenson:2014a,stevenson:2014b} report improved fits using a quadratic
model and so we considered both models in this work. We further extended our
investigation to include an observation-long exponential ramp model. This last
model is motivated by visual inspection of the light curve, which appears to
ramp up and flatten, as well as the asymptotic behavior it introduces which is
more physically motivated than an ever-increasing/decreasing trend.
We speculate that it could perhaps be caused by the same charge trapping
that causes orbit-long ramps, only operating on a much longer timescale.

If we look at specific spectral channels, rather than the white light curve,
a flux offset occurs between the 14\textsuperscript{th} and 15\textsuperscript{th}
orbits for the reddest wavelengths (see Figure~\ref{fig:spectral_channels}). This
moment in time corresponds to the HST visit change, during which the spacecraft
performed a full guide star acquisition at the beginning of the
15\textsuperscript{th} orbit, which placed the target spectrum $\simeq$0.1\,pixels
away from where it appeared during the first 14 orbits. This discontinuity
appears in the raw photometry mid-planetary transit and is potentially problematic
due to the fact that it can mimic a moon signal. Any model placing the moon at the
kink would be immediately suspect.

In the white light curve, a flux offset at this time is barely
noticeable but since the red spectral channels contribute to the white, then
we deemed it necessary to allow for an offset term in our three trend models.

Since only centroid position changes during the visit switch, whether the
visit-long trend be astrophysical or due a long-term instrumental effect,
there is no reason to expect a different functional form or function parameters
to become introduced at the instant of the visit change. For this reason, we
generally expect a smooth continuous function (such as the linear, quadratic
or exponential models) but with an offset term to account for any remaining
pixel sensitivity variations. We therefore did not consider models
described by two completely independent polynomials, for example, on either side
of the visit change.

Since the visit-long trend occurs on a long timescale, it is
inextricably mixed with the transits of the planet and possible moon. In such
a case, strong covariances are expected between the trend parameters and the
transit parameters and thus joint fitting is required. Our joint fits are
described later in Section~\ref{sec:fits}, but for now we point out that
all six visit-long trend models were regressed in conjunction with the
transit models considered (e.g. models M, P, T and Z), but independent of
one another. In this way, we can rank the different approaches based on their
Bayesian evidences, as well as the resulting associated physical parameters.

The results of these fits using the planet+moon (M) transit model may be seen
in Figure 
4 of the main text, with Bayesian evidences tabulated in Table 
1 in the main text. 
All of the fits are able to explain the previously noted decrease in brightness 
towards the end of the observations as being due to a moon transit, although 
the duration and depth of the event vary somewhat between the three trend 
models. In the exponential trend model the moon fully egresses before the end of the observation, 
while in the linear and quadratic fits the moon is still in the process of transiting. 
However, in virtually every case the dip remains discernible after detrending has been performed. 
Note that all of these models were found to be preferable to any other transit 
model (e.g. T, Z or P) attempted using the Bayes factor (we direct the reader to
Section~\ref{sec:fits} for a more detailed discussion of model comparison).

\subsubsection{Is the moon-like dip instrumental or astrophysical?}
\label{obs:isthemoonreal}

As noted earlier, there appears to be a decrease in flux in our WFC3 photometry
towards the end of our observations. Since we are primarily interested in the
possible existence of exomoons in this work, that decrease is of particular
importance as it will greatly affect photodynamical model fits, if real.

To assess whether this dip is instrumental or astrophysical in nature, we
considered three tests: i) inspection of the comparison star ii) inspection of
the centroids iii) a chromatic test. The chromatic test is described later
in Section~\ref{fits:colortest}, but we here describe i) and ii) in more detail than
possible in the main text.

The only other bright star with a full spectrum in the HST images is
KIC 4760469. This star was not observed by \kepler, but is listed as a
$0.84$\,$R_{\odot}$ $5555$\,K main sequence star in the KIC. Accordingly,
it is expected to be photometrically stable and provide a good
test for our correction algorithm. We therefore applied the exact same
routines to this star as was done for Kepler-1625.

As seen in Figure 
2 in the main text, the comparison star displays no obvious long-term trends
that might be attributed to instrumental systematics, apart from a gentle
upward slope at the beginning of the observation which is also seen in the
light curve for Kepler-1625b and is cleanly corrected with our detrending.
We used \multi\ to compare the linear, quadratic and exponential trend models
and found all three were similar in evidence, with the quadratic model
being slightly favored.

In terms of assessing if the moon-like dip associated with Kepler-1625b is
real or not, the relevant region is visible from BJD\,2,458,456 of
Figure~2 of the main text. The photometry appears quite stable at this time
and certainly no indications of an instrument-induced flux change.


The second check we performed was to look at the centroids. Column ($x$)
and row ($y$) centroids were computed from the optimal apertures of the
target and comparison star and are shown in Figure~\ref{fig:centroids}.
We highlight that the $y$ position has been decreased by 0.1 pixels
after the visit change (for both sources) to fit them on the same scale,
but the $x$ position has not been altered. Both the target and the
comparison star show nearly identical centroid behavior, as expected.

The moon-like dip occurs across several orbits and thus the only way
centroid variations could explain the dip would be an inter-orbit
centroid variation. We therefore take the median centroid position of
each orbit as binned points, shown by the black data in
Figure~\ref{fig:centroids}. Aside from the visit change shift, there are no
substantial changes in the inter-orbit centroid position, and certainly
not around the time of interest (highlighted by the vertical lines).

Comparing the centroids to the flux variations observed can also be
used to gauge how feasible it is that the moon-like is a product of
these clearly small centroid changes. From our later fits of the target,
we find that the visit-change flux offset is $(330\pm120)$\,ppm,
$(180\pm190)$\,ppm and $(220\pm140)$\,ppm for the linear, quadratic
and exponential models respectively. That corresponds to a decipixel
(0.1\,pixels) change in the $y$-position and about half that in the $x$
direction. Together then, this indicates that a decipixel shift in
centroid position may slightly affect the photometry at the
$\sim 200$\,ppm level.

Inspection of the centroid variations after the visit-change - where
the moon-like dip occurs - reveals a shift of approximately
one-tenth of a decipixel (see Figure~\ref{fig:centroids}). Since a
one decipixel shift is associated with a 200\,ppm photometric change,
we argue that it is highly unlikely that $\sim$500\,ppm moon-like dip is a
product of centroid shifts. However, we highlight that a quantative calculation
of the centroid-induced photometric change in this region is not possible
without knowledge of the functional form governing the residual inter-pixel
sensitivity.

The above considered the centroid-flux correlation as observed using
just the target, but the comparison star is also worth considering.
If we assume the comparison star is stable, a flux versus centroid
plot reveals much more information than that from above. However, we
caution that such a plot only maps the sensitivity at this part of
the detector and this may not necessarily be the same as that on
the target itself. Taking the second visit data only, which is the
region of interest, we computed a cross-correlation of normalized
intensity versus $x$ and $y$ centroid positions, as shown in
Figure~\ref{fig:centroids}.

No clear relationship is apparent from inspection of this plot.
If there is a functional relationship between flux and pixel
position, it appears to operate at a level below the noise of this data.
Certainly no linear relationship is detected, with the Pearson's
correlation coefficient being consistent with zero for both $x$ and $y$
positions. Computing a linear slope in both
cases implies that flux depends on $x$ and $y$ centroid position
as $(-140\pm180)$\,ppm/decipixel and $(+50\pm480)$\,ppm/decipixel
respectively. These numbers are consistent with the $200$\,ppm
change observed in the target for a decipixel shift during the visit
change, despite being located at a different part of the detector.

We can make some simplifying assumptions in order to have an
approximate estimate of the centroid-induced flux changes expected
around the moon-like event. These assumptions should not be treated
as truth, but rather as plausible and necessary for quantitative progess.
Although we don't know the true functional form, let's assume
that intensity indeed maps linearly with centroid position for small
changes (exploiting a Taylor expansion logic). We further assume that the
detector's behavior on the comparison star is representative of the
source. The $x$ range in the source's second visit is 0.25
decipixels, which would be associated with a $<50$\,ppm change using
the assumptions above. Similarly, the $y$ range is 0.1 decipixels,
implying a maximum variation of $<50$\,ppm. Accordingly, although this
is certainly a simplified model, it suggests a 500\,ppm flux change
is quite unlikely to arise from the centroid variations.

\subsubsection{Spectral analysis}
\label{obs:spectrum}

With a low-resolution transmission spectrum we may also attempt to characterize
the planet's atmosphere. A transmission spectrum, measured as the
changing transit depth as a function of wavelength, can potentially reveal 
molecular absorption features in the atmosphere, and with sufficient sensitivity it 
may also be used to infer the planet's mass, as atmospheric abundances and Rayleigh
scattering will be sensitive to the planet's surface gravity.

We split our optimal aperture in 10 even segments along the column direction
and corrected for the hook independently in each using our non-parametric
algorithm. We found that the tenth and reddest channel was quite unstable and thus
neglect it in what follows. Light curves from the other nine channels are shown in
Figure~\ref{fig:spectral_channels}, and final transmission spectrum is shown in 
Figure~\ref{fig:transmission_spectrum}.

We utilized our own MCMC code to explore parameter space and generate a best-fit 
transmission spectrum using the Exo-Transmit code \cite{kempton:2017, exotransmit_source1, 
exotransmit_source2, exotransmit_source3}. 
The generated high-resolution spectrum is binned at the appropriate wavelengths to
test against the data, which also includes the \kepler\ transit depth, for a total of
10 data points across the wavelength range, 9 of which are derived from the WFC3 spectrum.

The variable inputs are radius of the planet $R_{P}$, surface gravity of the planet $g_P$, 
and (optionally) a cloud deck atmospheric pressure. The code treats the atmosphere as opaque at 
pressures higher than the cloud deck pressure value, corresponding to greater depths 
and effectively increasing the radius of the planet. All variables have uniform priors. 
Three other inputs (stellar radius, temperature, and metallicity) are fixed, as discussed below.

$R_{P}$ was allowed to range from half to twice the radius of our best fit planet radius, while 
$g_{P}$ is restricted between 1 and 1000 m s$^{-1}$. The mass may then be inferred 
from the combination of surface gravity and planetary radius. The cloud top pressures 
(when applied) could range from 1 to $10^7$ Pa, allowing for virtually no cloud deck down to $\sim$ 
100 atmospheres. To speed up convergence the MCMC was initialized with reasonable 
first guesses for the radius (from the transit depth) and surface gravity of the planet 
(randomly chosen to be some value less than 30 m s$^{-1}$.

We fix the stellar radius at 1.793 $R_{\odot}$, as there is nothing in the transmission 
spectrum that can constraint the size of both the planet and the star. In addition, we fix 
the metallicity for the model at solar abundances and the planet temperature at 300K. 
Unlike other parameters in the Exo-Transmit code, for which any number can be specified, 
there are a narrow range of options to choose from for metallicity and temperature, 
owing to the fact that large files for abundances and temperature-pressure profiles 
must be used. While the Exo-Transmit authors state that interpolations between two files 
may be performed (for example, a 350K T-P profile could be obtained by interpolating 
between the 300K and 400K models), there is considerable uncertainty for the target 
with respect to both parameters; ExoFOP lists Fe/H for the target at 0.12 $\pm$ 
0.15 (i.e. consistent with solar), and the equilibrium temperature at 350K. 
The albedo of the target is of course unknown, but likely pulls the temperature 
down closer to 300K. Meanwhile, generating interpolated files at each MCMC step would 
be considerably more expensive computationally.

After some experimentation we opted to model the spectrum without use of a cloud deck.
Motivating the elimination of clouds is the fact that it will act to suppress 
molecular features in the spectrum, thereby confusing the situation; if the spectrum 
is consistent with a flat line, we cannot know whether the molecular features are 
suppressed because of a cloud deck, or because the planet is very massive. 
Eliminating the cloud deck thereby allows us to characterize what the spectrum is 
doing were it to be influenced by planet mass alone.

As the uncertainties are quite large across all wavelength bins, due to the 
faintness of the target, the system parameters derived from this test are poorly
constrained. We find the spectrum is consistent with a flat line / featureless atmosphere.
If we assume no clouds present, atmospheric absorption would be potentially detectable
only for very low surface gravity worlds, with sub-Saturn masses.

%% file: sections/fits.tex
\subsubsection{Stellar Parameters}

In our previous analysis \cite{teachey:2018}, our source for fundamental
stellar parameters came from \kepler\ DR25 \cite{mathur:2017}. Since that
time, data release two from \gaia\ has been released providing parallax
information which should be expected to yield an improved inference
\cite{luri:2018}. Although these parallax constraints have been incorporated
in a prior publication for \kepler\ planet hosts \cite{berger:2018}, that
work does not include stellar masses necessary for this work.

We therefore decided to use isochrone modeling to derive revised stellar
parameters including the \gaia\ parallax. To do this, we use the \isochrones\
package \cite{morton:2015} with Dartmouth tracks coupled to the \emcee\ Bayesian inference
algorithm \cite{dfm:2013}. The \gaia\ parallax of $(0.406 \pm 0.035)$\,mas
corresponds to a distance of $(2460 \pm 220)$\,pc. Following the
recommendations on the \gaia\ DR2 portal
(https://www.cosmos.esa.int/web/gaia/dr2), the global systematic offset
can be neglected since it is far less than the measurement error, and the
measurement error is not expected to be an underestimate given the target's
brightness. In addition to the parallax, we used the stellar effective
temperature, surface gravity and metallicity inferred by \cite{mathur:2017}
from spectroscopic constraints. Finally, we included the \kepler\ apparent
magnitude ($15.756$) with an uncertainty set to $0.1$\,mag \cite{huber:2016}.

Our revised stellar parameters yield an approximately Solar mass star
($1.04_{-0.06}^{+0.08}$\,$M_{\odot}$) with an enlarged radius
($1.73_{-0.22}^{+0.24}$\,$R_{\odot}$), implying that the star has evolved off
the main sequence. This is consistent with the prior classification
using the \gaia\ parallax \cite{berger:2018}, yielding physical
dimensions highly consistent with both previous estimates
\cite{mathur:2017,berger:2018}. The evolved state of the star means
that the its age can be constrained to be
$8.7_{-1.8}^{+1.8}$\,Gyr, some 4\,Gyr older than our Sun. This means
that although the present day luminosity is
$2.55_{-0.58}^{+0.72}$\,$L_{\odot}$, the star would have been 2.5 times
less luminous for most its life, meaning Kepler-1625b would have
received very close to Earth's present-day insolation during the
main-sequence lifetime, given its semi-major axis.

\subsubsection{Model Description}
\label{fits:description}

A transiting planet model represents a nested model of the more general
planet+moon model, where the moon mass and radius equals zero. For this
reason, a moon fit is guaranteed to provide a lower $\chi^2$, or (more
rigorously) a higher maximum likelihood. This basic fact forces exomoon
hunters to adopt methods able to account for model complexity, such as
Bayesian model selection, in order to make any progress, something long
advocated by the Hunt for Exomoons with Kepler (HEK) project since its
inception \cite{HEK1}.

In this work, we fit light curve models to the data using a normal
likelihood function with the \multi\ regression package
\cite{feroz:2009}. \multi\ is designed to estimate the Bayesian evidence
of any model attempted, using multimodal nested sampling (see
\cite{skilling:2004} and \cite{feroz:2008}) to conduct inference.
The Bayes factor between two models is then evaluated by taking the ratio
of two evidences. A by-product of this process is the parameter posteriors,
which are also useful checks when comparing models against one another.
 
Our light curve model is generated by \luna\ \cite{luna:2011}, a
photodynamical Fortran code for simulating planet-moon light curves.
In total, we consider four basic transit models, which we designate
as P, T, M, and Z. We describe these models in turn.

The planet-only model P is described by seven parameters for a lone planet:
the ratio-of-radii ($p=R_P/R_{\star}$), the stellar density
($\rho_{\star}$), the transit impact parameter ($b$), the time of
transit minimum of the second observed \kepler\ transit ($\tau_0$),
the orbital period ($P$) and two quadratic-law limb darkening
coefficients ($q_1$ and $q_2$; see \cite{kipping:q1q2}). However,
the presence of HST data meant chromatic differences were expected
and so we included three extra terms,
$p_{\mathrm{HST}}/p_{\mathrm{Kep}}$ to describe the ratio of the
HST-to-\kepler\ ratio-of-radii and two new limb darkening terms for
the WFC3-band, giving 10 parameters in total.

Model T is the same as model P except that each transit epoch (four in total)
is given its own unique time of transit minimum, thereby allowing
for timing variations. Duration variations or any other kind of
dynamical change are not modeled. Model T requires four extra free
terms but also two fewer (no orbital period and no $\tau_0$), thereby
giving a 12-parameter model.

Model M is the planet+moon model, which is similar to model P except that a moon
is included. As such, seven additional parameters are added:
the radius ratio, $R_S/R_P$; the mass ratio, $M_S/M_P$;
the orbital period, $P_S$; the semi-major axis, $a_{SP}/R_P$;
a term describing the orbital phase at time $\tau_0$, $\phi_S$; the orbital
inclination angle, $i_S$; and the longitude of the ascending node,
$\Omega_S$. These seven new parameters give a total of 17 terms in model
M (see \cite{HEK3} for details on these definitions).

Finally, model Z is identical to model M except that $R_S/R_P$ is fixed to 
zero and thus has one fewer free parameter. Model Z does not simply 
reproduce model T because moons can induce duration variations 
\cite{kipping:2009a} as well as impact parameter changes \cite{kipping:2009b},
both of which are modeled by Z but not T.

We highlight that the number of free parameters described above
only represent the transit-model parameters, and in practice the
total number of free parameters is higher due to the inclusion of
visit-long trend terms. We direct the reader to our previous
papers (e.g. \cite{HEK5}) for a description of the priors used.

It is important to note that model M also includes some constraints
on physically acceptable parameter combinations. The planet and
moon density can be derived as described in \cite{weigh:2010} and we
reject any samples for which these exceed 28\,g\,cm$^{-3}$ or drop
below 0.08\,g\,cm$^{-3}$ in a bid to remove physically unsound
combinations. This has two important consequences. First, by penalizing
a part of the parameter volume, particularly at small signal sizes
compatible with very marginal signals, model M will generally obtain
a lower Bayesian evidence than it otherwise would. By demanding the
model is physically sound, we are thus being more stringent in our
calculation of Bayes' factors when assessing the case for a moon.
Second, there is strong evidence for a timing offset in the HST
data, which requires a non-zero $M_S/M_P$ to explain for model M. Since
infinitesimal radii would lead to infinite densities, $R_S/R_P$ cannot
approach zero and is therefore forced to always take on a positive
value.

Without careful consideration, these positive values could be
misinterpreted as evidence for a moon. Instead, the correct
procedure should be to compare the evidences between the different
models. In order to weigh up the evidence for a moon-like transit,
we introduced the Z model which handles all of the moon-induced
timing effects but without the radius effect. Comparing the
evidences between models M and Z is the most direct way to
infer the case for the moon-like transit signature. Comparing
the evidences between models M and P is the most direct way
to evaluate the overall preference for the moon model over the
lone planet model.

We also take a brief aside to mention a subtlety with model T.
Model T is somewhat unphysical. There is a single lone planet
in the system which exhibits essentially arbitrary transit
times. In practice, something must be causing these timing
variations, be it a moon or another planet in the system.
Any such body would require at least six new parameters to
describe it and yet we here model the case with just (net) two
additional parameters. For this reason, model T is able
to do better than it really should in terms of the Bayesian
evidences, which generally penalize models for using extra
degrees of freedom. For this reason, it is always worth keeping
this point in mind when considering the T model and looking
at the likelihoods as well.

\subsubsection{The case for TTVs}
\label{fits:TTVs}

Resulting Bayesian evidences ($\mathcal{Z}$) from the joint fits are presented
in Table~1 of the main text,
along with the maximum likelihoods
($\hat{\mathcal{L}}$) amongst the derived (and finite) posterior samples.

We begin by first considering the evidence for transit timing variations.
As can be seen in Table~1 of the main text,
model T is consistently favored
over model P for all choices of the visit-long trend with $2\log K \simeq 4$,
indicating positive evidence for TTVs. Leaving the Bayesian evidences aside,
the $\Delta\chi^2$ between model P and model T ranges from 17.1 to 19.2. Since
model T uses just two more parameters than model P, the Akaike Information
Criterion would range from 13.1 to 15.2, which indicates a very strong case
for TTVs.

Figure~4 in the main text shows the maximum \textit{a posteriori} predicted time
of the transit, as inferred using a linear ephemeris P model fit to the \kepler-only data
(described in Section~\ref{sec:kepler}) with a dashed vertical line in the
right-hand-side panels. This time is clearly visually offset from the center
of the planetary transit signal irrespective of visit-long trend model. The
vertical dashed lines on those plots denote the location of the time of transit
minimum resulting from model T (and now including the HST data), and indicates
that the transit came in ${\simeq}78$ minutes earlier than expected.

Based on the \kepler\ data alone, we found that the P and T models had a Bayes
factor of approximately unity (see Section~\ref{sec:kepler} and
Table~\ref{tab:kepler_evidences}). Therefore, the first three transits observed appear
compatible with a linear ephemeris. It is only the new HST transit which
appears offset. However, including the HST epoch causes the maximum likelihood
linear ephemeris to change substantially. This is visualized in
Figure~\ref{fig:TTVs}, where one can see the first three \kepler\ epoch times
with the corresponding maximum \textit{a posteriori} linear ephemeris plotted in black.
The overall O-C diagram uses the updated maximum \textit{a posteriori} ephemeris from
model P using the HST data (marginalized over all of the visit-long trend
models). Similarly, each transit time plotted (and also presented in
Table~\ref{tab:ttvs}) is that from a Bayesian model averaged posterior across
the various visit-long trend models using model T (for details on the model averaging see
Section~\ref{fits:BMA}). It is clear from main text Figure \ref{fig:TTVs}
that predicted \kepler-based prediction is inconsistent with the HST epoch with a one-sided
$p$-value of $>3$\,$\sigma$. For reference, each \kepler\ transit midtime has an uncertainty 
on the order of 10 minutes and the standard deviation on linear ephemeris predictions 
is 25.2 minutes derived from posterior samples. On this basis, we consider there to be a strong
case for TTVs after introducing the HST epoch.


Against the maximum \textit{a posteriori} linear ephemeris, the TTVs have a semi-amplitude
of $\simeq 25$\,minutes. With four transit times alone, is it perhaps not
surprising that it is easy to find good solutions to the TTVs using an external
perturber, although the solutions are extremely degenerate. We used \ttvfaster\
\cite{TTVFaster} with an MCMC exploration of the parameter
space and found a wide range of plausible solutions assuming an external
perturber.

\subsubsection{Other timing effects}
\label{fits:TDVs}

An important point is that no matter how well a perturbing planet can explain
the observed TTVs, there are additional effects that an exomoon is expected
to impart on a light curve that a perturbing planet, in general, will not. These are
all the other photodynamical effects that may be seen in the transit of Kepler-1625b, 
such as transit duration variations \cite{kipping:2009a}, transit impact 
parameter variations \cite{kipping:2009b} and transits of the moon itself
\cite{luna:2011}. Therefore, the way to establish whether a putative TTV
is due to a moon or a perturbing planet is to look for these additional 
characteristic signatures expected from the presence of a moon.

Taking the dynamical signatures first, these are generally expected to be
much smaller than the TTVs \cite{kipping:2009a}, and accompanied
by larger measurement errors (for example, transit durations have twice the uncertainty
of transit times; \cite{carter:2008}). Given the fact that the TTVs
are only significant at the 3\,$\sigma$ level, we certainly do not expect,
nor do we observe, noticeable duration variations. Nevertheless, these
small effects are accounted for in a photodynamical model such as \luna\
and thus we can see if they lead to any improvement in the fits. Model Z
serves this purpose by fixing the moon radius to zero, but otherwise
describing a full three-dimensional moon orbit with six more terms
than model P.

It is instructive to compare the maximum likelihoods from models T and
Z. Although our formalism for model T only uses two extra free parameters,
as noted earlier in Section~\ref{fits:description}, this is somewhat
artificial and in reality timing variations would require a planet
described by six orbital parameters too. In this way, models T and Z
can be directly compared as having essentially the same number of free
parameters. Model Z consistently out-performs T in this regard. 
However, we consider these improvements as being necessary for any
successful moon model rather than being convincing evidence in isolation.
This is because model Z has greater flexibility than model T to explain
light curve changes and thus in many ways is guaranteed to lead to
an improvement. Nevertheless, the evidences show model Z consistently
out-performs T by around $2 \log K \simeq 4$, meaning there is some
evidence favoring the moon hypothesis over a planet perturbation model.
We caution that we have assumed planets do not induce short-period
TDVs and/or impact parameter changes here but in some rare cases
such changes have been detected (e.g. \cite{koi142,koi13}).

\subsubsection{The case for an exomoon}
\label{fits:exomoon}

So far we have established that a) Kepler-1625b exhibits an early
transit in the HST epoch indicating TTVs in the system, and b) a zero-radius
moon model, which explains TDVs and other effects in addition to TTVs,
leads to a modest improvement in the fits. A true exomoon would be
expected to exhibit both points, and so while this is not convincing enough on its
own, the exomoon hypothesis is certainly ``on-track''. The next step
is to consider the full planet+moon models.

We first note that model M, the full planet+moon model, is the favored
hypothesis in each and every visit-long trend model attempted (see
Table~1 of the main text), which already formally establishes the moon
hypothesis as the leading candidate solution. With a strong case for
TTVs already established (see Section~\ref{fits:TTVs}), and some
modest evidence for other moon-induced dynamical effects
(Section~\ref{fits:TDVs}), the mass signature of the putative moon is
measurable and significant. To build a compelling case for an exomoon,
we also need to detect the radius signature, since the dominant
dynamical effect (TTVs) could be plausibly caused by a perturbing planet
instead. Not only should we detect this radius signature, but that radius
signature must be consistent with the moon solution in terms of both
phase and physical parameters.

As a result of the strong TTV, model M has no other way to explain the timing
offset except for using the moon, which leads to a positive $M_S/M_P$, which in
turn demands a positive $R_S/R_P$ (due to our rejection of unphysical moon
densities; see Section~\ref{fits:description}). Accordingly, inspection of the
marginalized posterior distributions of $R_S/R_P$ is not a useful strategy
for evaluating the case for a detection of the moon's radius signature. The
most direct method would be to directly compare the Bayes factor between the
full moon model (i.e. model M) and an identical model for which the radius
effect is turned off (i.e. model Z).

In all visit-long trend models attempted, we find that model M is favored
over model Z. However, there is considerable range in the Bayes
factors which result (see Table~2 of the main text). This can be linked to
Figure 
4 in the main text, where one can see that the high Bayes factor
cases (e.g. linear) tend to correspond to cases where the moon transit
is noticeably larger. In all cases, the moon transit occurs towards the end
of the observations.

We would argue that a moon-like transit is immediately obvious even in the
non-detrended data (see Figure 
2 in the main text) and corresponds to
the location where all of the moon models place the signal. This location is
important because Kepler-1625b transits early in the HST epoch, for which we
would therefore expect a corresponding moon to be transiting late. In other
words, just combining the fact that we see an early transit along with the
extra dip observed in the non-detrended data already presents a strong case for
a self-consistent moon solution (and indeed our detailed modeling verifies
this statement). Although the phasing is indeed aligned as expected then, this
does not address whether the amplitude of the TTV (implying a certain moon
mass) is compatible with the amplitude of the moon-like transit (implying a
certain moon radius). We can therefore summarize the situation as being that
the best explanation to the data is an exomoon, driven by a timing offset
and moon-like dip with self-consistent phases, although the properties of
the moon (in particular the size) appear to be dependent upon the trend model
adopted and one (or all) of these may not be physically permissible. This
latter point forms the subject of investigation in
Section~\ref{fits:physics}. 

Before discussing whether the moon properties are physically sound or not,
we briefly highlight that the $\chi^2$ improvement in
going from model M to Z is quite high, ranging from 19.2 to 33.7 
(note that we have $n=1048$ data points). On this basis, we find
the SNR of the moon-like transits to be at least 19.

\subsubsection{Evaluating the physicality of putative system parameters}
\label{fits:physics}

We now have a TTV signal, modest evidence for other moon-induced dynamical
effects and a sizable $\chi^2$ improvement when including the transit of
a moon at the correct phase to explain the TTV. It is not yet clear, however, whether
the candidate moon actually has physically sound parameters, beyond falling
within the very generous density constraints imposed by our model (and
described in Section~\ref{fits:description}).

Each trend model clearly leads to distinct moon parameters, as evident
by the different moon depths in Figure 
4 of the main text. Not only is
the depth different, but the moon sometimes displays an egress feature and
other times the egress occurs after the observations have ceased, which would
require a greater planet-to-moon semi-major axis. Since the semi-major axis
multiplied by the moon-to-planet ratio dictates the TTV amplitude, these longer
semi-major axis solution necessarily require a smaller $M_S/M_P$. Furthermore,
these longer semi-major axis solutions tend to be correlated to the greater
moon transit depths as a result of the interplay with the trend model. This
leads to a situation where $R_S/R_P$ and $M_S/M_P$ are somewhat anti-correlated
between the various trend models, although admittedly there are just three
models under inspection here.

The differing planet+moon transit parameters between each trend model imply
different physical parameters then, with no guarantee of such parameters being
necessarily physically plausible. To investigate this, we have to go from
ratios to absolute dimensions. For the radius, this is easy. All solutions
lead to good constraints on $R_S/R_P$ and $R_P/R_{\star}$ and since the
stellar radius is known we can derive absolute sizes with reasonable
precision.

Deriving absolute mass solutions is far trickier, since although $M_S/M_P$ 
is well-constrained from the TTVs, $M_P/M_{\star}$ is found to be only 
weakly constrained photodynamically in all models. This can be seen in 
Figure~\ref{fig:masses}, by inspection of the dotted lines. In an attempt 
to shore-up this planetary mass measurement, we fed the absolute radius 
samples into the probabilistic empirical mass-radius relation \forecaster\ 
\cite{forecaster}, which are the dashed lines in that plot. Taking the 
product of the two PDFs (using a Gaussian kernel density estimator) 
leads to the solid black posteriors in main text Figure~\ref{fig:masses},
which represents one method of estimating the planetary mass from our data.

A second way to measure the planetary mass is to come from the exomoon side.
Using the moon radius, we feed the samples into \forecaster\ to get a moon
mass and then convert that into a planetary mass using the $M_S/M_P$
distribution inferred by our fits. This second planetary mass solution is
plotted in solid orange in Figure~\ref{fig:masses}. A physically
self-consistent solution would correspond to no tension between the two
distributions. Visual inspection of Figure~\ref{fig:masses}
reveals that the two estimates are consistent with one another for all
three trend models, with the only difference being support for lower
masses in the quadratic case.

To quantify the compatibility, we draw a random sample from one of the
distributions (the moon-forecast based solution) and evaluated its likelihood
using the product PDF from the other solution. We repeated this for all
available samples ($\sim 40000$) and then evaluated the mean likelihood,
which is reported in the top-right corners of Figure ~\ref{fig:masses}. In
agreement with simple inspection, the linear model appears to
provide the greatest degree of physical self-consistency. All three provide
broadly consistent mean likelihoods, none of which appear unphysical.

\subsubsection{Combining the two mass estimates}
\label{fits:finalmass}

Although the absolute dimensions of the system are constrained from
photodynamics and our stellar properties, we are
able to obtain more precise constraints by folding in the \forecaster\
results from the previous paragraphs. For each trend model,
we take the two planetary mass distributions (represented by the
black solid and orange solid lines in Figure~\ref{fig:masses})
and combine them to form a single planetary mass solution. This
is achieved by taking the product of two PDFs and then defining the
result as a likelihood function. We then sample from the function
with a simple Markov Chain until 40,000 samples have been computed.
The resulting planetary mass posterior is converted into a satellite
mass using the $M_S/M_P$ parameter from the model M fits, which can
be found in Table~2 of the main text.

\subsubsection{Bayesian Model Averaging}
\label{fits:BMA}

The moon models display subtle but important variations in the associated
parameters for each trend model. In contrast, we find the depths and transit
times from model T are highly stable. In order to compile a single posterior
for the transit times (presented in Table~\ref{tab:ttvs}), we decided to
marginalize over the three models using Bayesian Model Averaging.

The basic idea is that each model (in our case the three 
visit-long trend models) are assigned a weight based on their Bayesian
evidence, $\mathcal{Z}$ (which is also known as the marginal likelihood).
The odds-ratio between model $i$ and $j$ may be computed as

\begin{math}
O_{ij} = \exp( \log \mathcal{Z}_i' - \log \mathcal{Z}_j' )
\end{math}

And the final weights are now defined as

\begin{math}
w_i = \frac{ O_{i1} }{ \sum O_{i1} },
\end{math}

such that $\sum w_i = 1$. With the weights assigned, we draw a random
integer from a multinomial distribution using the weights vector to
choose a model. We then choose a random posterior sample from that
model and append it to a new array. After many iterations, we construct
a Bayesian Model Averaged posterior for model T.

We repeated this process on the nine spectral channels computed earlier
using model T with the three trend models. The resulting marginalized
ratio-of-radii is treated as our final spectral retrieval for
Kepler-1625b and is presented in Table~\ref{tab:spectrum}.

\subsubsection{Stability Analysis}

Moons are generally considered stable if they reside beyond the 
Roche limit and within $\sim$ 0.4895 $R_{Hill}$ for prograde moons
and 0.9309 $R_{Hill}$ for retrograde moons \cite{domingos:2006}.
Our solutions place the moon at $a_s = 0.265_{-0.081}^{+0.123}$, 
$0.275_{-0.095}^{+0.126}$, and $0.258_{-0.081}^{+0.111} \; R_{Hill}$ 
for the linear, quadratic, and exponential models, respectively.

However, satellites on inclined orbits may be less stable over the long
term. Our three moon solutions all suggest significant inclinations
with respect to the planet's orbital plane,
so we carry out a stability analysis using the analytical 
formula provided by \cite{donnison:2014}. Drawing from our joint model
posteriors we find the moon solution has a stable configuration for $73.1\%$ of the draws
with the linear model, $73.6\%$ of the time for the quadratic model,
and $78.3\%$ of the time for the exponential model. 

\subsubsection{Analysis of Residuals}

As an extra check, we here describe an analysis of the light curve residuals.
Strong time-correlated noise structure could potentially explain the moon-like
dip without invoking a satellite. In total, we considered six sets of
relevant residuals, the three trend models applied to the moon model, M,
and the same three applied to the zero-radius moon model, Z. These are shown
in Figure~\ref{fig:residuals}.

The M model residuals appear consistent with Gaussian noise, displaying no
obvious trends or time-correlated structure. In contrast, the Z models
consistently fail to explain the moon-like dip leading to a noticeable
excursion in the residuals at this time.

We binned the residuals into progressively larger bins and monitored the
effect this has on the root mean square (RMS). Gaussian residuals would
be expected to bin as $\sigma_n = \sigma_1 n^{-1/2} \sqrt{m/(m-1)}$,
where $n$ is the number of points binned, $m$ is the number of bins and
$\sigma_1$ is the RMS of the unbinned data.

Setting $\sigma_1$ to our standard photometric error, derived from our
hook correction procedure, reveals that this curve yields a close match
to the M model residuals (see lower panels of Figure~\ref{fig:residuals}).
If anything, the noise appears to behave even better than Gaussian suggesting
that we have slightly overestimated our standard uncertainty value.

In contrast, all three Z models show excess power at large bin sizes,
indicative of time-correlated structure in the residuals. This structure
must be caused by the moon-like dip since models M and Z are otherwise
identical.

In summary, our analysis of the residuals adds weight to the case that
the moon-like dip is a real feature in the data.

\subsubsection{Chromatic test}
\label{fits:colortest}

In Section~\ref{obs:isthemoonreal}, we discussed how inspection of the
comparison star's photometry and the centroids revealed no evidence to
suspect the moon-like dip is an instrumental artifact. A third test,
mentioned in that section, was to consider if the dip was chromatic
in nature. An exomoon transit might be expected to be slightly chromatic
due to the atmosphere, but large transit depth variations would be
indicative of a blend or some previously unexplained instrumental effect.
Accordingly, we here describe a brief investigation into this possibility.

Already we have discussed how spectral elements can be extracted by dividing
the white light curve aperture up into slices (see Section~\ref{obs:spectrum}).
As a result of dividing up the aperture in this way, the noise naturally
increases in each channel. For the sake of testing if the moon-like dip is
chromatic or not, we only require two different colors, allowing us to use
broader slices with higher precision. Accordingly, we decided to split up
the optimal aperture into two colors for a chromatic test.

When divided into ten slices, we found that the reddest and bluest channels
displayed less stable photometry, likely as a result of the sharp drop off
in WFC3's sensitivity towards the wavelength extremes. For this test then,
we elected to omit those most extreme channels and ultimately took the
central two-thirds of the aperture, split into two, as our two colors. This
led to a ``blue'' channel defined from columns 476 to 518 (inclusive) and
a ``red'' channel from 519 to 561, which corresponds to 1.2-1.4$\mu$m and
1.4-1.6$\mu$m respectively.

We first independently correct for the hooks using the same method as
described in Section~\ref{obs:hooks}. The white light curve led to a final
inter-orbit RMS of 375.5\,ppm, the blue and red channels here yielded
598.8\,ppm and 612.8\,ppm respectively. This is slightly better than the
naive $\sqrt{3}$ scaling one would expect if the product of the source and
WFC3 had a perfectly flat wavelength response, reflecting how the response
actually peaks in the center and drops off towards the edges.

With the chromatic photometry in hand, our objective is now to test whether
each channel is more consistent with either the model including or excluding
an exomoon transit; formally models M and Z.
To accomplish this, we took the maximum \textit{a posteriori} solution from model M
and plotted the light curve morphology with the trend parameters turned
off, thus creating a ``template'' of the moon solution. We then multiplied
this template by a trend model with unset trend parameters. These unknown
trend parameters were then fit by regressing (least squares) the
template multiplied by the trend model to the blue/red channel photometry.
By doing this, we allow for the fact that the visit-long trend model is
itself chromatic. Since there are three possible trend models considered in
this work, we tried all three and selected the best one, since it was clear
sometimes a particular trend model was a poor representation of the data.
It is interesting to note that the quadratic trend won out every time.

One complication with the above is that there is not a single moon model template,
but rather three - since we originally attempted three trend models and have
been unable to definitively select a single model. We therefore repeated the
above the templates resulting from M + linear, M + quadratic and M + exponential
trends. This was done for both the blue and red channel data, and in each case
we recorded the (six) resulting $\chi^2$ values of the (three) trend-regressed
templates. We then repeated the entire excercise using the Z model templates,
which do not allow for an exomoon transit. The results, including a comparison
of the $\chi^2$ values, is shown in Figure~\ref{fig:chromatic}.

In all six cases, the moon template provides a closer match to the red/blue
channel data than the Z model. Although the moon model appears favored in
every case, the red channel does appear to yield a more significant preference.

In general, one should expect the red channel to be less affected by stellar
activity, such as rotating star spots, plages, granulation and micro-flares.
If the blue channel, which is more sensitive to such effects, were to exhibit
a stronger preference for a moon-like dip, this could have been a cause for
concern. As it stands, however, the results appear to be consistent with an overall
preference for model M over Z.

We also fitted the two channels with independent Z and M models using the same
procedure as for the white light curve. In all six cases (three trends, two
channels), the moon model is favored over model Z with evidences of
$1.51\pm0.31$, $1.84\pm0.31$ and $3.78\pm0.31$ for the blue channel
and $6.68\pm0.32$, $2.59\pm0.32$ and $4.58\pm0.31$ for the red, where the
three numbers correspond to $2\log(\mathcal{Z}_M/\mathcal{Z}_Z)$ for the
linear, quadratic and exponential trends, respectively. We do note that the
solutions are less well-converged than before, as a result of the reduced
precision, with credible intervals inflating by up to a few times.

In conclusion, the moon-like dip occurring post-egress is statistically
favored in both the red and blue channels and appears consistent
with the solution derived from the white light curve alone.

\subsubsection{Predictions}

As noted in the main text, we consider that future observations will be
critical in assessing the true nature of Kepler-1625b and the reality
of the exomoon signature. To this end, we took 100 random samples from
our model M posteriors and used them to predict the morphology of the
next transit event in May 2019 (epoch 9). This prediction is shown in Figure~\ref{fig:future}.
The model suggests a high likelihood of observing a pre-transit feature 
due to the exomoon and thus we strongly encourage observations around this time.


%% file: tables/keplertable.tex
\begin{table}
\centering
\caption{
\textbf{Kepler-only fits}. P for planet model. T for planetary TTV model. Z for a zero-radius moon model. M for moon model.
}
\vspace{2mm}
\begin{tabular}{lll} 
\hline
\textit{model} & $\log\mathcal{Z}$ & $\log\hat{\mathcal{L}}$ \\ [0.5ex] 
\hline
$\mathrm{P}$ & $4924.84 \pm 0.07$ & $4950.26$ \\
\hline
$\mathrm{T}$ & $4924.38 \pm 0.08$ & $4954.38$ \\
\hline
$\mathrm{Z}$ & $4927.53 \pm 0.08$ & $4956.42$ \\
\hline\hline
$\mathrm{M}$ & $4925.34 \pm 0.08$ & $4959.59$ \\
\hline\hline
$\mathrm{M} - \mathrm{P}$ & $0.50 \pm 0.11$ & $9.33$ \\
$\mathrm{M} - \mathrm{T}$ & $0.96 \pm 0.11$ & $5.22$ \\
$\mathrm{M} - \mathrm{Z}$ & $-2.19 \pm 0.11$ & $3.17$ \\ [1ex]
\hline 
\label{tab:kepler_evidences}
\end{tabular}
\end{table}

%% file: tables/spectrum.tex
\begin{table}
\centering
\caption{\textbf{Transmission spectrum.}
Marginalized ratio-of-radii derived from a Bayesian model averaged joint-posteriors
of the linear, quadratic and exponential HST detrending models, using the
averaged (``AVG'') \kepler\ data detrending model.
}
\vspace{2mm}
\begin{tabular}{llll} 
\hline
$\lambda$ & $\Delta \lambda$ & $R_P/R_{\star}$ & $\sigma_{R_P/R_{\star}}$ \\ [0.5ex] 
\hline
0.65  &	0.200 & 0.06102 & 0.0008 \\
1.135 & 0.027 & 0.06438 & 0.0015 \\
1.192 &	0.025 & 0.06198 & 0.0015 \\
1.252 &	0.027 & 0.05770 & 0.0017 \\
1.310 & 0.024 & 0.06157 & 0.0015 \\
1.368 &	0.028 & 0.05960 & 0.0014 \\
1.426 &	0.026 & 0.05990 & 0.0017 \\
1.485 &	0.027 & 0.05924 & 0.0019 \\
1.543 &	0.023 & 0.06019 & 0.0020 \\
1.600 & 0.027 & 0.05910 & 0.0026 \\ [1ex]
\hline 
\label{tab:spectrum}
\end{tabular}
\end{table}

%% file: tables/ttvtable.tex
\begin{table}
\centering
\caption{\textbf{Transit timings.}
Marginalized transit times derived from a Bayesian model averaged joint-posteriors
of the linear, quadratic and exponential HST detrending models, using the
averaged (``AVG'') \kepler\ data detrending model.
}
\vspace{2mm}
\begin{tabular}{lll} 
\hline
Epoch & $\tau$ & O-C [mins] \\ [0.5ex] 
\hline
-2 & $55469.2037_{-0.0049}^{+0.0048}$ & $-21.9_{-7.1}^{+7.1}$ \\
0 & $56043.9715_{-0.0040}^{+0.0040}$ & $+15.9_{-5.8}^{+5.6}$ \\
1 & $56331.3337_{-0.0051}^{+0.0049}$ & $+3.7_{-7.2}^{+7.0}$ \\
7 & $58055.5563_{-0.0014}^{+0.0013}$ & $-0.4_{-1.8}^{+1.7}$ \\ [1ex]
\hline 
\label{tab:ttvs}
\end{tabular}
\end{table}